\documentclass[10pt,a4paper]{book}

\usepackage[dvips]{graphicx}
\usepackage{amsmath,graphics,subfigure,float}
\usepackage{bm,setspace,layout,color}
\begin{document}

\begin{titlepage}
\begin{center}

$\left. \right.$ \vspace{3cm}

\textsc{\LARGE Quantum Effects In Low Temperature Bosonic Systems} \\[3cm]

\textsc{\Large Jose Reslen} \\[3cm]

\textsc{\LARGE University College London} \\[3cm]

\textsc{\Large Ph.D. in Physics} \\[3cm]

\textsc{\Large \today} 

\end{center}

\newpage

\begin{center}
I, Jose Reslen, confirm that the work presented in this thesis is my
own. Where information has been derived from other sources, I confirm
that this has been indicated in the thesis.
\end{center}

\newpage

\textsc{\Large Thesis Abstract} \\[0.5cm] 

In the first part, we investigate the effect of long range particle exchange
in ideal bosonic-chains. We establish that by using the Heisenberg 
formalism along
with matrix product state representation we can study the evolution as well as
the ground state of bosonic arrangements while including terms beyond
next-neighbour hopping. The method is then applied to analyse the quench
dynamics of condensates in a trapping potential and also to study the emergence
of entanglement as a result of collision in boson chains.
In the second part, we study the ground state as well as the dynamics of
1D boson-arrangements with local repulsive interactions and nearest-neighbour
exchange using numerical techniques based on time evolving block decimation
(TEBD). We focus on the development of quantum correlations between the
terminal places of these arrangements. We find that long-range
entanglement in the ground state arises as a result of intense boson
tunnelling taking place across the whole chain in systems with appropriate
hopping coefficients. Additionally, we identify the perturbations
necessary to increase the entanglement between the end sites above their 
ground state values. In the final part, we study the wave function of a 
kicked condensate using a perturbative approach and compare the results 
obtained in this way with numerical simulations.

\newpage

\textsc{\Large Acknowledgements} \\[0.5cm] 

Most of the work has been supervised by Prof. Sougato Bose. 
The results shown in chapter \ref{cucu} were produced as part of 
a collaboration project with Prof. Tania Monteiro and Dr.
Charles Creffield. This Ph.D. has been funded by an EPSRC 
Dorothy-Hodgkin-Postgraduate-Award scholarship. The original
manuscript of the thesis was greatly improved by the
observations of the examiners, Prof. Andrew Ho and Prof.
Jacob Dunningham.

\newpage

\textsc{\Large Frequently Used Abbreviations} \\ 

EEE = End-to-End Entanglement \\
BH = Bose-Hubbard \\
PTH = Perfect Transmission Hopping \\
CH = Constant Hopping \\
MPS = Matrix Product States \\
TEBD = Time Evolving Block Decimation \\[0.5cm]

\textsc{\Large Note About Units}  

Throughout this work we measure energy in units of the
recoil energy $E_R$, an energy reference very common in
optical lattice experiments. In most cases, we explicitly
indicate the energy units, otherwise, it should be
assumed that energy in being measured in terms of the
recoil energy. Similarly, we use the dimensionless
parameter $\frac{t E_R}{\hbar}$ as a measure of time.  
We have chosen not to give units to variables that
represent imaginary time, because such variables do not
have a direct physical meaning. They are given in
arbitrary units.\\[0.5cm]

\textsc{\Large Note About Graphs} 


In all our simulations of bosonic chains, we always consider
chains of unit filling, that is, the number of bosons
is equal to the number of sites.

\end{titlepage}

\tableofcontents

\listoffigures


\chapter*{Introduction}

\addcontentsline{toc}{chapter}{Introduction}

Weakly interacting systems can be studied using few-component models, which
describe the physics of a small number of particles isolated from their environment.
Such an approach has been successful in reproducing, thoroughly or partially,
a large variety of physical phenomena studied since the establishment of
quantum mechanics over the past century. However, particles in real
systems, specially strongly correlated systems, interact with each other and develop 
a long-scale coherence that causes deviations from the predictions of simple models. 
As a result, understanding the physics of highly correlated systems has become the 
focus of contemporary physics.

Bosons and fermions obey different statistical properties
that determine the behaviour of compound systems, specifically, bosons
can occupy the same quantum level while fermions cannot. Nowadays, 
the tremendous sophistication of cooling
techniques in optical lattices allows a closed-form study of atomic and
molecular systems in combination with optical interactions. As a result,
physicists have been able to probe not only single particle physics in
weakly interacting phases, but also the arising and taking over of
highly correlated states of matter. This has prompted a major interest
in strongly interacting systems whenever the resources to observe
many-body effects under controlled circumstances are now available using
state-of-the-art technology, which, at the same time, has revolutionized 
the way as scientists approach both theory and experiment. Indeed, 
while a couple of decades ago the characteristics of the sample under 
study depended almost entirely on its inherent physical composition, 
today it is possible to create samples with desired properties and 
characteristics using optical lattices. One of the most outstanding 
achievements in experimental physics occurred in 1995 with the observation of 
boson condensation in optical latices of ${}^{87}Rb$ and ${}^{23}Na$ reported
by Anderson {\it et al.} \cite{BEC1} and Davis {\it et al.} \cite{BEC2}
respectively. Such observations have since been the subject 
of intense theoretical and experimental investigation. Certainly, the reason
for this growing interest is twofold. On the one hand, people are interested
in practical applications, while on the other hand, there is a compelling
desire to scrutinise the canonical framework that sustains contemporary physics. 
It is because of these reasons, and others which we will point out further ahead, 
that we have opted for concentrating on bosonic models.

Systems of bosons display several unique characteristics, but it was the
observation of superfluidity in ${}^4He$ that ultimately triggered the scientific
desire to understand the physics behind the many-body effects of bosonic
systems. Ever since, superfluidity has been explained in terms
of the Bose-Hubbard (BH) model, which describes a system of bosons with
hopping and repulsion. An alternative version of the BH model, known as 
the Hubbard model, can be used to study fermions, but here we focus
on the BH model unless otherwise stated. Notably, it has
been recently shown by Ho {\it et al.} \cite{Ho} that the phases of the Hubbard 
model can be simulated by a version of the same Hamiltonian with attractive, 
rather than repulsive interactions. On the other hand, it was 
Fisher {\it et al.} \cite{Fisher} in 1989 who first unified
the fragmented knowledge available at the time and established a consistent
theoretical framework for the BH model. In that work, the
BH model is studied both in the absence and presence of disorder, and 
parametric phase diagrams depicting the different phases of the model are
discussed. The standard BH model shows two basic phases, the Mott insulator
and the superfluid. The first is characterized, among other features, by the
existence of an energy gap. The transition from insulator to superfluid is
found to be mean field in character and universality properties are also
discussed. Conversely, it is found that in the presence of disorder a new 
phase in between the insulator and the superfluid exists. This is called the
Bose-glass and is similar to the Mott insulator, but has no gap. 
Importantly,
it is argued that in the presence of disorder the transition to the superfluid 
is always from the Bose-gas, and never directly from the Mott insulator. 
This work settled the basis for future approaches to the BH model, but
the lack of numerical methods and additional experimental applications
prevented further advancements for a short time, even though additional 
investigations were carried out in 
subsequent years. It was not until 1999 with the article of 
Jaksch {\it el at.} \cite{Cirac}
that an experimental proposal to verify the BH model in optical lattices
was published. It was shown that both the Mott insulator as well as the 
superfluid regimes were reachable in optical lattice experiments. Three years later
the transition in a gas of ${}^{87}Rb$ was observed in the experiment of 
Greiner {\it et al.} \cite{Bloch}, using cooling techniques previously
employed in Bose-Einstein condensation. In this experiment the phases
were identified from absorption images after ballistic expansion of 
atoms. Certainly, while a gas in a Mott insulator phase projects a 
Gaussian distribution with non-visible coherent features, images from the
superfluid gas display fringes, which are interpreted as a signature of
Bose-condensation of momentum wave-functions. A variety of experiments
has been taking place ever since the presentation of this pioneering work.
Here we would like to mention the work by Stoferle {\it et al.} \cite{Stoferle}, 
where the phases of the gas are probed spectroscopically. Once the gas has been
trapped and cooled in a magneto optical trap (MOT), an optical excitation is
sent through the sample in the form of a shaking optical potential. 
The absorption profile is worked out from the absorption images after
ballistic expansion. In this way, the Mott insulator can be identified 
from the peaks of the absorption profile, which indicate a coincidence
between the energy of the optical excitation and the insulator gap.
Curiously, absorption images from the superfluid phase indicate 
optical absorption stronger than the one observed in the Mott phase.
This unexpected behaviour does not match the phase diagram of the
Hubbard-model, since the superfluid is essentially gapless. This
may indicate that strong many-body effects supersede particle tunnelling
in sections of the phase diagram where the superfluid is due to exist. 
Likewise, the marked absorption profile has been recently used
to probe electromagnetically induced transparency (EIT) in Mott
insulators, as reported by Schnorrberger {\it et al.} in 
reference \cite{EIT}. Another interesting effect observed in
optical lattices is the condensation of fermions in the form of
Cooper pairs reported by Regal {\it et al.}\cite{Regal} and 
Bourdel {\it et al.} \cite{Bourdel}
in ${}^{40} K$ and ${}^6 L$ respectively. In these experiments 
the interaction among particles is varied using a Feshbach resonance,
which can be induced using a magnetic field applied directly on
the sample. In the repulsive regime, the fermionic atoms couple 
in dimers, forming weakly bounded bosonic-molecules that can undergo Bose-Einstein
condensation. In the attractive regime, on the other hand, fermions
couple in Cooper pairs, and then the pairs condense
in the lowest energy level. In this latter case experimental detection
is challenging as the fermionic nature of particles prevents ballistic
expansion, therefore alternative techniques are implemented.
In the same way, equally revealing experiments in optical lattices have been 
reported over the past years, a few of which we include in our 
references \cite{Chang,Spielman,Best,Fertig,Feld,Muller,Marek,Hayes}.

Simultaneously to the development of cooling techniques and the
increasing experimental efforts in optical lattices, there have been 
noticeable advancements regarding many-body numerical methods. As it 
is well known, the description of real quantum systems require 
exponentially large resources. As a result, the use of numerical approaches 
becomes essential. One of the main breakthroughs came in
1992 with the work of White \cite{White3,White4}, and the introduction of the
density matrix renormalization group (DMRG) method to calculate
the ground state of many-body systems. Since its
introduction, DMRG has been extensively applied, with diverse emphases
and enhancements, to the study of many-particle configurations in multiple 
scenarios and it is considered one of the most efficient and reliable 
numerical methods to date. Moreover, following the ideas underlying DMRG  
such as the description of the system using matrix product states (MPS),
another method was proposed by Vidal in 2003 to simulate both real and
imaginary time evolution of slightly entangled systems \cite{vidal1}. 
The method was later introduced as time evolving block decimation 
(TEBD) \cite{vidal2}. These works encouraged further research 
in the area in subsequent years such as implementations in infinity
systems, known as iTEBD \cite{vidal3}, the use of disentanglers to increase the 
simulation efficiency \cite{vidal4} and applications in two
dimensions using the so called multiscale entanglement 
renormalization ansatz (MERA) \cite{vidal5}, among other contributions by 
Vidal's group. Equally important studies have been carried out by the 
group of Verstraete {\it et al.} (\cite{Verstraete} and references therein),
who have utilized MPS to describe mixed states. They also have applied 
MPS to simulate the master equation and have introduced collateral 
methodologies based in what they call matrix product operators (MPO), 
in contrast to MPS. 
Similarly, in the work by Hartmann {\it et al.} \cite{Prior,Exact} it has been shown 
that in certain circumstances the use of Heisenberg operators has advantages 
over the usual approach. Additionally, a method based in DMRG that can be used
to simulate time evolution of many-body systems, therefore known as tDMRG, 
was introduced by White and Feiguin \cite{White} shortly after Vidal's 
seminal paper.

This development in numerical methods along with the increasing availability of
computing technology has provided the tools to explore highly correlated
systems. However, the application of such numerical methods to relevant
physical models is by no means straightforward. In fact, each problem 
possesses its own set of complications and handicaps. The first works 
regarding the use MPS-alike methods in
highly correlated systems often considered, among other systems, the Hubbard 
model (as for example in \cite{White1}), which
needs a supporting space smaller than the BH model. Not long ago
a complete numerical study of the BH model in one dimension
including next-neighbour interactions was undertaken by 
Kuhner {\it et al.} \cite{Kuhner,White2} using DMRG, but the
first application of TEBD in BH chains came with the work of 
Daley {\it et al.} \cite{Daley}, where the currents resulting from
a density gradient in a 1D bosonic arrangement are studied in the presence of an
impurity in the centre of the chain. The impurity works as a 
switch (transistor) that can be used to control the flux of particles in a process
that resembles the phase interference effect underlying EIT. The 
authors implemented an enhanced version of TEBD that uses the conservation
of the total number of bosons to improve the efficiency of the
simulation. The currents are characterized in terms of the system
parameters and very interesting results are shown, although the
authors report a number of sensitive issues regarding the behaviour
of TEBD, which seemed to reproduce inconsistent results under
specific circumstances. A similar approach was explored in the paper
of Hartmann and Plenio \cite{Hartmann}, but this time the currents
resulting from a difference in the phases of two adjacent 
chains are the focus of study. Namely, a chain  prepared in a 
Mott insulator state is connected to a chain of equal size prepared 
in the superfluid state. Particles migrate from the insulator towards 
the superfluid generating a bosonic current that is simulated using 
a symmetry-enhanced TEBD. Similarly, the method was used by Mishmash 
{\it et al.} \cite{Mishmash1,Mishmash2} to simulate the evolution of dark 
solitons in a chain of ultracold atoms. Interestingly, simulations show 
that soliton waves lose coherence and eventually vanish as a result of 
non-linear effects induced by the BH Hamiltonian. In a different
investigation, Muth {\it et al.} \cite{Muth} analyse the phase
diagram of the BH model in the presence of disorder using
both TEBD and iTEBD. TEBD has also been utilized by Mathey {\it et al.}
in reference \cite{Mathey} to identify supersolid phases in
1D boson-mixtures. Additionally, TEBD has been used to simulate 
the response of a bosonic system to a sudden displacement of the
confinement potential in the letter by Danshita and Clark \cite{Danshita1}.
In this work a first principle approach to the
experiment of Fertig {\it et al.} \cite{Fertig} was proved successful. A
similar paper by Montangero {\it et al.} \cite{DFazio} employed tDMRG to 
explain the observations of the same experiment.

As can be seen, TEBD has turned out to be a very useful tool in the study of
bosonic systems with a growing interest from the scientific
community to apply the method to diverse situations and scenarios. 
Motivated by such enthusiasm, in this work we have applied the method 
and its underlying ideas to the study of {\it entanglement} in boson
chains. As it is now well known, entanglement is the main resource
of quantum information processing (QIP) \cite{nielsen},  and as such 
it has received  much attention and analysis. In the context of 
many-body problems, entanglement has been  studied extensively in 
arrangements of two-level systems such as spins, qubits and 
fermions \cite{lorenzo,Bose1,Stockton,ResQui,Tsomokos}  
(for a complete
review of entanglement in many body systems, including bosonic
systems, see reference \cite{Review}), where the size of the local 
Hilbert space of each site is bounded by a small integer. 
Boson chains, however, do not necessarily offer the same 
advantage, as the associative nature of bosons demands a broader 
spectrum of states in order to fully characterize the quantum state. 
Studies of entanglement in bosonic systems have 
focused on diverse kinds of entanglement \cite{Moeckel,Tian}, 
but here we focus specially on 
the entanglement shared between the end sites of the chain.
It has been already demonstrated by Campus-Venuti {\it et al.} \cite{lorenzo} 
and Eisert {\it et al.} \cite{Eisert} that distant places of a quantum chain 
can be entangled without the need for a direct interaction among them.
In addition, the problem of calculating the entanglement between the terminals of
a boson chain has been attacked before in the works of
Romero-Isart {\it et al.} \cite{Oriol}, where the dynamics is reducible to 
a single-particle propagation, and Plenio {\it et al.} \cite{Plenio} using
Gaussian states. Conversely, the situations analysed here include
a number of effects that do not allow the reduction of the
problem as mentioned just before, and therefore the use of TEBD 
becomes essential. We found that in the standard BH model the 
entanglement between distant places of the chain decreases as 
the chain-length augments, but we also found that the same entanglement
can be made to increase using a slightly modified version of
the Hamiltonian with variable, rather than fixed hopping constants.
Our results are explained in terms of the tunnelling profile displayed
by the particles along the chain. We argue that in the case where a 
form of strong 
and resilient entanglement arises, the chain undergoes a special kind of 
fluidity enhancement in which particle tunnelling takes place across the
whole length of the chain and not only inside localized clusters
of sites. We hope that our results provide a basis upon which
further advancements could be made in the same direction.
In this document we also discuss how to implement 
alternative numerical methods based on MPS that can be used in a 
variety of situations where the Hamiltonian is sufficiently regular to 
allow an explicit solution of the Heisenberg equations of motion
for the operators. The methodology proposed is then employed to
simulate the propagation of bosons with emphasis on
the amount of entanglement generated during the evolution. 
As a complement to our studies of the Hubbard model we also
present an analytical development of a system consisting of atoms under
the action of a periodic kicking. In this part we use the Gross-Pitaevskii
equation to generate the dynamics of atoms, which are considered
as a single cloud and not individually as in the BH model. In a
sense, we can say that this study has been inspired by the works of 
Zhang {\it et al.} \cite{Raizen1} and Liu {\it et al.} \cite{Raizen2}, 
where basically the same phenomenon is analysed following the experimental 
realization of Moore {\it et al.} \cite{Moore} in ultracold atoms. 

From our point of view, finding the conditions for the emergence
of entanglement between distant places of a boson chain is
an important contribution, not only because long range correlations are 
crucial in quantum information protocols, but also because such
correlations give insight about the system phenomenology. Another
merit of the work is the difficulty associated with some of the 
numerics that we present below, especially those regarding entanglement. 
This is because long range entanglement, especially
entanglement between extreme places, inevitably involves all the
intermediate degrees of freedom in between the boundaries, and so one
needs to synchronise a large number of correlated processes that 
derive from our use of MPS instead of a number-of-particles basis. 
On the other hand, 
we consider that the method that we introduce further 
ahead will prove useful in the study of dynamical models as it
does not depend so heavily on the amount of entanglement in the
system as the same TEBD or DMRG, which makes it suitable to
perform long-time simulations, although our method does not apply 
to the same wide spectrum of problems as the previously 
mentioned do.

This document is organized as follows, in chapter \ref{entla} we include
a review of some of the concepts and methods that are to be used in 
subsequent chapters such as entanglement and matrix product states. 
Chapter \ref{imple} sketches our numerical approach and discusses 
several programming issues that proved sensitive while coding
our algorithms. Chapter \ref{recho} deals with the application of MPS in
circumstances where the regularities of the Hamiltonian allow 
an elegant application of this representation. Chapter \ref{mede} 
introduces the BH Hamiltonian and shows how TEBD works specifically 
for this model. In chapter \ref{cart} we apply the method to find the 
ground state as well as the dynamics of the BH model under different 
circumstances. A number of schemes are proposed and analysed. We then 
proceed to present an analysis of the dynamics of cold atoms driven by 
a periodic kicking in chapter \ref{cucu}. After this, we summarize and 
present our conclusions.


\chapter{Preliminary concepts}

\label{entla}

\section{Schmidt decomposition}

Given a {\it pure} quantum state of a system made up of
many individual components, it is possible to write the
quantum ket of the system as a sum of product of states 
in the following manner \cite{nielsen},

\begin{equation}
|\psi \rangle = \sum_i \lambda_i |\nu_i \rangle |\mu_i \rangle,
\label{Schmidt}
\end{equation}

where kets $| \nu_i  \rangle$ and $| \mu_i  \rangle$ form 
orthonormal sets of vectors corresponding to complementary
subspaces. Namely,

\begin{eqnarray}
\langle \nu_i | \nu_j \rangle = \delta_i^j, \nonumber \\
\langle \mu_i | \mu_j \rangle = \delta_i^j, \nonumber \\
\langle \nu_i | \mu_j \rangle = 0,
\end{eqnarray}

where $\delta_i^j$ is the Kronecker delta.
These kets are known as the {\it Schmidt vectors}. Similarly, 
the set of $\lambda_i$ are the {\it Schmidt coefficients}. 
The Schmidt coefficients are positive real numbers that 
satisfy the condition,

\begin{equation}
\sum_i \lambda_i^2 = 1.
\end{equation}

The process of writing the state as in equation (\ref{Schmidt})
is known as the {\it Schmidt decomposition}. Very important
consequences follow from the Schmidt decomposition. For instance,
it can be shown that the reduced density matrices describing the
states of complementary subsystems share the same set of
eigenvalues, which are equal to the square of the Schmidt 
coefficients. Similarly, there are two important characteristics 
that we want to remark. 
First, the Schmidt decomposition depends on how the original system
is divided into complementary subspaces. This division may or
may not be related to the actual geometry of the system.
Second, the Schmidt decomposition for a given partition is
in general {\it not unique}. For instance, given a particular
set of Schmidt vectors we can obtain a different set of 
Schmidt vectors for the same partition by applying unitary
operations on the original vectors. This however does not
produce any change on the coefficients, which remain as
positive numbers. In a sense, the topology of the decomposition
is preserved after unitary operations, but this is only true
when such operations take place in the subspaces that support
the Schmidt vectors. One way of getting the Schmidt vectors
is by simple inspection. Obviously, this is not at all practical 
when we are dealing with intricate quantum states. The standard
method to get Schmidt vectors is by diagonalizing the reduced
density matrices corresponding to the subspaces defined by
the partition. As we will see, this property is at the heart of
the numerical techniques that we will use to study our models.

\section{Entanglement characterization}

Even though the fundamental concepts of quantum mechanics were already well 
established and accepted by the scientific community by 1930,
it has not been until recent times that the concept of entanglement has
started to receive attention. Among other reasons, it is because nowadays 
people are 
quite interested in what aspects of the physical systems are purely
``quantum''. 
Entanglement is a characteristic associated exclusively with quantum states. 
Classical representations of physical systems do not contain any form of 
entanglement whatsoever. It is for this reason that entanglement provides a 
measure of how efficient a physical system can be and how much of the 
state provides quantum resources that can be potentially used. 
This is why quantum entanglement is now a new branch of physics with a 
promising future. 
Quantum entanglement is the main resource of several highly efficient tasks
proposed in QIP such as superdense coding, quantum state teleportation
and quantum cryptography. Quantum entanglement is also at the heart
of the quantum computer. In many senses, the quantum computer can be
more efficient than its classical parallel. Let us take for example
one of the simplest tasks of a computer: generate random numbers.
Notably, this apparently simple operation carries some difficulties
for a classical computer, as such a machine is essentially deterministic. 
Usually, random distributions are generated using recursive algorithms 
that always introduce deviations. For a quantum computer, on the other
hand, the generation of a random distribution would result naturally by
performing measurements over an equivalent state superposition. This
simple example illustrates the usefulness of quantum states in
a practical scenario. Furthermore, it is the concept of entanglement
what captures the degree of utility of a quantum state.
As a matter of fact, entanglement has proved to be
a rich field of theoretical investigation from which very interesting
results have been derived \cite{Jacob1,Jacob2,Virmani}.
The progress in experimental physics has been interesting but not equally
dynamic, although a number of experiments involving entangled states have 
been carried out with relative success \cite{Marek}. 
So far, entanglement appears to be consistent with experimental 
observations, but practical implementations using entanglement as a 
resource remain challenging \cite{Hayes}.

Entanglement in a pure bipartite system can be consistently characterized from 
the reduced density matrix of any of the component subsystems. Let us call 
$\hat{\rho}_A$ the reduced density matrix of subsystem $A$, $\hat{\rho}_B$ with 
an analogous meaning, then entanglement between subsystems $A$ and $B$ is given 
by the von Neumann entropy,

\begin{equation}
S = -tr(\hat{\rho}_A \log_2 \hat{\rho}_A)=-tr(\hat{\rho}_B \log_2 \hat{\rho}_B).
\label{VonNeumman}
\end{equation}

Furthermore, $S$ can be calculated directly from the eigenvalues of either 
density matrix,

\begin{equation}
S =  - \sum_i \lambda_i \log_2 (\lambda_i).
\label{marek}
\end{equation}

It can be verified that $S=0$ for a separable state. In fact, when the state
is separable the reduced density matrix contains only a single eigenvalue
equal to one so that the logarithmic function in (\ref{marek}) causes 
the whole expression to vanish. 
It can also be verified that the maximum value displayed by $S$ is
$\log_2[d]$, where $d$ is the dimension of the smallest subsystem. Von Neumann
entropy provides a reliable estimation of the amount of entanglement shared 
between subsystems $A$ and $B$ as long as the whole $AB$ system remains in
a pure state. In most cases von Neumann entropy is an operational criterion,
which means it can be calculated from the expression that gives the quantum
state. Such is the case when the state is given in terms of a discrete basis.
Then, in order to get the entanglement, the reduced density matrix must be 
found and $S$ is computed from the eigenvalues of such matrix. 
This procedure can be considerably difficult to apply when
the state is given in a continuous basis. In this case the equivalent of
finding the eigenvalues of the reduced matrix corresponds to solving a second order
differential equation. Therefore, it is
more convenient to use a criterion such as the trace of $\hat{\rho}_A^2$ which
can be obtained from direct integration. The amount of entanglement is in this
way given by how much the trace deviates from one. Similarly, more entanglement
criteria can be worked out, but $S$ is well established as the most consistent
measure of entanglement for bipartite pure states. Among the conditions that 
a good entanglement measure, say $E$, should satisfy, we find \cite{Virmani,Jacob3},

\begin{itemize}

\item $E$ must be positive

\item $E$ must be zero for separable states

\item $E$ must be invariant under unitary transformations performed on 
subsystems $A$ and $B$. Hence, we can understand why $S$ depends
entirely on the eigenvalues of the reduced matrices, precisely because the 
eigenvalues are invariant under unitary transformations. Additionally, 
because reduced density matrices of subsystems that correspond to 
complementary partitions of a pure general state share the same set of 
eigenvalues, $S$ is independent of which reduced matrix one chooses to 
make the calculation. This is not the case, however, when the state of 
the whole system is mixed, since the eigenvalues of reduced matrices 
obtained from reducing a bigger matrix may be different.
 
\item $E$ must not increase under local unitary operations and classical 
communications (LOCC). Actually, it can be shown that this property
implies the previous one in the absence of classical communications.
Such classical communications make reference to the sharing of 
information between parts $A$ and $B$ using classical resources 
such as telephones or computer networks.

\item There must be maximally entangled states. This implies that
$E$ establishes an ordering of elements in the Hilbert
space. This condition cannot be fully incorporated for
multi-particle entanglement-measures and therefore the statement
should be considered only in the bipartite realm.

\end{itemize}

The fact that reduced density matrices of mixed states do not 
share the same set of eigenvalues as in the 
case of pure states prevents a direct generalization of
results such as equation (\ref{VonNeumman}). One of the main 
breakthroughs in the subject of entanglement characterization is 
due to Peres and members of the Horodecki family \cite{Peres,Horodecki}
who simultaneously came up with the idea of {\em partial transpose}, 
which we now present. A general separable mixed state of systems 
$A$ and $B$ can be written as,

\begin{equation}
\hat{\rho} = \sum_{i} c_i \hat{\rho}_i^A \otimes \hat{\rho}_i^B,
\hspace{0.3cm} c_i \ge 0, \hspace{0.3cm} \forall i
\end{equation}

So, if we swap matrices $\hat{\rho}_i^B$ by their corresponding 
transpose matrices, the expression above would still be a valid 
density matrix since the transpose matrices of $\hat{\rho}_i^B$ are 
density matrices themselves. As a consequence, the new density matrix 
describing the whole system is a positive operator, that is, all its 
eigenvalues are positive. This operation can be generalized to the 
case when matrix $\hat{\rho}$ is not separable. In such circumstance
transposing subsystem $B$ would involve swapping the indices of the 
big density matrix, namely,

\begin{equation}
\hat{\eta}_{\{i_1,i_2\},\{ j_1,j_2 \}} = \hat{\rho}_{\{i_1,j_2\},\{ j_1,i_2 \}},
\end{equation}

where we have used $\hat{\eta}$ as the partial transpose of matrix 
$\hat{\rho}$. Crucially, in this case
we cannot argue that $\hat{\eta}$ is a valid density matrix and
a positive operator, therefore, showing that $\hat{\eta}$ is not positive
provides evidence that the state is entangled. Consequently, in order to
know whether the state is entangled, it suffices to get the partial
transpose and see if any of its eigenvalues turn out to be negative.
On the other hand, it is also good to comment that this criterion does
constitute a necessary rather than sufficient condition for entanglement.
It could be that $A$ and $B$ are entangled albeit $\hat{\eta}$ having
a positive spectrum. It is possible to take one step forward and formulate
a quantitative expression for entanglement from the partial transpose 
criterion. Vidal and Werner \cite{LogN} propose {\it Log-negativity}, which
bounds the amount of distillable entanglement in $\hat{\rho}$. By definition, 
distillable entanglement makes reference to the pure state entanglement that 
can be extracted from $\hat{\rho}$. 
Log-negativity is an extension of the simpler {\em negativity}, the sum 
of the negative eigenvalues of the transpose. Formally, Log-negativity 
can be written as,

\begin{equation}
E_N = log_2 \left( tr \sqrt {\hat{\eta}^{\dagger} \hat{\eta}} \right) = log_2 \left( 1 - 2 \sum_{\lambda_i < 0} \lambda_i  \right),
\label{logn}
\end{equation}

where $\lambda_i$ are the eigenvalues of $\hat{\eta}$. According to 
\cite{LogN}, Log-negativity is an additive quantity, which makes it
a very useful measure. Also, the fact that it is given in terms of
the logarithm function allows a comparison with von Neumann entropy
in systems of zero mixture. In this case, it has
been shown that Log-negativity provides a greater estimation of
entanglement than von Neumann entropy. We want to emphasise that
von Neumann entropy can only be applied to measure the entanglement
of pure states. In this thesis, the entanglement of mixed states will 
be calculated using the definition Log-negativity presented above. 
In a sense, Log-negativity is a
quantification of the Peres-Horodecki criterion. One positive consequence
of this relation, is that when a quantum state is shown to be
entangled according to the Log-negativity criterion, then it can
be shown that the entanglement contained in such state can be
distilled, that is, it can be transformed into pure state entanglement
through a distillation process. As it has been mentioned before, 
Log-negativity sets an upper bound for the amount of distillable entanglement
of a given state. Hence, we can say this entanglement measure
establishes the degree of usefulness of a quantum state. Sometimes
it is interesting to see the relation between entanglement and
correlations. To this end, it is important to highlight that 
entanglement is a property associated {\it exclusively} with the
state, while correlations require the intervention of an observable.
It is widely accepted that entanglement is a resource more exotic than
correlations. There can be correlations without entanglement but
there cannot be entanglement without correlations. In the models
studied in this thesis, correlations play an important role in
defining the phases of the system. We explore the development
of entanglement in situations where correlations are known to
induce highly intricate states. As we will see, this can derive
in very entangled states that we try to characterize from
the physical features displayed by the system.

\section{Matrix product state representation}

\label{MPS}

Initially, when quantum mechanics began to be used as an accurate
theory capable of delivering insight in atomic systems, fixed 
bases sufficed to provide an operational platform to perform, most
of the time, analytical calculations. Even when Dirac introduced the
interaction picture, a method in which
basis kets and operators evolve according to the dynamics generated
by the integrable part of the Hamiltonian while the quantum state
evolves according to a non-diagonal term, it was intended to be
used mostly as a complement of the existing quantum pictures, namely,
Heisenberg's and Schrodinger's. The idea of a dynamical basis has
been brought recently, partially as a result of the insight obtained
in the process of understanding entanglement. Let us think of
dynamics as a unitary operation that is applied on the initial state.
If we want to make things easy and use a basis in which the evolved
state remains simple at all times we would rather employ a basis that
does not change, or changes little at least, when unitary operations
are applied on the quantum state. For someone who is familiar
with the formalism of entanglement, the idea of Schmidt vectors is 
likely to come to mind in this specific situation. Indeed, Schmidt vectors
do not change when unitary operations are applied on the
system. As a consequence, entanglement
between complementary subsystems is invariant under local
unitary operations. Nevertheless, the evolution operator acts
globally and therefore induce changes on every Schmidt vector along 
the system. However, when the evolution operator can be split, at
least approximately, into non-overlapping semi-local operators, the
idea of using Schmidt vectors as basis vectors becomes feasible.
Such is the case for those spin and boson chains where 
hopping takes place only among next neighbours. In such a scenario
every Schmidt decomposition determines a splitting of the chain. Schmidt
vectors describe the state of a subset of the chain. In general, the 
Hamiltonian can be written as,

\begin{equation}
\hat{H} = \sum_{k} \left( \hat{A}_{k,k+1} + \hat{B}_k \right),
\end{equation}

while the evolution operator is given by 

\begin{equation}
\hat{U} = e^{-i \delta t\hat{H} } \approx \prod_k e^{-i \delta t \left( \hat{A}_{k,k+1} + \hat{B}_k \right) } = \prod_k \hat {U}_k.
\label{uni}
\end{equation}

\begin{figure}
\includegraphics[width=1.\textwidth]{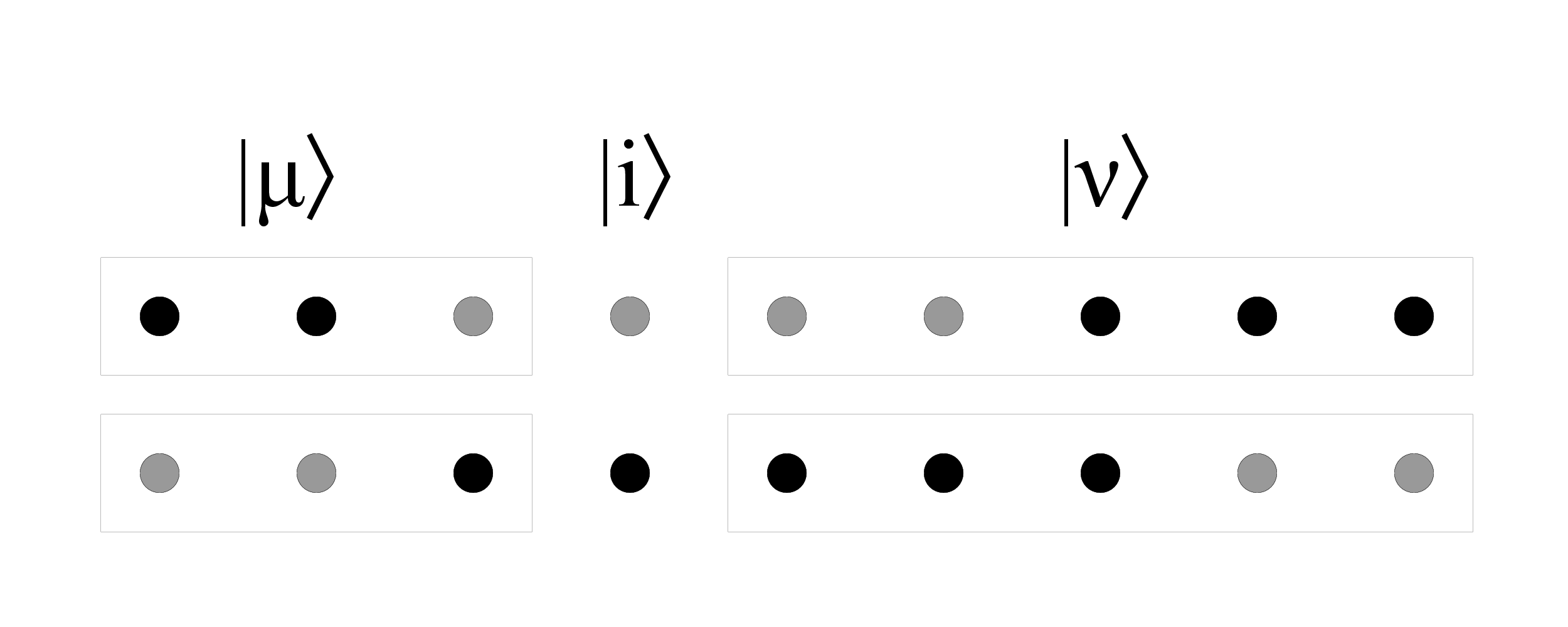}
\caption{\label{split_d} Example of MPS. $|\mu \rangle$ and $|\nu \rangle$  
are Schmidt vectors from the decompositions at the sides of place $i$.} 
\end{figure}

In this way, unitary operations applied on pairs of neighbour
sites do not affect all the Schmidt decompositions. This suggests that
Schmidt vectors can form a basis suitable for state description, one
in which the process of state updating is efficient. To see how Schmidt
vectors can be used, suppose that the chain is in a pure state.
To see how the state can be written in matrix product state (MPS)
representation, let us focus on one site of the chain. Initially,
in order to describe the state, we use, on the one hand, a local 
representation for such a site, say $|i\rangle$, given in a 
{\it standard basis}, for 
instance spin orientation or occupation number, on the other hand,
to describe the state outside this site, we use Schmidt vectors
obtained from splitting the system to both sides of the place in
consideration (figure \ref{split_d}). So, making explicit reference to
a local basis at site $n$ the quantum state reads,

\begin{equation}
| \psi \rangle = \sum_{\mu \nu i} \lambda_{\mu}^{[n-1]} \lambda_{\nu}^{[n]} \Gamma_{\mu \nu}^{[n]i} |\mu \rangle^{[1,n-1]} |i \rangle^{[n]} |\nu \rangle^{[n+1,N]}.
\label{kirko}
\end{equation}

In this way written, the quantum state appears as a superposition
of states in the basis of the Schmidt vectors and the local basis
of site $n$. We have deliberately specified the Hilbert spaces of
every ket as a superscript. Tensor $\Gamma_{\mu \nu}^{i[n]}$ contains the
components that describe the state in the new basis while coefficients
$\lambda_{\mu,\nu}$ are just the Schmidt coefficients, necessary in
this case to guarantee that vectors $|\mu \rangle$ and 
$|\nu \rangle$ are normalized.

One important property of this tensor decomposition is that
it is actually possible to establish a relation among
the $\Gamma$'s  and the Schmidt vectors via induction. Indeed,
vectors $|\nu \rangle$ can be expanded in terms of a basis
$|j\rangle^{[n+1]}$ and Schmidt vectors $|\xi \rangle^{[n+2,N]}$
in this way,

\begin{equation}
|\nu \rangle^{[n+1,N]} = \sum_{j\xi} \Gamma_{\nu \xi}^{[n+1]j} \lambda_{\xi}^{[n+1]} |j\rangle^{[n+1]} |\xi \rangle^{[n+2,N]} 
\label{korko}
\end{equation}

In fact, according to \cite{vidal1}, $|\mu \rangle$ and $|\nu \rangle$
can be written in terms of the state tensors associated 
with the subsets $[1,...,n-1]$ and $[n+1,...,N]$ and the standard
basis in the following form,

\begin{equation}
|\mu \rangle^{[1,n-1]} = \sum_{\alpha,...,\kappa} \left( \sum_{i_1,...,i_{n-1}} \Gamma_{1 \alpha}^{[1]i_1} \lambda_\alpha^{[1]}...\Gamma_{\kappa \mu}^{[n-1]i_{n-1}}  |i_{1},...,i_{n-1} \rangle \right),
\end{equation} 

and,

\begin{equation}
|\nu \rangle^{[n+1,N]} = \sum_{\gamma,...,\omega} \left( \sum_{i_{n+1},...,i_N} \Gamma_{\nu \gamma}^{[n+1]i_{n+1}}...\lambda_{\omega}^{[N-1]} \Gamma_{\omega 1}^{[N]i_N}  |i_{n+1},...,i_N \rangle \right),
\end{equation} 

where $N$ is the number of sites in the chain. Additional
relations can be derived, in particular we would like
to mention that the standard coefficients can be put in 
terms of the coefficients of the decomposition in the 
following way,

\begin{equation}
c_{i_1,i_2,...,i_N} = \sum_{\alpha,\beta,...,\omega} \Gamma_{1\alpha}^{[1]i_1}\lambda_\alpha^{[1]}\Gamma_{\alpha \beta}^{[2]i_2}...\lambda_\omega^{[N-1]}\Gamma_{\omega 1}^{[N]i_N}.
\end{equation}

These coefficients enable us to write the quantum state back
in the conventional basis,

\begin{equation}
| \psi \rangle = \sum_{i_1,i_2,...,i_N} c_{i_1,i_2,...,i_N} | i_1,i_2,...,i_N \rangle, 
\end{equation}

so that the representation of the state in terms of Schmidt coefficients
and tensors, to which we refer to as canonical decomposition, 
fully characterizes the quantum state. 

The canonical decomposition is very convenient
to study numerically the time evolution of the chain. Every time
that a semi-local unitary operation as in equation (\ref{uni}) 
is applied, the canonical representation is to
be updated only for the elements involved directly
with the transformation. For example, if a unitary
transformation is applied on sites $4$ and $5$, we
can see that for instance $\Gamma_{\mu \nu}^{[7]i_7}$
does not need to be updated, since the topology of the
decompositions associated with the tensor, namely 
those in which $|\mu \rangle^{[1,6]}$ and $|\nu \rangle^{[8,N]}$
are involved, is not affected by the operation. 
For the case of vectors $|\mu \rangle^{[1,6]}$ 
particularly, a unitary operation on sites $4$ and $5$ 
induce a change on every vector, but the new vectors are
Schmidt vectors of the evolved state with the same
Schmidt coefficients. So, the new Schmidt decomposition
preserves exactly the same configuration of elements with
exactly the same coefficients of the initial decomposition.
Therefore, no change takes place on the elements of the decomposition 
in places far away from where the semi-local unitary 
transformation operates. Following the same analysis one can
establish that a unitary operation on two neighbour sites
$n$ and $n+1$ generates changes only in $\Gamma^{[n]}$,
$\Gamma^{[n+1]}$ and $\lambda^{[n]}$. This property constitutes
the basis of efficient simulation: as the cost of updating the
state depends only on the local characteristics of the system,
numerical routines of low memory consumption and fast execution
can be implemented using the canonical decomposition. 
Crucially, the factor
that determines the speed of the simulation is the number of
Schmidt vectors in the single value decomposition. Systems with few 
vectors can be updated through relatively few
computational steps \cite{vidal2,Kollath}. Moreover, the
canonical decomposition is also known as matrix product
states (MPS) representation, and extensive documentation can
be found under this denomination \cite{Verstraete}. To see how the canonical
decomposition can be updated after a semi-local unitary
matrix operates, let us first focus on the simplest case where
a one-site unitary operation, $\hat{U}^{[n]}$, is applied on an
arbitrary site $n$.
In this case we can use the expression in equation (\ref{kirko})
to apply the transformation directly,

\begin{equation}
| \psi' \rangle = \hat{U}^{[n]} |\psi \rangle = \sum_{\mu \nu} \sum_{i} \lambda_{\mu}^{[n-1]} \lambda_{\nu}^{[n]}  U_i^{i'[n]} \Gamma_{\mu \nu}^{[n]i}  |\mu \rangle^{[1,n-1]} |i \rangle^{[n]} |\nu \rangle^{[n+1,N]},
\label{nightw}
\end{equation}

so that the effect of the unitary matrix is reflected only in the
coefficients $\Gamma^{[n]}$. The updated canonical coefficients are identical
to the originals, with only one exception,

\begin{equation}
{\Gamma'}_{\alpha \beta}^{[n]i} = \sum_{k} U_i^{k[n]} \Gamma_{\alpha \beta}^{[n]i}. 
\end{equation}

Furthermore, this transformation does not increase the number of elements 
necessary to describe the state and the updating procedure can be coded 
easily. 

Let us now assume that a two-site transformation, $\hat{U}^{[n,n+1]}$, is 
applied. In order to carry out such operation, the state must be written
with explicit reference to the local basis of sites $n$ and
$n+1$. This is done by inserting (\ref{korko}) in (\ref{kirko}),
which results in,

\begin{equation}
| \psi \rangle = \sum_{\mu \nu \xi} \sum_{ij}\lambda_{\mu} \lambda_{\xi} \Gamma_{\mu \nu}^{[n]i} \lambda_\nu \Gamma_{\nu \xi}^{[n+1] j}|\mu \rangle |i \rangle |j\rangle |\xi \rangle.
\end{equation}

Therefore, the updated state reads,

\begin{equation}
|\psi' \rangle = \hat{U}^{[n,n+1]}| \psi \rangle = \sum_{\mu \nu \xi} \sum_{ij}\lambda_{\mu} \lambda_{\xi} U_{ij}^{i'j'} \Gamma_{\mu \nu}^{[n]i} \lambda_\nu \Gamma_{\nu \xi}^{[n+1]j}|\mu \rangle |i \rangle |j\rangle |\xi \rangle.
\label{rora}
\end{equation}

As discussed before, this transformation induces changes in 
tensors $\Gamma_{\mu \nu}^{[n]i}$, $\Gamma_{\nu \xi}^{[n+1]i}$ and
$\lambda_\nu$ as a result of the changes generated in the
Schmidt decompositions of the splitting 
$[1,n-1]:[n,N]$ and $[1,n]:[n+1,N]$. Consequently, the next
step consists in finding the reduced density matrix of
subsystem $[n+1,N]$, from which new updated Schmidt vectors
can be obtained as eigenvectors:

\begin{equation}
\hat{\rho}'^{[n+1,N]} = tr_{[1,n]} |\psi'\rangle \langle \psi'|.
\end{equation}

As a result of the reduction, we are left with a density matrix 
spanned by the local basis of site $n+1$ and Schmidt vectors 
$| \xi \rangle$. Crucially, the cost involved in manipulating such 
density matrix, that is, storage and diagonalization, is 
proportional to the number of Schmidt vectors in the state. In 
this way, the astronomical memory requirements associated with
the standard basis can be avoided, as long as the number
of Schmidt vectors remains small. New coefficients ${\lambda'}_{\nu}$
can be identified as the square roots of the eigenvalues of the 
reduced density matrix, whose eigenvectors can be used to get the 
new tensor ${\Gamma'}_{\nu \xi}^{[n+1]j}$ directly from equation (\ref{korko}).
Finally, the new tensor ${\Gamma'}_{\mu \nu}^{[n]i}$ comes from
projecting the Schmidt vectors over the whole state given by
equation (\ref{rora}) \cite{vidal1,vidal3}.

In order to exemplify how the canonical decomposition can be
used to represent the state, let us focus on a system of three qubits.
Suppose that the state of the system is given by,

\begin{equation}
\phi = |110\rangle.
\end{equation}

Now, if we take the first qubit and consider the other two as the rest 
of the system, we can see there is only one Schmidt vector to the right, 
namely, $|\nu\rangle = |10\rangle$. Then, using the convention introduced in 
equation (\ref{kirko}) the state can be written as,

\begin{equation}
|\phi \rangle = \lambda_1^{[1]} \Gamma^{[1] 0}_{1,1} |0\rangle |\nu \rangle +	
\lambda_1^{[1]} \Gamma^{[1] 1}_{1,1} |1\rangle |\nu \rangle,
\label{loboh}
\end{equation}

with $\lambda_1^{[1]} = 1$ (the Schmidt coefficient) and, 

\begin{eqnarray}
\Gamma^{[1] 0}_{1,1} = 0, \nonumber  \\ 
\Gamma^{[1] 1}_{1,1} = 1. \nonumber
\end{eqnarray}

Note that the sub indices of $\Gamma^{[1]}$ make reference to the Schmidt 
vectors to the left and right of the first site. In this case there is only
one vector to the right while for the left we imagine there is an ancillary 
state. Similarly, for the second site we can see the Schmidt vectors to
the right and left are $|\mu \rangle = |1\rangle$ and $|\nu \rangle = |0\rangle$ 
respectively. The state can therefore be written as,

\begin{equation}
|\phi \rangle = \lambda_1^{[1]} \lambda_1^{[2]} \Gamma^{[2] 0}_{1,1} |\mu \rangle |0\rangle |\nu \rangle +	
\lambda_1^{[1]} \lambda_1^{[2]} \Gamma^{[2] 1}_{1,1} |\mu \rangle |1\rangle |\nu \rangle,
\label{sombrero}
\end{equation}

with,

\begin{eqnarray}
\Gamma^{[2] 0}_{1,1} = 0, \nonumber  \\ 
\Gamma^{[2] 1}_{1,1} = 1, \nonumber
\end{eqnarray}

and $\lambda_1^{[2]}=1$. Finally, the coefficients of the third site can be extrapolated from,

\begin{equation}
|\phi \rangle = \lambda_1^{[2]}  \Gamma^{[3] 0}_{1,1} |\mu \rangle |0\rangle  +	
\lambda_1^{[2]} \Gamma^{[3] 1}_{1,1} |\mu \rangle |1\rangle,
\label{picas}
\end{equation}

with,

\begin{eqnarray}
\Gamma^{[3] 0}_{1,1} = 1, \nonumber  \\ 
\Gamma^{[3] 1}_{1,1} = 0, \nonumber
\end{eqnarray}

and as the reader may have guessed, $|\mu\rangle = |11\rangle$. In the 
same manner, we can obtain the canonical decomposition of the following
less trivial state,

\begin{equation}
\phi = \frac{|001\rangle + |011\rangle}{\sqrt{2}} = |0\rangle \left\{ \frac{|0\rangle + |1\rangle}{\sqrt{2}}   \right\}  |1\rangle.
\end{equation}

This state can be written with explicit reference to the
coordinates of the first qubit exactly as in equation (\ref{loboh}),
but with,

\begin{eqnarray}
\Gamma^{[1] 0}_{1,1} = 1, \nonumber  \\ 
\Gamma^{[1] 1}_{1,1} = 0, \nonumber
\end{eqnarray}
 
and,

\begin{equation}
| \nu \rangle =  \left\{ \frac{|0\rangle + |1\rangle}{\sqrt{2}}   \right\}  |1\rangle.
\end{equation}

Similarly, the state also adopts the form of equation (\ref{sombrero}),
but with the following important changes,

\begin{eqnarray}
\Gamma^{[2] 0}_{1,1} = \frac{1}{\sqrt{2}}, \nonumber  \\ 
\Gamma^{[2] 1}_{1,1} = \frac{1}{\sqrt{2}}, \nonumber
\end{eqnarray}

and $|\mu \rangle = |0\rangle$ and $|\nu \rangle = |1\rangle$. Likewise, 
it can be seen that for the third site the coefficients of equation 
(\ref{picas}) are given by,

\begin{eqnarray}
\Gamma^{[2] 0}_{1,1} = 0, \nonumber  \\ 
\Gamma^{[2] 1}_{1,1} = 1, \nonumber
\end{eqnarray}

and the Schmidt vector is,

\begin{equation}
| \mu \rangle =  |0\rangle \left\{ \frac{|0\rangle + |1\rangle}{\sqrt{2}}  \right\}.
\end{equation}

More complex representations can be derived when there are partitions 
with more than one Schmidt vector in the decomposition.
We hope that these simple examples allow the reader to grasp an idea about
the mechanics associated with writing a quantum state using MPS.

\section{Perfect transmission hopping}

In a chain of bosons with constant chemical potential and zero repulsion 
the Hamiltonian can be written as,

\begin{equation}
\hat{H} =  \sum_{k=1}^{N-1} {J_k (\hat{a}^{\dagger}_{k+1} \hat{a}_k  + \hat{a}^{\dagger}_k \hat{a}_{k+1})}, 
\label{PBH}
\end{equation}

where $\hat{a}_k^{\dagger}$ and $\hat{a}_k$ are the usual bosonic
operators with standard commuting rules, namely,

\begin{equation}
[ \hat{a}_k,\hat{a}_l ] = 0, \hspace{0.5cm} \left [ \hat{a}^{\dagger}_k,\hat{a}^{\dagger}_l \right ] = 0, \hspace{0.5cm} \left [ \hat{a}_l,\hat{a}^{\dagger}_k \right ] = \delta_k^l,
\label{comm}
\end{equation} 

and $N$ is the number of sites in the chain. One question of interest 
in several branches of physics is
whether it is possible to dynamically transfer an arbitrary quantum
state from one end of the chain to the other with {\it maximum} 
fidelity. In other words, whether it is possible to have a
perfect transmission channel. This problem has been extensively
studied and here we limit ourselves to outline the main results of
more specialized investigations presented elsewhere 
\cite{Pater,Yung,PT,Nori,Clark}. 
Perfect transmission was first studied in spin chains due to the
interest prompted by the novel scheme shown in reference \cite{Bose1}. 
In this reference, spin chains were proposed as alternative channels to
transfer quantum information encoded in the state of the spins. 
In this context, it became important to know under which specific
circumstances a spin chain could transmit a state without the 
state being corrupted by the underlying dynamics. It turned out
that chains with constant coefficients could be used as efficient
transmission channels only in chains of maximum 3 sites. However,
it was also found that chains with variable hopping coefficients
could be made efficient if the right hopping coefficients were
chosen. Such coefficients could be identified by noticing the 
parallel between the dynamics of the chain and the physics of the
angular momentum. In this analogy, states on the
ends of the chain correspond to eigenstates of angular momentum
with large eigenvalues, while states in the centre turn out to
be analogous to eigenvectors associated with the smallest eigenvalues.
It was shown that in a system governed by 
Hamiltonian (\ref{PBH}) perfect transmission can be accomplished 
by choosing what we call perfect transmission hopping (PTH),

\begin{equation}
J_k^{PTH} = \frac{\lambda}{2}\sqrt{k(N-k)},
\end{equation}

where $\lambda$ is a constant that fixes the time scale (for simplicity
we chose $\lambda=2$ for numerical simulations). This stands
in contrast to the widely used constant hopping (CH), given simply by,

\begin{equation}
J_k^{CH} = 1.
\end{equation}

When inserted in Hamiltonian (\ref{PBH}), PTH induces a mirror-reflection of 
the initial state with respect to the chain centre at a time, 

\begin{equation}
t=\frac{T}{2}=\frac{\pi}{\lambda}. 
\end{equation}

This property constitutes the basis of perfect state transmission between 
the chain terminals. Independently of how many bosons initially occupy 
either chain end, they all turn up at the opposite end in a PTH chain with 
all repulsion constants set to zero \cite{Plenio}. 

Although the dynamics of a PTH chain with no repulsion has been
already studied, there are several research extensions of interest.
On the one hand, it is important to know if chains with variable
hopping coefficients and repulsion can display perfect transmission. 
If they cannot, it would be interesting to know how the repulsion
hampers the state transmission. Similarly, if transmission with
repulsion is possible, it would be important to know the specific
circumstances under which this phenomenon can be observed. Works
on this direction have produced very interesting results \cite{Nori}.
In addition to these research areas, here we also focus on one 
alternative approach. In fact, in sections ahead we intend
to see the problem from a rather different perspective. 
Certainly, it is quite valid to ask how PTH affects physical phenomena 
in which transport does not participate straightforwardly. From our
viewpoint, this constitutes an important aspect to study in chains
with efficient transmission, as the regularities of the Hamiltonian 
that lead to perfect transmission could give rise to interesting 
coherent processes. Similarly, it should be mentioned that very
promising experimental proposals to realize PTH in optical
lattices have been discussed in reference \cite{Clark} by the same
group that proposed the realization of the Hubbard model in
optical latices. Therefore,
such kind of hopping profile should be considered as a feasible
alternative and not just as a theoretical idealization.

\chapter{Implementation and numerics}

\label{imple}

The first aspect to be handled when addressing the construction
of a program using the formalism presented in previous sections, is memory
allocation. It is indeed possible to implement the algorithm
using standard programming tools, that is, vectors and
matrices with permanent memory attributes. However, the way
in which the elements of the canonical decomposition depend 
on the Schmidt vectors makes the canonical tensors vary
in size according to the number of Schmidt vectors in the
decomposition. Therefore, it is preferable to adopt a
programming style in which the dynamical nature of the
algorithm can be handled more appropriately. FORTRAN 95 offers
several tools that can be conveniently adjusted to approach
the dynamical nature of the method. On the one hand, it
allows one to actually define the geometry of the objects
employed to store data. This means that in addition
to vectors and matrices, tensors of whatever shape and
dimension can be used. For instance, we can define a vector
in which every component is a matrix, or a matrix in which
every element is made of a complex number and a real vector. 
The other tool is dynamical allocation. The elements that
we use to store numbers are not necessarily fixed in size, 
instead, we can make them bigger or smaller according to 
the simulation requirements. In FORTRAN 95, dynamical allocation
can be implemented using either allocatable objects or pointers.
Pointers work very much as allocatable objects, but they have
two useful additional properties. First, they can be alternatively
used as aliases of other objects, and second, they can be
passed from one routine to another as arguments. Modules
are another useful tool. Variables defined in a module can be 
used in any routine by just including the module in the routine's
headlines. In our program for example, the module {\it bskt} contains
the tensors of the canonical decomposition. In order to define an
object such as $\lambda_\alpha$, we must first generate the {\it type} 
that supports the object. This is done through the lines,

\begin{verbatim}
type dl
real(dp), dimension(:), pointer :: ld 
end type dl
\end{verbatim}

Tensor $\lambda_\alpha$, is then written as an element defined
by the rules of type  dl,

\begin{verbatim}
type(dl), dimension(N) :: LAM
\end{verbatim}

\begin{figure}
\includegraphics[width=1.\textwidth]{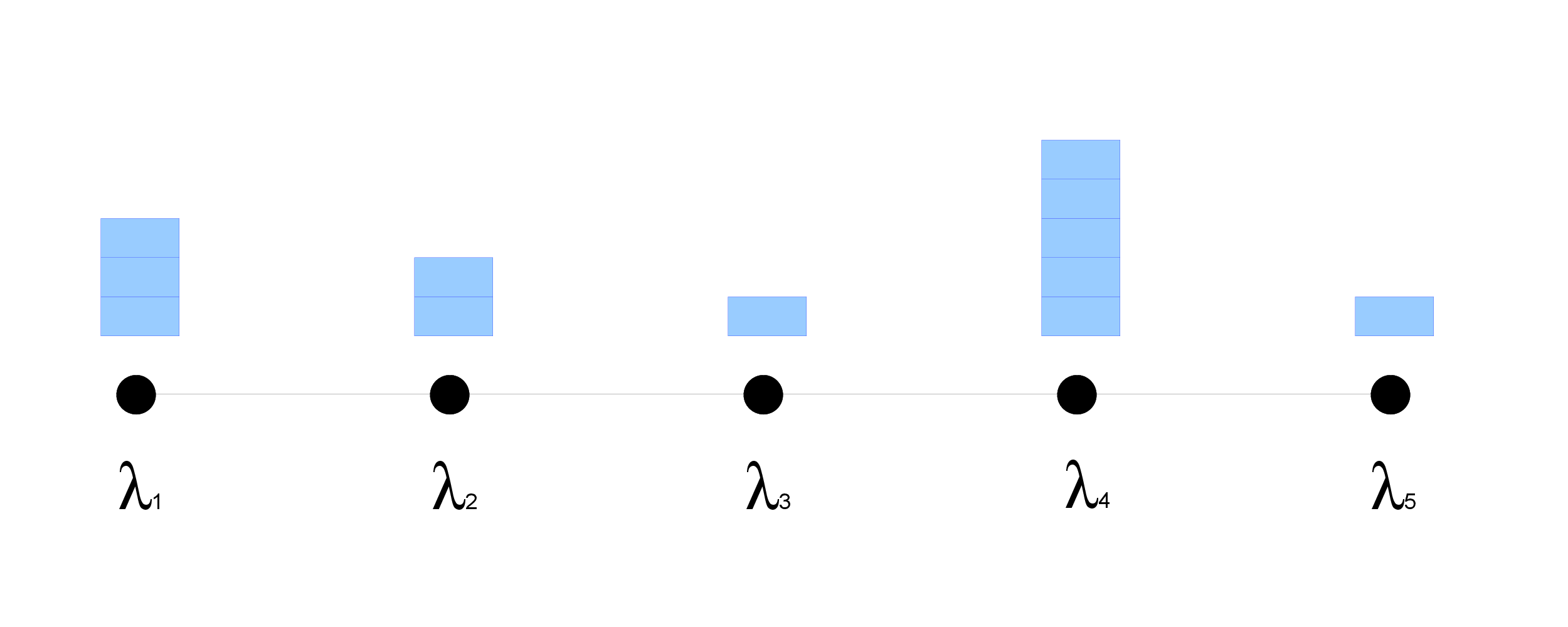}
\caption{ Sketch of memory allocation of LAM(i)\%ld} 
\label{lamb}
\end{figure}

\begin{figure}
\includegraphics[width=1.\textwidth]{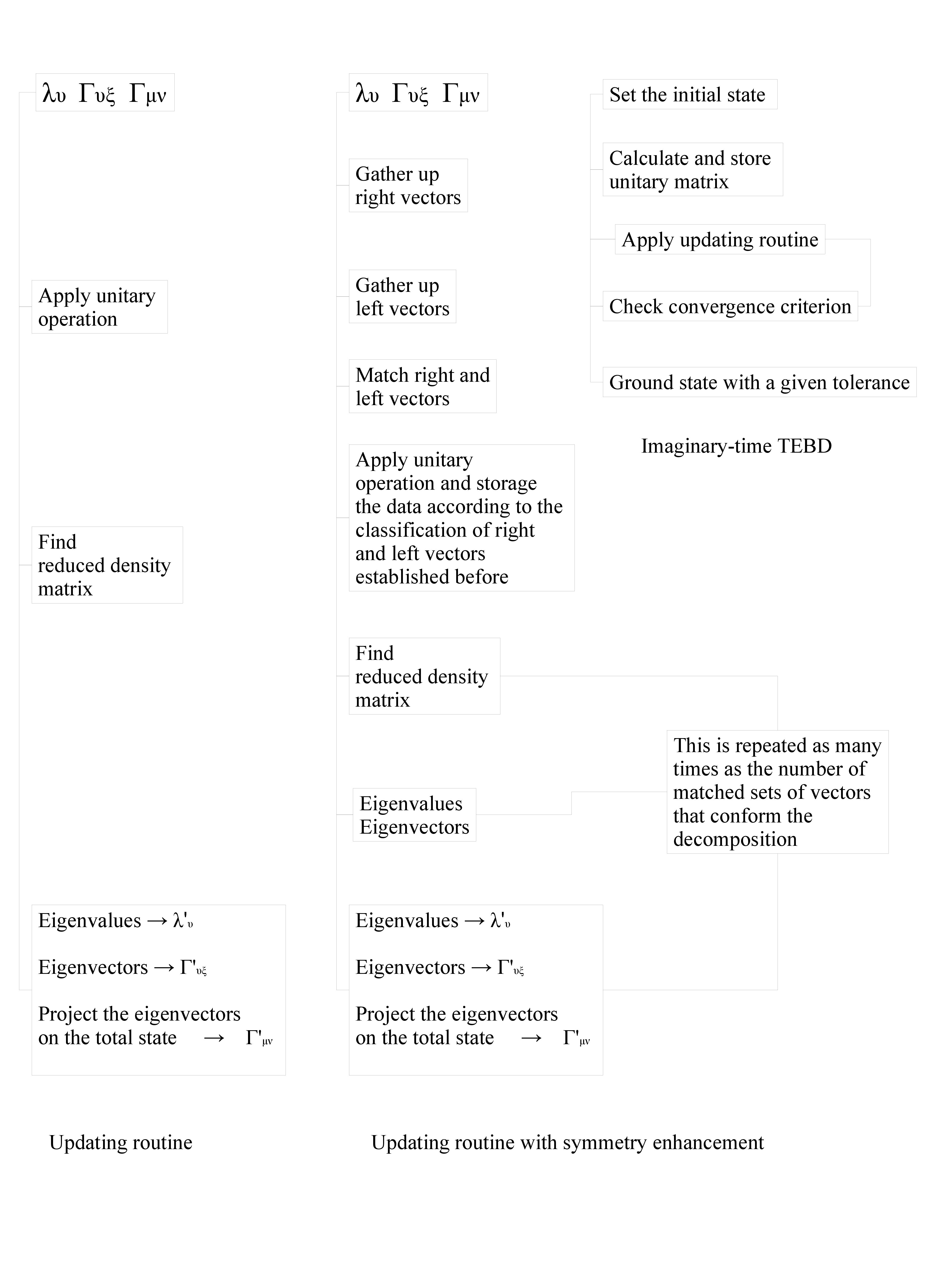}
\caption{Flux diagrams.} 
\label{fluxdia}
\end{figure}

This means that there is one pointer associated to every
site of the chain. Each pointer is independent and can be given
whatever memory one needs to allocate (figure \ref{lamb}). So, if
for example the Schmidt decomposition in site 5 has 3 coefficients,
then in order to allocate that specific amount of space we write,

\begin{verbatim}
allocate(LAM(5)%ld(3))
\end{verbatim}

and then proceed to specify every component of LAM(5)\%ld. Similarly,
tensor $\Gamma_{\alpha \beta}^{[n]i}$ is an object that can be 
allocated in two dimensions, namely those associated with $\alpha$
and $\beta$. Every component of this allocatable arrangement, on
the other hand, is a complex vector of fixed size. The components of
such vectors correspond to a local basis, in this case given in terms
of the label $i$, whose maximum value is set to be $M$. Therefore, we must 
utilize two types to define the variable, namely,

\begin{verbatim}
type fg
complex(dpc), dimension(M) :: gf
end type fg
\end{verbatim}
\begin{verbatim}
type dg
type(fg), dimension(:,:), pointer :: gd
end type dg
\end{verbatim}

the variable itself is defined the same way as LAM was defined before, 

\begin{verbatim}
type(dg), dimension(N) :: GAM
\end{verbatim}

that is, one storage unit per site. For this variable, allocation takes
place only in the pointer section. For instance,

\begin{verbatim}
allocate(GAM(2)%gd(3,5))
\end{verbatim}

and then, if we want to give numerical values to the tensor we would 
write, for example,

\begin{verbatim}
GAM(2)%gd(1,3)%gf(2) = some complex number
\end{verbatim}

In this way, no memory space is wasted through undefined memory slots.
When using memory allocation, special care must be taken in processing
allocatable units. If a variable is allocated, then it must be
deallocated if for some reason we want to allocate it again, for example
as the decomposition is updated, otherwise the memory associated with
the variable will not be accessible any more and one can easily run out
of virtual memory in a long simulation. The most challenging part of
a time evolving block decimation (TEBD) program, is the updating
routine in which a two-site unitary transformation is applied and
some tensors must be worked out (figure (\ref{fluxdia})). 

In a program that makes use of symmetries, we must first organize
the vectors of the decomposition in groups of elements sharing
the same symmetry, which at the same time involves finding the
symmetry of the vectors themselves. Once this is done, we 
apply the updating procedure to sets of vectors with complementary
symmetry, in such a way that one global symmetry is preserved.
Finally, the symmetry associated with the new tensors must be
found and attached in order to be able to access this important
information in the next run of computations. The reason why 
we implement symmetry enhancement is because the cost of 
updating the decomposition in small subspaces is lower
than the one of updating all the elements at once by 
diagonalizing a single large matrix.

Similarly, in a problem where the quantum state is symmetric
with respect to the centre of the chain, we can simulate the
dynamics by updating the coefficients only on one half of the
chain and extrapolating some coefficients in the middle. This
involves using the fact that in a symmetric chain with an even
number of sites the reduced density matrix of the two central
sites has a degenerate spectrum and eigenvectors corresponding
to the same eigenvalue are in fact complementary.

One additional feature of FORTRAN 95 which hugely facilitates 
memory management are {\it linked lists}. They basically consist
in variables that can be given extra allocatable space without 
destroying the information already contained in the variable. Note
that this is not the case for pointers, since every time that
a pointer is reallocated it must be previously deallocated and 
so any associated content is automatically lost. A linked list can 
be made of any valid memory unit such as vectors, matrices, pointers
and even types. In the updating routine, for instance, we use
linked lists to store the Schmidt vectors as we do not know in 
advance how many vectors conform the decomposition of the
updated state.

We also would like to comment about one element of the research
with challenging numerically aspects. As it has been discussed, the
calculation of entanglement in mixed states involves dealing with
the density matrix. Here, such matrix comes from reducing the pure state
of the whole system. In order to
find the entanglement between the ends, or end-to-end entanglement (EEE),
the reduced density matrix in the standard basis must be computed using 
the canonical coefficients. To see how the reduction takes place, let us 
write the quantum state of the chain in the following way \cite{vidal1},

\begin{equation}
|\psi \rangle = \sum_{\alpha_1,\alpha_2,...,\alpha_{N-1}} | 1 \alpha_1 \rangle^{[1]} \lambda_{\alpha_1}^{[1]} | \alpha_1 \alpha_2 \rangle^{[2]} \lambda_{\alpha_2}^{[2]}...\lambda_{\alpha_{N-1}}^{[N-1]}|\alpha_{N-1}1 \rangle^{[N]},
\end{equation}

where, 

\begin{equation}
|\alpha \alpha' \rangle^{[n]} = \sum_{i} \Gamma_{\alpha \alpha'}^{[n]i} |i\rangle.
\label{mist}
\end{equation}

Consequently, the reduced density matrix looks like,

\begin{eqnarray}
&& \hat{\rho}_{1N} =  \nonumber \\
&& \sum_{ \binom{\alpha_1,\alpha_2,...,\alpha_{N-1},}{{\alpha'}_1,{\alpha'}_2,...,{\alpha'}_{N-1}}
}   |1 \alpha_1 \rangle^{[1]} \langle {1 \alpha'}_1 |^{[1]} \lambda_{\alpha_1}^{[1]}\lambda_{{\alpha'}_1}^{[1]} \langle \alpha_1 \alpha_2 |{\alpha'}_1 {\alpha'}_2 \rangle^{[2]} \lambda_{\alpha_2}^{[2]}\lambda_{{\alpha'}_2}^{[2]}  
\langle \alpha_2 \alpha_3 |{\alpha'}_2 {\alpha'}_3 \rangle^{[3]} 
... \nonumber  \\
&& ...\langle \alpha_{N-2} \alpha_{N-1} |{\alpha'}_{N-2} {\alpha'}_{N-1} \rangle^{[N-1]} \lambda_{\alpha_{N-1}}^{[N-1]}\lambda_{{\alpha'}_{N-1}}^{[N-1]} |\alpha_{N-1} 1 \rangle^{[N]} \langle {\alpha'}_{N-1} 1 |^{[N]}  \\
&& =\sum_{\alpha_1,{\alpha'}_1,\alpha_{N-1},{\alpha'}_{N-1}
}   |1 \alpha_1 \rangle^{[1]} \langle {1 \alpha'}_1 |^{[1]} M_{ \{ \alpha_1 \alpha_{N-1} \}  \{ {\alpha'}_1 {\alpha'}_{N-1} \} } |\alpha_{N-1} 1 \rangle^{[N]} \langle {\alpha'}_{N-1} 1 |^{[N]}. \nonumber
\end{eqnarray}

Most of the programming work is devoted to write matrix 
$M_{\{\alpha_1 \alpha_{N-1}\} \{ {\alpha'}_1 {\alpha'}_{N-1}\}}$ 
in a way that can be efficiently stored and manipulated. The storage 
matter is related to conservation properties. In the case of 
$\hat{\rho}_{1N}$, conservation splits the operator in representations 
holding 
$S_{\alpha_1} + S_{\alpha_{N-1}}  = S_{{\alpha'}_1} + S_{{\alpha'}_{N-1}}$,
where $S_{\alpha}$ is the symmetry associated with ket $|\alpha \rangle$.
The same splitting applies for 
$M_{\{ \alpha_1 \alpha_{N-1} \} \{ {\alpha'}_1 {\alpha'}_{N-1} \}}$.
In order to compute this matrix, one first calculates the products 
$\lambda_{\alpha_1}^{[1]}\lambda_{{\alpha'}_1}^{[1]} \langle \alpha_1 \alpha_2 |{\alpha'}_1 {\alpha'}_2 \rangle^{[2]}$
and creates a temporary support matrix
$ M_{\{ \alpha_1 \alpha_{2} \} \{ {\alpha'}_1 {\alpha'}_{2} \}}$.
This computation can be accelerated by exploiting the fact that
in a state like (\ref{mist}) a symmetry restriction bounds the
indices through 

\begin{equation}
S_{\alpha} + S_{\alpha'} + S_{i} = S_{\text {Global Simmetry}},   
\label{sim}
\end{equation}

therefore, to compute an inner product such as,

\begin{equation}
\langle \alpha \alpha' | \beta \beta' \rangle,
\end{equation}

one just takes the indices in the kets and finds the corresponding
local coordinate $i$ using equation (\ref{sim}). Only if the coordinates
coincide the associated contribution is stored using linked lists.
In the next run of computations, for every pair of indices 
$\alpha_3,{\alpha'}_3$ one sweeps over the range of values of the
indices $\alpha_2,{\alpha'}_2$ and finds the product 
$\lambda_{\alpha_2}^{[2]}\lambda_{{\alpha'}_2}^{[2]} \langle \alpha_2 \alpha_3 |{\alpha'}_2 {\alpha'}_3 \rangle^{[3]}$. 
Again, only if the inner product is not zero the contribution is 
added up. After this is completed, we are left with a support matrix 
$ M_{\{ \alpha_1 \alpha_{3} \} \{ {\alpha'}_1 {\alpha'}_{3} \}}$.
The process goes on until we finally get the matrix with the 
end indices. 

Improved performance can be achieved if spatial symmetry is taken
into consideration. This is done in a very similar way than with the
canonical coefficients, that is, when the computations come to the 
middle of the chain, we group up complementary representations from
each side of the chain and perform our operations in subspaces,
always keeping the total number of bosons fixed. 

Once the reduced density matrix for the chain terminals has been
worked out, we use it to calculate the logarithmic negativity.
This involves performing a series of rearrangement operations, 
as for example, the explicit calculation of the
partial transpose matrix. This process alone carries a technical
issue of particular trickiness that we want to comment about. As
we already pointed out, the conservation of the total number of 
particles allows us to split large matrices into several smaller
representations that we can manage more easily. As a consequence,
what we call reduced density matrix is actually a set of
matrices, each one corresponding to a different quantum number.
This can be understood by thinking that every matrix matches 
a state of the chain with a complementary quantum number. Therefore,
there are as many matrices as number of bosons plus one, as the
no-boson possibility must be accounted for too. The splitting is
not only practical but often necessary, because the amount of 
information that we can handle using a computer is limited.
Nevertheless, this splitting may be broken if the state is
subject to non-unitary transformations. This is precisely what
happens when we try to compute the partial transpose. As a matter
of fact, the transposition operation mixes subspaces with 
different quantum numbers. The good news is that the transpose
matrix keeps some kind of regularity that allows us to
split the matrix into non-interacting subspaces that we can
address individually. Hence, before starting with the
operations that determine the new matrix, we first establish
the subspaces that conform the new splitting, and then
proceed to fill every conforming matrix with the corresponding
elements. Once this is completed, finding the eigenvalues and
computing the logarithmic negativity from equation (\ref{logn})
is straightforward.

These are the numerical issues that in our opinion require special 
care and analysis before they can be efficiently implemented in a 
program. Even though these aspects of our investigation do not shed by
themselves any physical insight, we have decided to include this
discussion as a way of presenting a complete view of the problem
as well as its collateral issues.

\chapter{Matrix product states in ideal boson chains}

\label{recho}

In a boson chain of $N$ sites and $M$ bosons in which particle
exchange can potentially take place among any two places, the
Hamiltonian of the system reads,

\begin{equation}
\hat{H} =  \sum_{k=1}^N R_{k,k} \hat{a}^{\dagger}_{k} \hat{a}_k +  \sum_{k=1}^{N} \sum_{l=k+1}^N {\left( R_{k,l} \hat{a}^{\dagger}_{k} \hat{a}_l  + {R^*}_{k,l} \hat{a}^{\dagger}_{l} \hat{a}_k \right ) } 
\label{ideal}
\end{equation}

In such a way that $R_{k,l}$ represents the strength of the 
hopping between sites $k$ and $l$. The creation $\hat{a}_k^{\dagger}$ 
and annihilation $\hat{a}_k$ operators follow the standard commuting 
rules for a discrete model given by equation (\ref{comm}).

The Hamiltonian above is quadratic and can be decoupled in order
to get the ground state. Additionally, the Heisenberg equations of
motion for the creation operators produce a complete set of
differential equations that can be written as,

\begin{equation}
\frac{d}{d t}
\left(
\begin{array}{c}
\hat{\alpha}_1 \\
\hat{\alpha}_2 \\
\hat{\alpha}_3 \\
\vdots \\
\hat{\alpha}_N \\
\end{array}
\right)
=
-i\left(
\begin{array}{ccccc}
R_{1,1} & R_{1,2} & R_{1,3} & \cdots & R_{1,N} \\
 {R^*}_{1,2}& R_{2,2} & R_{2,3} & \cdots & R_{2,N} \\
{R^*}_{1,3}& {R^*}_{2,3} & R_{3,3} & \cdots & R_{3,N} \\
\vdots & \vdots & \vdots & \ddots & R_{N-1,N} \\
{R^*}_{1,N} & {R^*}_{2,N} & {R^*}_{3,N} & {R^*}_{N-1,N} & R_{N,N} \\
\end{array}
\right)
\left(
\begin{array}{c}
\hat{\alpha}_1 \\
\hat{\alpha}_2 \\
\hat{\alpha}_3 \\
\vdots \\
\hat{\alpha}_N \\
\end{array}
\right)
\label{eq:three},
\end{equation}

or equivalently,

\begin{equation}
\frac{d \mbox{ \boldmath $\hat{\alpha}$ } }{d t} = -i \hat{R} \mbox{\boldmath$ \hat{\alpha}$},
\label{hei}
\end{equation}

where $\hat{\alpha}_k$ describes the corresponding creation operator 
in the Heisenberg picture, 

\begin{equation}
\hat{\alpha}_k = e^{-i t \hat{H}} \hat{a}^{\dagger}_k e^{i t\hat{H}}.
\end{equation}

These equations are complemented by the initial conditions, 

\begin{equation}
\hat{\alpha}_k(t=0)=\hat{a}_k^\dagger.
\end{equation}

The state, on the other hand, is given in terms of the Heisenberg
operators by,

\begin{equation}
|\psi(t)\rangle = \frac{1}{\sqrt{\prod_j^N n_j!}} \prod_{k}^N \hat{\alpha}_k^{n_k} |0 \rangle. 
\end{equation}

The way in which particles are distributed across the chain is determined
by the constants $n_j$.

From equation (\ref{hei}) it is in fact possible to obtain the ground
state of the system through imaginary time analysis (equation (\ref{sile})). 
In doing so, the ground state is found to be,

\begin{equation}
| G \rangle = \left ( \sum_{k=0}^N c_k \hat{a}^{\dagger}_k \right )^M |0 \rangle,
\label{expt}
\end{equation}

where coefficients $c_k$ are the components (possibly complex) of the 
ground eigenvector of matrix $\hat{R}$, so that, 

\begin{equation}
\sum_{k=1}^N |c_k|^2 = 1.
\end{equation}

From equation (\ref{expt}) we can extract some information.
Nevertheless, because the description of the state demands exponentially
growing resources which scale with both $N$ and $M$, calculations involving
non-local correlations can be fairly challenging. In order to set state
(\ref{expt}) in a way that can be easily handled, we will apply
unitary operations to the state so as to simplify it as much as possible.
The idea is to reduce $|G\rangle$ to an expression that can be written
using MPS. Then, if the unitary operations only involve transformation
to first neighbours, the updating method presented in previous sections
can be applied. This would provide us with a complete description of
the state that we can efficiently use.

We first operate locally on individual sites using the transformation,

\begin{equation}
\hat{U}_k^{(1)} = e^{-i \theta_k \hat{a}_k^{\dagger} \hat{a}_k}.
\end{equation}

In this equation $\theta_k$ is the phase of the complex number $c_k$.
As a result of this transformation we get,

\begin{equation}
\hat{a}_k^{\dagger} \rightarrow e^{-i \theta_k} \hat{a}_k^{\dagger}.
\end{equation}

In such a way that the complex phase is cancelled out.
Additionally, any operator different from $\hat{a}_k$ remains
unaffected by the transformation. Once this has been done
for every operator $\hat{a}_k$, the new coefficients $c_k$ in
equation (\ref{eq:three}) are real.

Subsequently, we apply unitary operations involving first neighbours 
in the following fashion,

\begin{equation}
\hat{U}_k^{(2)} =e^{-i \phi_k \left(  \frac{1}{2 i} (\hat{a}_{k+1}^{\dagger} \hat{a}_{k}-\hat{a}_{k}^{\dagger}\hat{a}_{k+1}) \right)} = e^{-i\phi_k \hat{Q}_k }.
\label{tras}
\end{equation}

In physical terms, operator $\hat {Q}_k$ is known as the {\it current}, and its eigenvalues indicate how many bosons circulate in between sites
$k$ and $k+1$.
As a result of this unitary operation, the pair 
$\{\hat{a}_k^{\dagger},\hat{a}_{k+1}^{\dagger}\}$ transforms as,

\begin{eqnarray}
&& c_{k+1} \hat{a}_{k+1}^{\dagger} + c_{k} 
\hat{a}_{k}^{\dagger} \rightarrow \\
&& \left ( c_{k+1} \cos \left( \frac{\phi_k}{2} \right) - c_{k} 
\sin \left ( \frac{\phi_k}{2}  \right )  \right ) \hat{a}_{k+1}^{\dagger} + \left ( c_{k+1} \sin \left( 
\frac{\phi_k}{2} \right) + c_{k} \cos \left ( \frac{\phi_k}{2}  \right )  \right ) \hat{a}_{k}^{\dagger}. \nonumber
\end{eqnarray}

Therefore, we can cancel operator $\hat{a}_{k+1}^{\dagger}$ by just
choosing an angle $\phi_k$ satisfying,

\begin{eqnarray}
\tan \left( \frac{\phi_k}{2} \right) = \frac{c_{k+1}}{c_{k}}.
\end{eqnarray}

Consequently, in order to reduce the state to one single mode
operating on the vacuum, we apply the operations (\ref{tras}), 
starting from $k=N-1$, to every couple of consecutive operators. Note that 
every time a transformation acts on any pair of modes, the 
coefficient that accompanies the creation operator that is 
not taken out is affected, but it always remains real. 

When the reduction is completed, we are left with a state as
(up to an overall constant),

\begin{equation}
|g \rangle = \frac{1}{\sqrt{M!}} \left ( \hat{a}_1^{\dagger} \right )^M |0 \rangle,
\end{equation}

which can be easily written using MPS. The next step consists in
applying the inverse operations in reverse order to the state
using the method presented in section \ref{MPS}. From now on, we
will refer to this reduction operation as {\it state folding}.

There is also the case when we must deal with numerous summations acting 
on the vacuum. For instance, in chains with an initial state given by
bosons arranged on different positions. In this kind of situation
state folding can be utilized too. Let us consider the case when just two
summations get involved so that the state reads,

\begin{equation}
\left ( \overbrace {\sum_{k=1}^N c_k \hat{a}_k^{\dagger}}^{S_2} \right )^{M_2} \left ( \overbrace {\sum_{l=1}^N z_l \hat{a}_l^{\dagger}}^{S_1} \right )^{M_1} |0\rangle,
\end{equation}

where $M_{1}$ is the number of bosons originally allocated on one
site of the chain and $M_{2}$ has an analogous meaning.
We can use the standard technique to fold $S_1$, but it is worth
saying that in the process the coefficients of $S_2$ are also
affected. Next, we can fold the new $S_2$ from $\hat{a}_N^{\dagger}$ 
just until $\hat{a}_2^{\dagger}$,
since folding $\hat{a}_2^{\dagger}$ in $S_2$ would unfold $S_1$.
As a result, the folded state must be written making explicit
reference to such last folding operation in the form,

\begin{equation}
|g \rangle = \left( \hat{a}_1^{\dagger} \right )^{M_2} e^{-i \phi_1 \hat{Q}_1} \left (  \hat{a}_1^{\dagger} \right )^{M_1} |0 \rangle.
\label{eq:six}
\end{equation}

Consequently, $|g\rangle$ can be written in MPS by first translating 
$|M_1,0,...,0\rangle$ into MPS and then
applying $e^{-i \phi_1 \hat{Q}_1}$. In this way, we are left with
a canonical decomposition that can be used to apply the last operation
in the row. With this in mind we set down the state with explicit 
reference to the local coordinates of the first position and then
apply $\left( \hat{a}_1^{\dagger} \right )^{M_2}$ in a very 
straightforward way. In doing so we get,

\begin{equation}
|g\rangle = \sum_{\gamma=1}^{N_\gamma} \Gamma_{1\gamma}^{[1]j}\lambda_\gamma^{[1]} \sqrt{(j+M_2)!} |j + M_2\rangle  |\gamma \rangle^{[2-N]}.
\label{eq:seven}
\end{equation}

In latter equation we have made use of the fact that the number of
bosons in the local basis of the first site $|j+M_2\rangle$ is 
determined by the number of bosons in the complementary Schmidt vector
$|\gamma \rangle$ and therefore there is only one relevant coordinate
that describes the local basis. This explains the lack of a summation
symbol for the label $j$. From the expression above the canonical 
coefficients of state $|g\rangle$ can be directly obtained, namely,

\begin{eqnarray}
& {\lambda'}_\gamma^{[1]} = \sqrt{(j+M_2)!} \lambda_\gamma^{[1]}
+ \text{normalization}, & \\
& {\Gamma'}_{1\gamma}^{[1] j+M_2} = \Gamma_{1\gamma}^{[1]j}. &
\end{eqnarray}

However, because this is not a unitary transformation, it is important 
to show that the canonical decomposition obtained in this way is
consistent. It is not difficult to see that the canonical tensors
attached from the third site onwards do not suffer any modification
whatsoever since this section of the chain is made of complementary 
partitions which contain only one Schmidt vector. This is because 
there is no boson at all between the third and the last position. 
We can see that such is indeed the case by noticing that the operations 
performed on the vacuum involved only the first and second places, 
thereby only these two positions can hold any non-vanishing population.
On the other hand, we know that the tensor elements in positions
one and two depend directly on the Schmidt vectors
$\{|\gamma_i\rangle^{[1]}\}$. These vectors all have well defined 
quantum numbers on account of particle conservation. Therefore,
when we apply $\left( \hat{a}_1^{\dagger} \right )^{M_2}$ we are 
actually lifting the boson occupation which means that the resulting 
vectors are valid Schmidt vectors which are orthogonal to each other.

\section{Quench in trapped systems}

\begin{figure}
\includegraphics[width=0.35\textwidth,angle=-90]{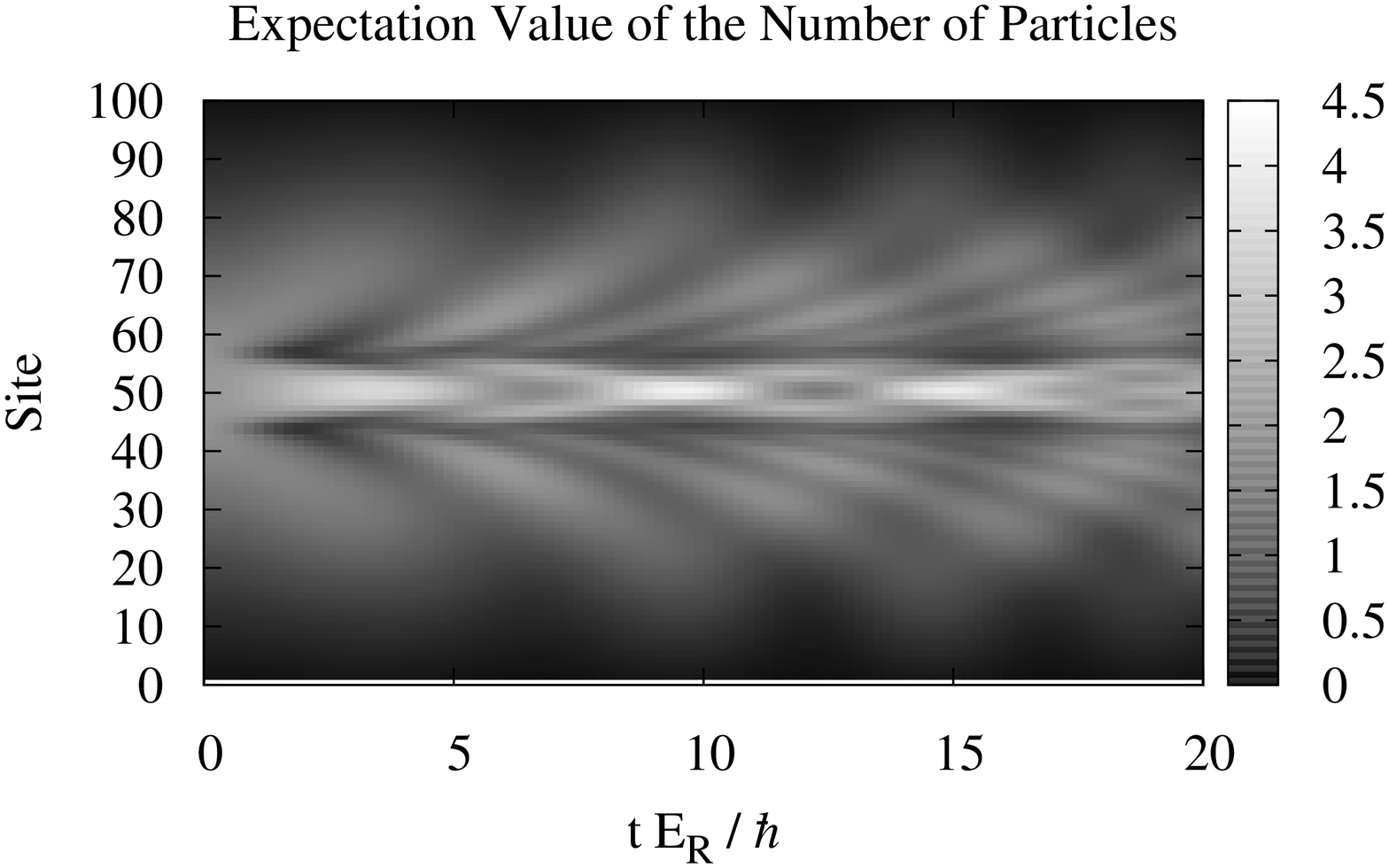}\includegraphics[width=0.35\textwidth,angle=-90]{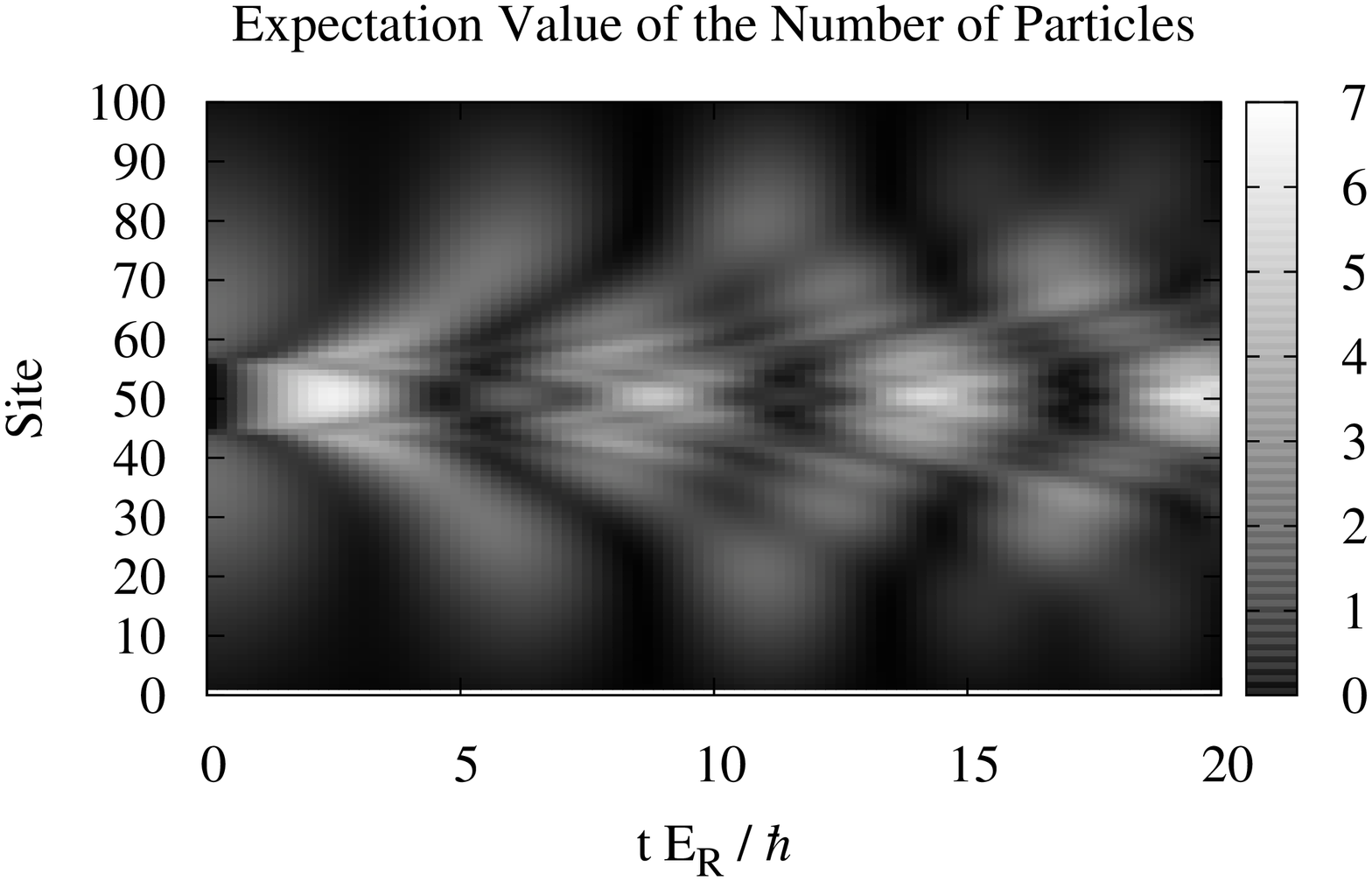}
\includegraphics[width=0.35\textwidth,angle=-90]{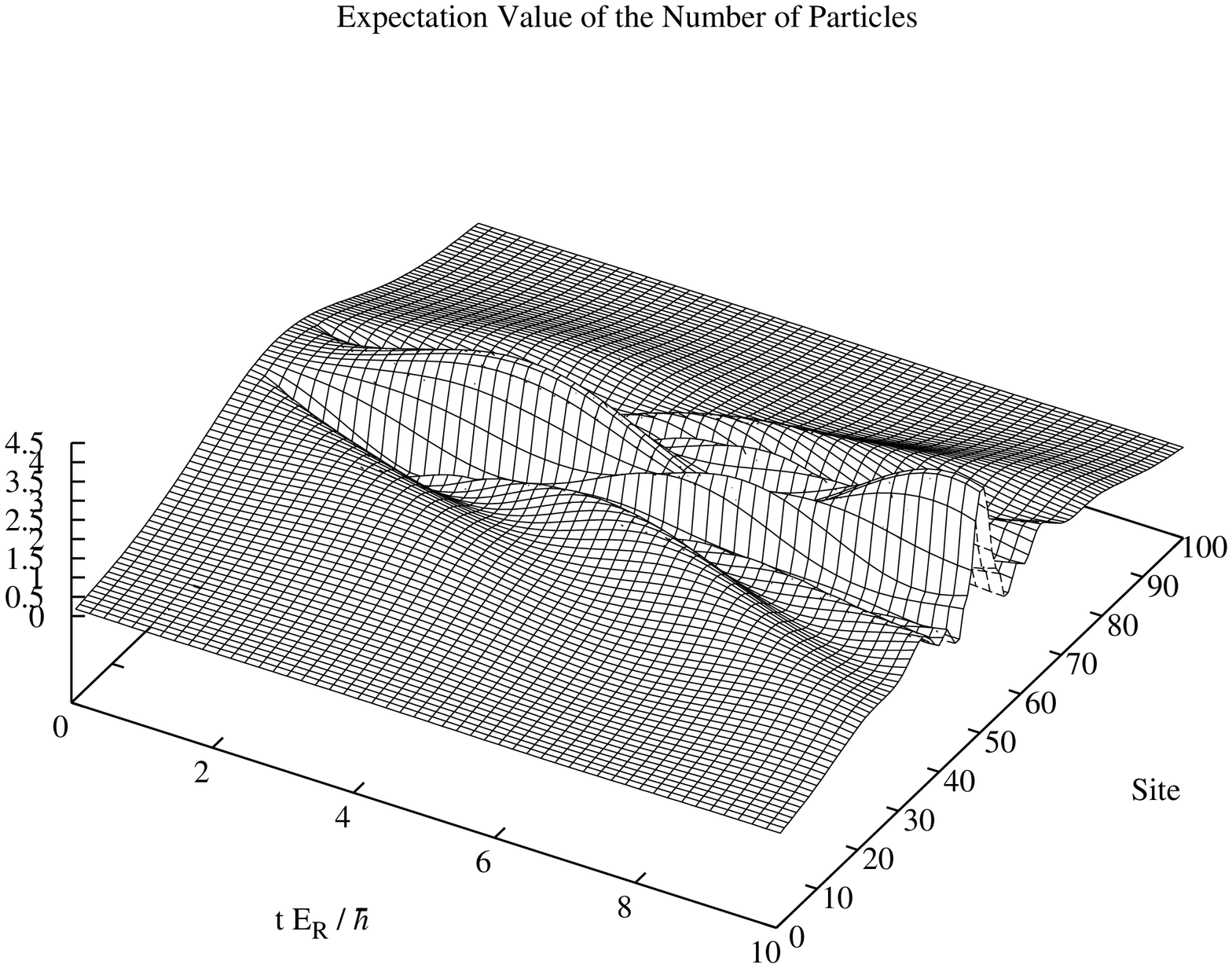}\includegraphics[width=0.35\textwidth,angle=-90]{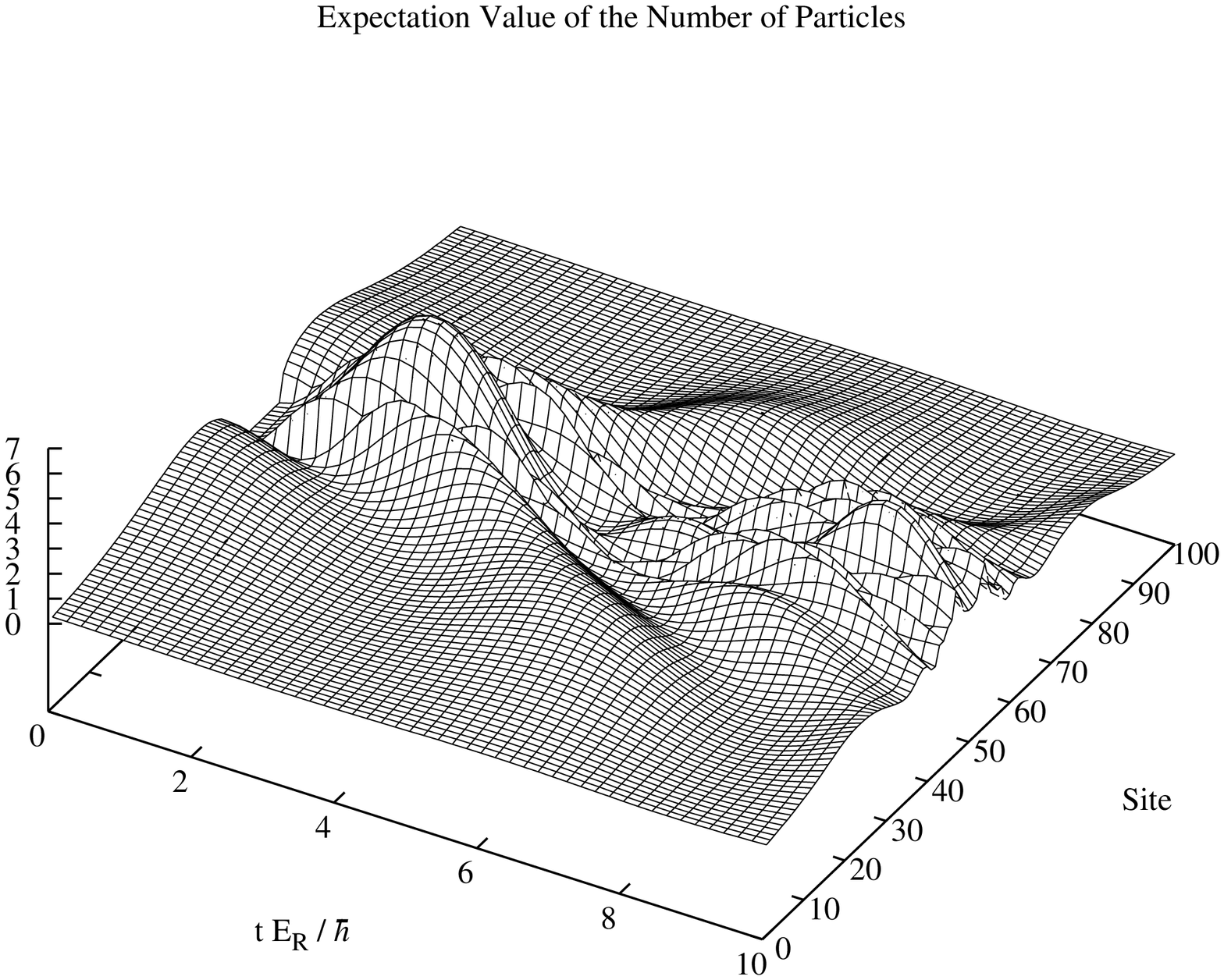}
\caption{Expectation value of the number of particles. Top plots are grey 
maps of the lower
plots. Dynamics is generated by turning a potential barrier in the
middle on (left) and off (right).} 
\label{quench}
\end{figure}

We now show how the method presented in the previous section 
is capable of delivering results in challenging scenarios. 
In what follows, we assume that energy is
given in terms of the recoil energy,

\begin{equation}
E_{R}=\frac{\hbar k^2}{2 m},
\end{equation}

where $k$ is the wavelength of the confining laser and $m$ is the
atomic mass. Let us
consider a chain of 100 sites and 100 bosons which has been
cooled down to the ground state in the presence of a trapping 
potential given by,

\begin{equation}
R_{i,i} = \Omega (i-50)^2,
\end{equation}

with $\Omega = 0.00046E_R$, a reasonable experimental factor according
to reference \cite{Danshita1}. Additionally,
we assume that particle exchange can take place among any two places
in the chain, not only next neighbours. The intensity of the hopping 
is proportional to the off-diagonal matrix elements,

\begin{equation}
R_{i,j} = \frac{\Xi}{|i-j|},
\end{equation}

for every $i \ne j$ and with $\Xi = 0.3E_R$. 
This choice of hopping can be justified in chains with
long-range exchange effects, such as expected in situations where
the Coulomb potential plays an important role. In a first set of 
simulations, we find the ground vector of matrix $\hat{R}$ and
insert the coefficients in equation (\ref{expt}). Dynamics is 
generated by instantaneously turning on (off) a very high potential 
barrier in the middle of the chain. This barrier is written as,

\begin{equation}
R_{i,i} = 1000E_R \hspace{0.5cm} \text{for} \hspace{0.5cm} 45<i<55.
\end{equation}

Next, we find the ground state of the system including the potential
barrier so as to generate a distribution with two condensates at
the sides of the wall and then turn the potential barrier off.
Figure \ref{quench} shows the behaviour of the expectation value of 
the number of 
bosons for the above mentioned configurations. As a result of the 
quench very characteristic wave patterns are generated depending 
on how the quench takes place. In both cases, however, fleeing  
waves are generated after a particle bulk assembles in the chain
centre. In the first case, when the barrier is turned on before
the condensate is released, we can see that a large part of the 
condensate remains in the middle of the chain, just where the 
barrier stands. The barrier domain seems to determine a zone
from which bosons can hardly escape. It is reasonable as the
high potential difference between sites inside and outside the
wall zone energetically hinders any quantum jump. In contrast, the
pattern generated in the case when the dynamics is generated by
switching the potential barrier off is more compatible with 
uniform expansion. Nevertheless, in both cases we can appreciate
the effect of the inclusion of long range hopping terms in the 
Hamiltonian. When the bosonic waves travel outwards they gradually
lose particles. However, this loss does not translate in dissipation,
instead, a new boson bulk reassembles around the centre of the chain 
and a new expansion starts again. This effect is clearer in the case
of expansion without intermediate potential, but it also takes place
in the other case analysed. Such profile is in contrast with 
the behaviour of travelling packets of chains with next-neighbour 
hopping only. In the latter case the changes in the form of the
wave pattern develop locally around the boson bulk, and the effect
of particle exchange between distant places of the chain is much
less pronounced.

\section{Entanglement as a result of collision}

\begin{figure}
\begin{center}
\includegraphics[width=1.\textwidth,angle=0]{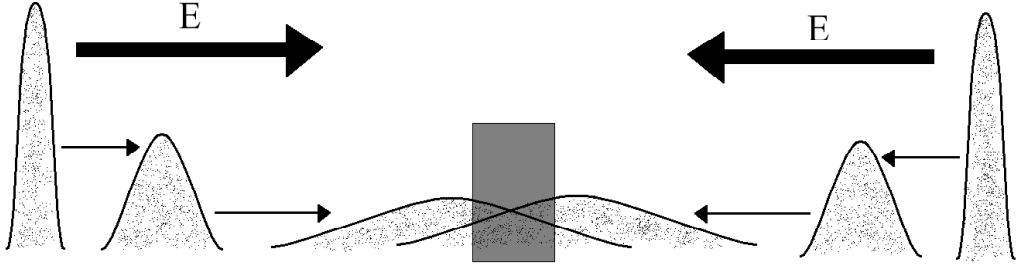}
\end{center}
\caption{Sketch of a bosonic collision} 
\label{scotish}
\end{figure}

In this section we want to show how the two-sum folding method
can be used in time dependent problems. Let us suppose that in
a boson chain we have two very well localized boson clouds at
the ends. We then use a physical mechanism to accelerate the clouds
and make them interact with each other. After this collision
process, the scattered bosons are collected and taken back to
the chain terminals. Because of the interaction between the
boson packets, the collected particles at the ends are strongly
entangled, as discussed in \cite{Zoller,Schmuser,Law,Singh}. 
Certainty, this entanglement would be potentially useful for
multiple quantum information procedures. Before focusing on 
entanglement, however, we will look at how the bosons get
transmitted from one end of the chain to the other in the
presence of a perturbation in the central part. In such a
system the Hamiltonian would be ideally given by equation 
(\ref{ideal}) with a matrix $\hat{R}$ that can be written
as,

\begin{equation}
\hat{R} = \hat{J}_x + \epsilon e^{-\beta \hat{J}_z^2},
\label{Rmatrix}
\end{equation}

where $\hat{J}_{x,y,z}$ are the standard angular momentum 
operators. $\epsilon$, on the other hand, represents the 
intensity of a symmetrically localized perturbation in the
middle of the chain and $\beta$ measures the spread 
of such perturbation. One reason to put matrix $\hat{R}$ in
terms of the angular momentum operators is that in this way
we can highlight the analogy between perfect transmission in
a boson chain and the physics of angular momentum. In fact, 
in the absence of a term proportional to $\epsilon$, this
would be basically a particle spinning around the x-axis.
The perturbation has been deliberately chosen 
in the form of an exponential in order to generate a decaying
profile as going from the middle of the chain towards the terminals.
This form is also very convenient to perform analytical calculations 
as we will show.
In addition to this formulation of matrix $\hat{R}$, we must
establish a relation among the state kets in the Hilbert 
space and the actual operators. This can be worked out by
just comparing the effect of the angular momentum on the
corresponding kets with the predicted behaviour of the
operators in the perfect transmission scenario. From this
we can infer,

\begin{equation}
|j,m \rangle \rightarrow \hat{a}_{j-m+1}^\dagger.
\label{cochito}
\end{equation}

In this expression we used the standard notation for the 
angular momentum kets, that is, $|j,m\rangle$ are the 
eigenstates of the z-component operator $\hat{J}_z$. The
quantum number $j$ is related to the total number of
sites in the chain by the identity,

\begin{equation}
j=\frac{N-1}{2}.
\end{equation}

In this way, the evolution of the Heisenberg operators can
be studied by following the dynamics of the associated kets.
In a problem where the physics of the system is dictated by
equation (\ref{Rmatrix}) the wave function acquires the following 
form after half a period evolution,

\begin{equation}  
|\psi(t=\pi) \rangle = e^{-i \pi ( \hat{J}_x + \epsilon e^{-\beta \hat{J}_z^2})}|j,j \rangle. 
\label{bulg}
\end{equation}

Here we have implicitly assumed that the length of the angular
momentum $j$ is an arbitrary integer which accounts for $N$ being 
odd. We have focused on the evolution of $|j,j\rangle$ because that
is the ket relevant to operator $\hat{a}_1$ and therefore the one
which contains information about the transmission of bosons initially
prepared on the first site of the chain. Making use of time dependent
perturbation theory we can expand the latter expression in the following
way,

\begin{eqnarray}
& |\psi(t=\pi) \approx  e^{-i\pi \hat{J}_x} |j,j \rangle - i \epsilon |\epsilon \rangle,& \nonumber \\
& |\epsilon \rangle =  \int_0^\pi dt e^{-i(\pi-t)\hat{J}_x} e^{-\beta \hat{J}_z^2} e^{-i t \hat{J}_x} |j,j \rangle, &  
\label{guille}
\end{eqnarray}

of course, this approximation holds as long as the amplitude of the 
dynamics generated by the integral in the second term remains small 
compared to the unperturbed evolution. For this to happen it is important
not only that $\epsilon$ is a little fraction but also that the dynamics
develops out of resonance. Nevertheless, in this particular case 
where we restrict ourselves to well defined time intervals, the emergence 
of cooperative resonances is quite improbable. So, the integral above 
can be manipulated as,

\begin{equation}
|\epsilon \rangle = \int_0^\pi dt e^{-i(\pi-t)\hat{J}_x} e^{-\beta \hat{J}_z^2} e^{i (\pi-t) \hat{J}_x} e^{-i\pi \hat{J}_x} |j,j \rangle.
\end{equation}
 
From the properties of the angular momentum we know 

\begin{equation}
e^{-i\pi \hat{J}_x} |j,j \rangle = |j,-j \rangle. 
\end{equation}

Further simplifications apply by noticing that the three first 
exponentials from left to right underline a unitary transformation. 
As a result we can write,

\begin{equation}
|\epsilon \rangle = \int_0^\pi dt e^{-\beta (\cos (t) \hat{J}_z - \sin (t) \hat{J}_y )^2 }|j,-j \rangle.
\end{equation}

Because the operator on the integral is squared, we do not have a 
way to translate the expression into a sum of kets. Therefore, we 
rather use the well known result from integral calculus,

\begin{equation}
\int_{-\infty}^\infty dx e^{-(a x^2 + b x)} = \sqrt{\frac{\pi}{a}} e^{\frac{b^2}{4a}},
\end{equation}

to put $|\epsilon \rangle$ in a more operational fashion,

\begin{equation}
|\epsilon \rangle = \frac{1}{\sqrt{4\pi\beta}} \int_{-\infty}^\infty dx e^{-\frac{x^2}{4 \beta}} \int_0^\pi dt e^{i x (cos(t) \hat{J}_z -sin(t)\hat{J}_y )} |j,-j \rangle.
\end{equation}

In this way, we can now work out how the exponential operator acts on the
state by making use of Schwinger's alternative representation of the angular
momentum \cite{sakurai}. For this we define a couple of independent sets of bosonic operators
(contrary to the operators that describe the bosons on the chain, these do not 
have a direct physical correspondence),

\begin{eqnarray}
\left [ \hat{a}_+,\hat{a}_+^{\dagger} \right ] = 1, \hspace{0.5cm} \left [ \hat{a}_-,\hat{a}_-^{\dagger} \right ] = 1, \nonumber \\
\left [ \hat{a}_+,\hat{a}_-^{\dagger} \right ] = 0, \hspace{0.5cm} \left [ \hat{a}_-,\hat{a}_+^{\dagger} \right ] = 0.
\end{eqnarray}

It can be shown that these operators form a structure that correctly
reproduces the angular momentum through the following equivalence
transformations,

\begin{eqnarray}
& \hat{J}_+ = \hat{a}_+^{\dagger} \hat{a}_-, & \nonumber \\ 
& \hat{J}_- = \hat{a}_-^{\dagger} \hat{a}_+, &\nonumber \\
& \hat{J}_z = \frac{1}{2}(\hat{a}_+^{\dagger} \hat{a}_+ - \hat{a}_-^{\dagger} \hat{a}_-), & 
\end{eqnarray}

along with,

\begin{equation}
|j,m \rangle = \frac{(\hat{a}_+^{\dagger})^{j+m} (\hat{a}_-^{\dagger})^{j-m}}{\sqrt{(j+m)!(j-m)!}} |0_+,0_- \rangle,
\label{damx}
\end{equation}

for the state kets. In terms of the bosonic representation the perturbed
state reads,

\begin{eqnarray}
&& |\epsilon \rangle = \frac{1}{\sqrt{4\pi\beta (2 \pi) !}} \int_{-\infty}^\infty dx e^{-\frac{x^2}{4 \beta}} \nonumber \\ 
&& \int_0^\pi dt e^{i x \left ( \frac{cos(t)}{2} \left ( \hat{a}_+^{\dagger} \hat{a}_+ - \hat{a}_-^{\dagger} \hat{a}_-  \right )      - \frac{sin(t)}{2i} \left ( \hat{a}_+^{\dagger} \hat{a}_- - \hat{a}_-^{\dagger} \hat{a}_+  \right )\right )} 
\left( \hat{a}_-^{\dagger} \right )^{2j} |0_+,0_- \rangle.
\label{mach}
\end{eqnarray}

We now group terms in the second integral as indicated below,

\begin{equation}
e^{i \hat{A} \phi} \left( \hat{a}_-^{\dagger} \right )^{2j} |0_+,0_- \rangle = \left ( e^{i \hat{A} \phi}  \hat{a}_-^{\dagger}  e^{-i \hat{A} \phi}   \right)^{2j} |0_+,0_- \rangle,
\label{erth}
\end{equation}

where,

\begin{equation}
\hat{A} = x \left ( \frac{cos(t)}{2} \left ( \hat{a}_+^{\dagger} \hat{a}_+ - \hat{a}_-^{\dagger} \hat{a}_-  \right )      - \frac{sin(t)}{2 i} \left ( \hat{a}_+^{\dagger} \hat{a}_- - \hat{a}_-^{\dagger} \hat{a}_+  \right )\right ),
\end{equation}

and $\phi$ is just an auxiliary parameter which can be set to $\phi = 1$ 
whenever is convenient. This identity could be verified, for instance, 
just expanding the right hand side of equation (\ref{erth}) and noticing
that,

\begin{equation}
e^{-i \hat{A} \phi} |0_+,0_- \rangle = |0_+,0_- \rangle,
\end{equation}  

which can be seen if the exponential on the left is expanded in power
series. It can be seen, therefore, that finding an analytical expression
for the evolved operator is structurally analogous to the formulation
of the free bosonic model of section \ref{recho}. In fact, here we also 
get a system of equations that looks like,

\begin{equation}
\frac{d}{d \phi}
\left(
\begin{array}{c}
\hat{\alpha}_+ \\
\hat{\alpha}_- \\
\end{array}
\right )
=
-i
\left(
\begin{array}{cc}
-\frac{x }{2} \cos (t)& i \frac{x }{2} \sin (t)\\
 -i \frac{x }{2} \sin (t)& -\frac{x }{2} \cos (t)\\
\end{array}
\right )
\left(
\begin{array}{c}
\hat{\alpha}_+ \\
\hat{\alpha}_- \\
\end{array}
\right ),
\end{equation}

along with the initial condition,

\begin{eqnarray}
\hat{\alpha}_+(\phi=0) = \hat{a}_+, \nonumber \\
\hat{\alpha}_-(\phi=0) = \hat{a}_-,
\end{eqnarray} 

and the usual definition $\hat{\alpha}_{\pm} =  e^{i \hat{A} \phi}  \hat{a}_{\pm}^{\dagger}  e^{-i \hat{A} \phi} $. Therefore the complete expression for the evolved
operator is,

\begin{equation}
\hat{\alpha}_- = e^{i\frac{x}{2}\cos(t)} \left ( \sin \left ( \frac{x \cos(t)}{2}  \right ) \hat{a}_+^{\dagger}  +  \cos \left ( \frac{x \cos(t)}{2}  \right ) \hat{a}_-^{\dagger}  \right ).
\end{equation}

Inserting this result in equation (\ref{mach}) we obtain,

\begin{eqnarray}
&& |\epsilon \rangle = \frac{j}{\sqrt{(2j)!4\pi\beta}} \int_{-\infty}^\infty dy e^{-\frac{j^2 y^2}{4 \beta}}  
\int_0^\pi dt e^{i j^2 x \cos(t)}  \label{eq:four} \\
&& \left ( \hat{a}_+^{\dagger} \sin \left ( \frac{j y \sin(t)}{2} \right )  + \hat{a}_-^{\dagger} 
\cos \left (\frac{j y \sin(t)}{2} \right )    \right )^{2 j}  |0_+,0_- \rangle, \nonumber
\end{eqnarray}

where we have performed the variable change $x=jy$. Expanding the binomial
and applying the boson operators on their respective vacuum states we get,

\begin{eqnarray}
&& |\epsilon \rangle = \frac{j}{\sqrt{4 \pi \beta}} \sum_{k=0}^{2j} \sqrt{\binom{2 j}{k}}
\int_{-\infty}^\infty dy e^{-\frac{j^2 y^2}{4 \beta}} \int_0^\pi dt e^{i j^2 y 
\cos(t)} \nonumber \\
&& \sin \left ( \frac{j y \sin(t)}{2} \right )^{2j-k} \cos \left ( \frac{j y 
\sin(t)}{2} \right )^k |{2j-k}_+,k_- \rangle.
\label{supre}
\end{eqnarray}

Next, we approximate the trigonometric functions by the first term
in their series expansion,

\begin{eqnarray}
& \sin \left ( \frac{j y \sin(t)}{2} \right ) \sim \frac{j y \sin(t)}{2}, & \nonumber \\
& \cos \left ( \frac{j y \sin(t)}{2} \right ) \sim 1. & 
\end{eqnarray}

This approximation is valid as long as the negative quadratic exponential
in the first integral of equation (\ref{supre}) suppresses the contribution
of the trigonometric functions for large arguments. Such is indeed the 
case for a wide range of values of $\beta$ and $j$, since the argument of
the first exponential decreases quadratically while the argument of the second
exponential scales linearly. After some algebraic manipulations we arrive to,

\begin{eqnarray}
&& |\epsilon \rangle = \frac{j}{\sqrt{ 4 \pi \beta}} \sum_{k=0}^{2j} \sqrt{\binom{2 j}{k}}
\int_{-\infty}^\infty dy \left( \frac{j y}{2} \right )^{2j-k} e^{-\frac{j^2 y^2}{4 \beta}} \nonumber \\
&& \left \{ \int_0^\pi dt e^{i j^2 y \cos(t)} \sin (t)^{2j-k} \right \} |{2j-k}_+,k_- \rangle.
\end{eqnarray}

The integral in curly brackets can be worked out in terms of Gamma and Bessel 
functions. So we are left with,

\begin{equation}
|\epsilon \rangle = \frac{j}{\sqrt{4 \beta}} \sum_{k=0}^{2j} \sqrt{\binom{2 j}{k}}
 \Gamma \left ( \frac{2j - k}{2} + \frac{1}{2} \right ) \int_{-\infty}^\infty dy  
e^{-\frac{j^2 y^2}{4 \beta}} \left( \frac{y}{2} \right )^{\frac{2j-k}{2}}  
J_{\frac{2j-k}{2}}(j^2 y )  |{2j-k}_+,k_- \rangle.
\label{donde}
\end{equation}

We now use the first term in the series expansion of the Bessel functions
around $x=0$, 

\begin{equation}
J_{\alpha}(x) \approx \frac{1}{\Gamma(\alpha +1)} \left( \frac{x}{2}  \right )^\alpha,
\end{equation}

which provides a good estimation in the range of small arguments. We expect
this to be a good approximation since the negative exponential in (\ref{donde})
attenuates contributions from large values of the argument. The integral is
therefore reduced to,

\begin{equation}
|\epsilon \rangle = \frac{j}{\sqrt{4 \beta}} \sum_{k=0}^{2j} \sqrt{\binom{2 j}{k}}
\frac{\Gamma \left ( \frac{2j - k}{2} + \frac{1}{2} \right )}{\Gamma \left ( \frac{2j - k}{2} + 1 \right )} \left ( \frac{j}{2} \right )^{2j-k}  \int_{-\infty}^\infty dy e^{-\frac{j^2 y^2}{4 \beta } } y^{2j-k} |{2j-k}_+,k_- \rangle.
\end{equation}

From the integral above we can infer that to first order the contributions 
corresponding to odd positions all vanish. Consequently, through direct
integration and some re-indexing we obtain,

\begin{equation}
|\epsilon \rangle = \sqrt{\pi} \sum_{q=0}^j {  \sqrt{\binom{2 j}{2 q}}  \frac{\Gamma \left ( j - q + \frac{1}{2} \right )}{\Gamma \left ( j - q + 1 \right )}} 
\left ( \frac{\beta}{2} \right )^{j-q} (2j-2q-1)!!   |{2j-2q}_+,2q_- \rangle.
\end{equation}

Additionally, we utilize the following identity,

\begin{equation}
\Gamma \left ( n + \frac{1}{2} \right ) = \frac{\sqrt{\pi}}{2^n} (2n-1)!!,
\end{equation}

which helps simplify the whole expression to,

\begin{equation}
|\epsilon \rangle = \sum_{q=0}^j {  \sqrt{\binom{2 j}{2 q}}  \frac{\Gamma \left ( j - q + \frac{1}{2} \right )^2}{\Gamma \left ( j - q + 1 \right )}} \beta^{j-q} |{2j-2q}_+,2q_- \rangle.
\end{equation}

Now we can use equations (\ref{damx}) and (\ref{cochito}) along with
the original perturbative expansion in equation (\ref{guille}) to establish
a relation for the bosonic operators,

\begin{equation}
\hat{\alpha}_1(t=\pi) =
\hat{a}_N^\dagger     -i \epsilon \sum_{q=0}^{j} \sqrt{\binom{2 j}{2 q}}
\frac{\Gamma \left ( j - q + \frac{1}{2}\right )^2}
{\Gamma \left ( j - q + 1 \right )} 
\beta^{j-q} \hat{a}_{2q+1}^\dagger.
\label{ian}
\end{equation}

\begin{figure}
\begin{center}
\includegraphics[width=0.6\textwidth, angle=-90]{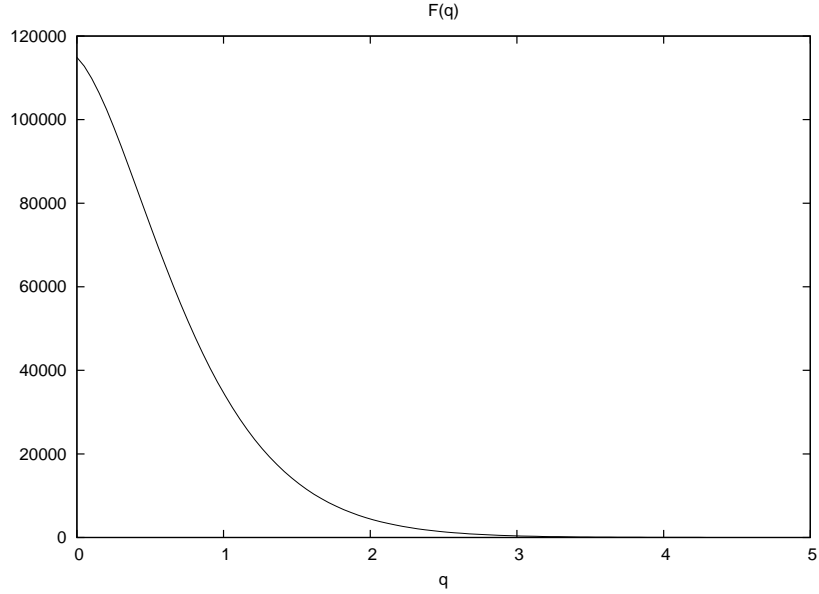}
\caption{Behaviour of function $F(q)$ for $j=5$ and $\beta=5.5$(arbitrary units).} 
\label{cere}
\end{center}
\end{figure}

In order to illustrate our results, we have depicted in figure \ref{cere}
the continuous function,

\begin{equation}
F(q) = \sqrt{\frac{\Gamma(2 j + 1 )}{\Gamma(2j-2q + 1) \Gamma(2q+1)}} 
\frac{\Gamma \left ( j - q + \frac{1}{2}\right )^2}
{\Gamma \left ( j - q + 1 \right )} \beta^{j-q},
\end{equation}

which underline the discrete function in the sum of equation (\ref{ian}). 
Equation (\ref{ian}) shows that when $\beta=0$ the evolution is determined 
by the integrable part of (\ref{bulg}) and therefore only the
creation operator $\hat{a}_N^{\dagger}$ survives. This means that
particles are efficiently transferred across the chain terminals.
For intermediate values of $\beta$, on the other hand, the distribution
function $F(q)$ gets delocalized and particles spread across the chain.
The most interesting case, however, occurs for large values of
$\beta$. Such as it is shown in figure \ref{cere}, the function is
highly localized around values of $q$ corresponding to 
$\hat{a}_1^{\dagger}$. As a result, bosons will not be spread
all over the chain any more, instead, only the terminals will
be macroscopically occupied. As this happens only for large values
of $\beta$, we conclude that this is characteristic of chains with
tightly localized perturbations around the centre. One can compare 
this situation with the dynamics of an individual 
particle in one dimension in the presence of a potential barrier.
From this parallel we can think that when the particles get over
the middle of the chain the perturbation simply acts as a thin 
wall that causes reflection without 
altering the wave packet shape. As a result, reflected particles 
preserve the coherence necessary to drift back into their 
original chain terminal while transmitted particles follow their
predetermined path. This special feature is of potential usefulness 
in scenarios in which particle localization plays a crucial role, 
such as the one we now focus on.

Let us now consider a situation in which particles are originally
prepared in a separable state with all the bosons distributed
evenly between both chain terminals. We use the PTH profile to 
induce coherent particle transmission but we also take into 
account the interaction of the boson clouds around the centre of
the chain (figure \ref{scotish}). Formally, the evolution is 
dictated by equation (\ref{ideal}) with the following coefficients,

\begin{equation}
R_{i,i+1} = \frac{1}{2} \sqrt{i(N-i)} 
\label{yin}
\end{equation}

and,

\begin{equation}
R_{\frac{N}{2},\frac{N}{2}} = R_{\frac{N}{2}+1,\frac{N}{2}+1} = \mu.
\label{yan}
\end{equation}

\begin{figure}
\includegraphics[width=0.35\textwidth,angle=-90]{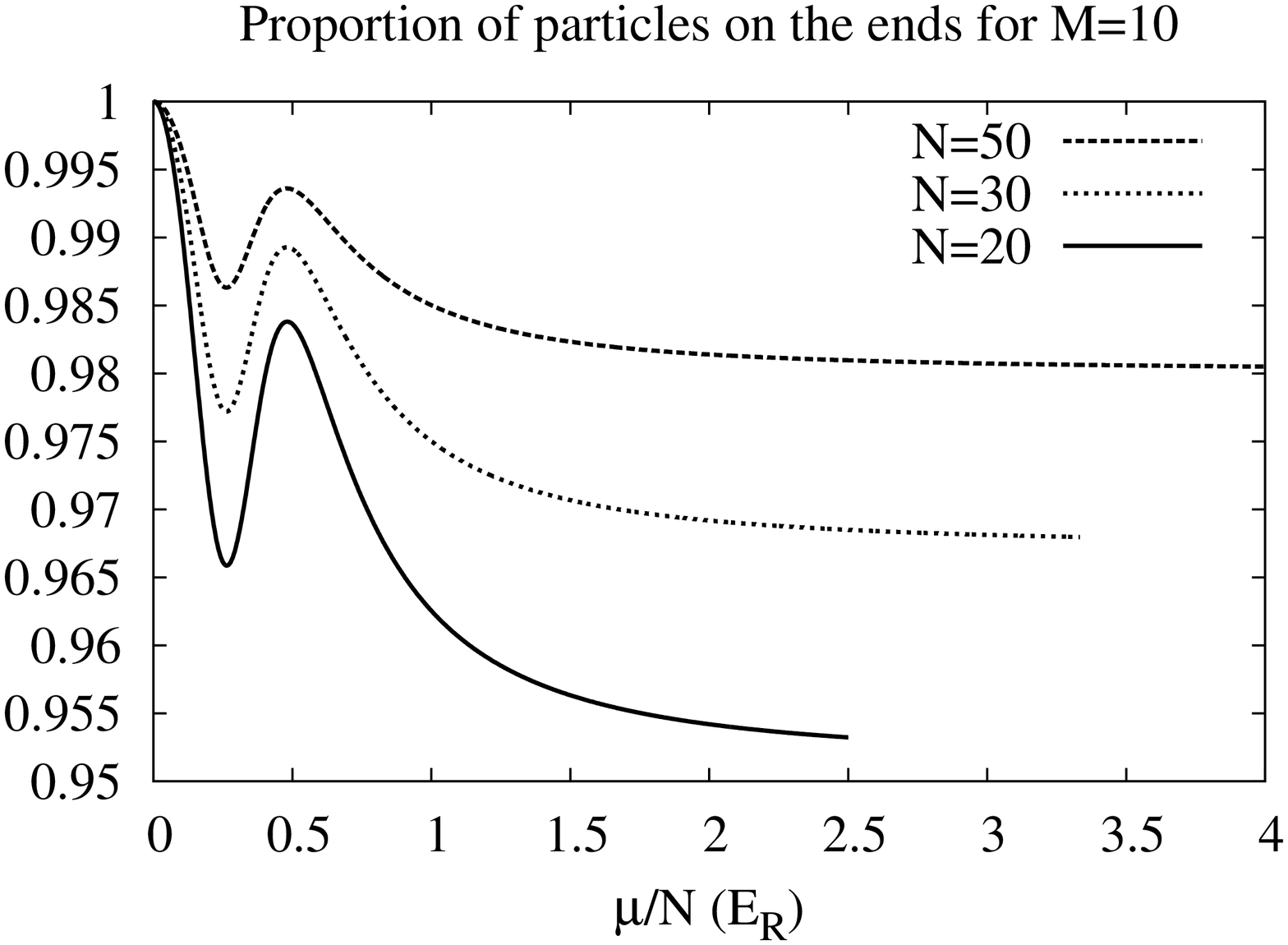}\includegraphics[width=0.35\textwidth,angle=-90]{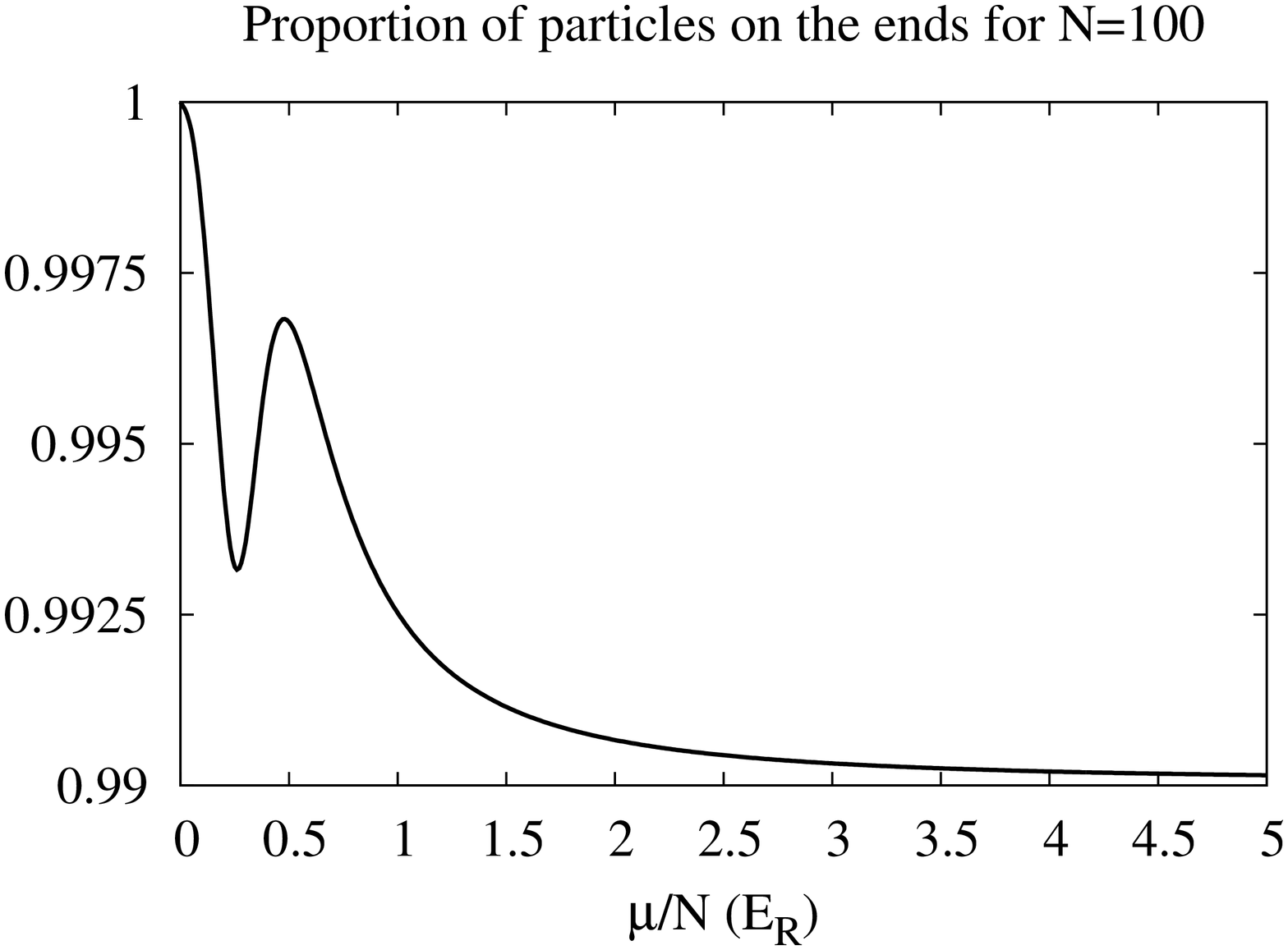}
\caption{Terminal occupation after half a period time. On the left we fixed
the number of bosons and vary the length of the chain. On the right we
maintain the number of sites fixed. In this situation the curves corresponding
to different number of bosons collapse into one single characteristic curve.
} 
\label{mateos}
\end{figure}

Here we assume that the number of sites across the chain is even, since
our algorithms are designed for such particular configurations. Hence,
we assume that the interaction takes place in two central sites, but in
a chain with $N$ odd, a highly localized perturbation can be modelled 
in one single central place. Recall also that the previous analysis
corresponds to chains of odd size, but we expect that our results
are robust against this small discrepancy since the process we study
is quite general and straightforward. Evolution is simulated
applying the two-sum folding method to the state,

\begin{equation}
|\psi(t=\pi) \rangle = \hat{\alpha}_1(t=\pi)^{\frac{M}{2}} \hat{\alpha}_N(t=\pi)^{\frac{M}{2}} |0\rangle,
\end{equation}

where operators $\hat{\alpha}_1$ and $\hat{\alpha}_N$ come from solving
equation (\ref{ideal}) for the coefficients in (\ref{yin}) and 
(\ref{yan}). After the state is written in MPS, we find the
reduced density matrix of the ends of the chain. 
This matrix must be written in the basis of the number of
particles in order to get the logarithmic negativity associated with
the state (sections \ref{entla} and \ref{imple}).

We first observe how efficient the collection process is. 
Figure \ref{mateos} depicts the proportion of particles on the ends once 
the process has taken place, namely,

\begin{equation}
\frac{\left \langle \hat{a}_1^{\dagger} \hat{a}_1 \right \rangle_{t=\pi} + \left \langle \hat{a}_N^{\dagger} \hat{a}_N \right \rangle_{t=\pi} }{2 N}.
\end{equation}

From both plots we can see that the particle proportion is very close
to 1 which means that almost all the bosons are collected on the
chain ends along the domain of $\mu$. This corroborates the
analytical results previously found. In figure \ref{mateos} we use the derived
variable $\frac{\mu}{N}$ because it appropriately rescales the plots
in suitable proportions and puts all the results in a similar scale.
This allows us to see that the collection efficiency notably improves 
when the size of the chain goes up, which is reasonable as in this way the 
perturbation gets more localized in relation to the rest of system.
For a chain of 100 places, collection efficiency remains above $99\%$
independently on the amount of bosons involved. This would be a
very positive aspect in a scattering experiment, where it is important
to have well localized boson clouds in order to measure local
observables. 

\begin{figure}
\includegraphics[width=0.7\textwidth, angle=-90]{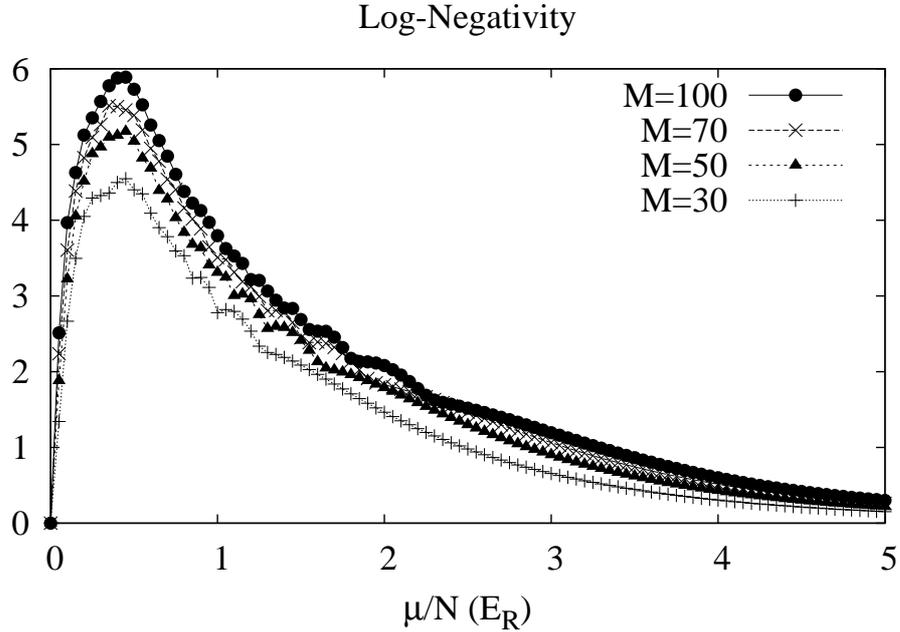}
\caption{End-to-end entanglement (EEE) of a chain of 100 sites for $t=\pi$.}
\label{nome}
\end{figure}

Once the bosons settled down in the chain terminals, the system is
very entangled as a result of the collision. We can understand this entanglement
as coming mainly from the quantum correlations among the dynamical degrees
of freedom at the moment of the interaction, as it is
in the centre of the chain that the bosons become more speedy. However,
such entanglement is difficult to use straight away, not only because 
the particle distribution is very diffuse, but also because it is
difficult to distinguish between highly overlapping boson clouds.
Therefore, it is better to wait until the system comes back to
a zero momentum state, where two static boson bulks lie well separated
from one another and entanglement becomes accessible. From figure \ref{nome}
we can see that the entanglement generated by this mean is quite
substantial for optimal parameters. Additionally, it steadily grows 
according to the number of particles involved in the collision. 
When the interaction parameter is $\mu=0$, the boson packets just
get across each other and no quantum exchange takes place. Under this 
circumstance the possibilities of any entanglement emergence are null. 
Nevertheless, as $\mu$ is slowly turned up, entanglement grows quickly.
In this case the boson clouds interact gently, but this small perturbation
over the integrable path effectively scrambles the degrees of freedom 
and entangle the boson waves. Notice that the entangling process takes
place mainly in the middle of the chain, when the packets are half the
way between the chain terminals. Once the interaction is over, however,
the particles will travel towards one of the chain corners, depending 
on whether they were transmitted or reflected in the collision. 
Interestingly, due to the absence of interaction the entanglement 
between the boson bulks does not build up during the travelling, 
instead, it is kept constant. 

On the right side of the graph in figure \ref{nome} we can observe how
entanglement decays for large values of the interaction constant.
Logically, when the interaction is infinity, bosons just bounce back
in the middle of the chain and return to the original terminal. As
a result, there is no way for places in opposite sides of the chain to
``communicate'' and neither correlations nor entanglement do arise. 
However, the entanglement decay displayed by the system for large
$\mu$ is much slower than the pace at which entanglement soars
close to $\mu=0$.

\begin{figure}
\includegraphics[width=0.7\textwidth, angle=-90]{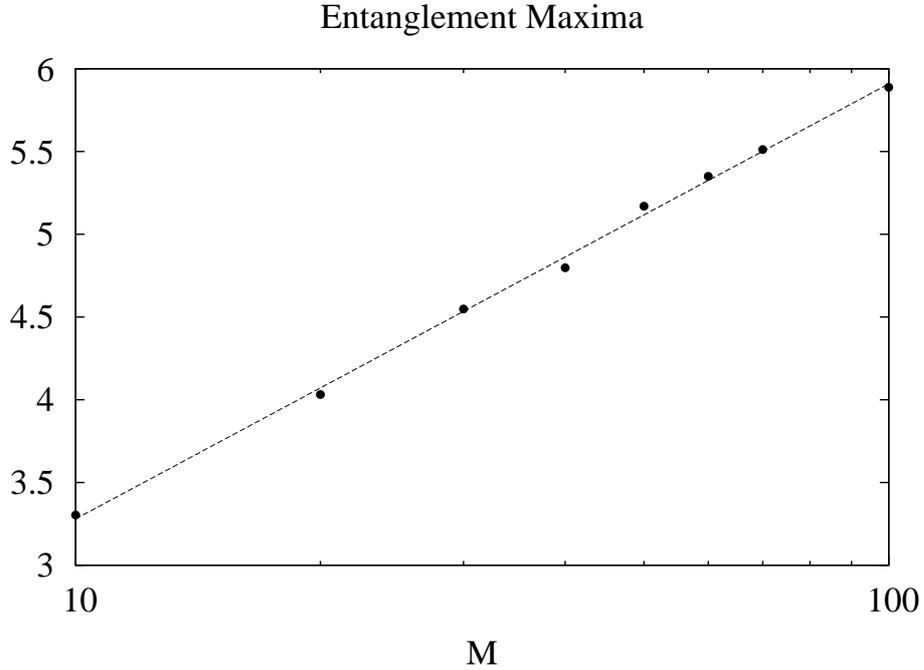}
\caption{Logarithmic dependence of the maxima of figure \ref{nome}.} 
\label{dejes}
\end{figure}

Figure \ref{nome} shows that the best entanglement production
takes place in the range $0.25 < \frac{\mu}{N} < 0.75$. Additionally,
the positions of the maxima are confined to a small interval, making
clear that the process is to some extent generic. Likewise, the values
of the maxima keep a logarithmic relation to the number of bosons, such
as can be appreciated in figure \ref{dejes}. Numerical fitting yields,

\begin{equation}
E_{MAX} = 0.65 + 0.79 log_2 (M).
\end{equation}

This basic behaviour can be explained by formulating the problem
in a slightly simplified way. As the entangling mechanisms are highly
localized in the middle of the chain, we can assume that the amount
of entanglement generated at the moment of the collision is equivalent
to the entanglement between the ends when the process finishes. In
a chain without interaction, the state would be given by,

\begin{equation}
|\psi \left ( t=\frac{\pi}{2} \right ) \rangle  = \hat{\alpha}_1 \left ( t=\frac{\pi}{2} \right )^{\frac{M}{2}}    \hat{\alpha}_N \left ( t=\frac{\pi}{2} \right )^{\frac{M}{2}} | 0 \rangle = \hat{\alpha}^M | 0 \rangle = \left ( \sum_{k=0}^N c_k \hat{a}^{\dagger}_k \right )^M |0 \rangle.
\end{equation}

This result comes from the fact that the particle distribution is
completely symmetric for a time equal to a quarter period. Consequently,
in a chain with no static interaction the Heisenberg operators became
identical and the state can be written as a single sum acting on the
vacuum state. As the focus of the analysis is entanglement, we can
operate on the state using unitary transformations that preserve the
amount of entanglement contained in the system. Additionally, because
most of the particles end up on the ends of the chain and also because
entanglement arises mainly on the interacting zone, we can assume that
the amount of entanglement between the halves of the chain represents
a good estimation of the entanglement between the terminals at the end of
the process. Therefore, we apply the one-sum folding technique 
introduced in section \ref{recho} but this time we do not reduce
the expression to one single creation mode. Instead, we operate on
pairs of operators belonging to different chain halves, starting with
the pairs of operators on the ends and going towards the operators
in the centre of the system. The result of this reduction can be written
simply as,

\begin{equation}
\text{Norm} \times (\hat{a}_{\frac{N}{2}}^{\dagger} + \hat{a}_{\frac{N}{2}+1}^{\dagger})^M |0\rangle.
\end{equation}

Entanglement between the chain halves results from expanding the
binomial,

\begin{equation}
\sum_{k=0}^M \sqrt{\frac{1}{2^M}\frac{M!}{(M-k)!k!}} |M-k \rangle |k \rangle,
\end{equation}

which provides us with the Schmidt coefficients of a symmetric partition,

\begin{equation}
s_k =\sqrt{\frac{1}{2^M}\frac{M!}{(M-k)!k!}} \approx \left ( \frac{2}{M \pi} \right )^{\frac{1}{4}} e^{-\frac{1}{M} \left ( k - \frac{M}{2}   \right )^2 },
\end{equation}

so that the coefficients can be approximated by a normal Gaussian. In
this auxiliary model Log-negativity is given by \cite{LogN},

\begin{equation}
E_{N} = 2 \log_2 \left ( \sum_{k=0}^M {s_k} \right ) \approx 2 \log_2 \left ( \left ( \frac{2}{M \pi} \right )^{\frac{1}{4}}\int_{-\infty}^{\infty} {dx  e^{-\frac{x^2}{M}}} \right ).
\end{equation}

Finally, after some direct calculations we obtain,

\begin{equation}
E_N \approx \log_2 (2 \pi) + \log_2 (M) = 2.65 + log_2 (M).
\end{equation}

Which positively verify the logarithmic behaviour of entanglement 
previously underlined by the numerical analysis. The discrepancy in the
numerical factors comes as a result of the approximations involved 
in the derivation.

In summary, we have shown how our folding technique can be used to
simulate the ground state as well as the dynamics of highly entangled
systems in an efficient way. Here we chose to perform computations in 
problems of physical relevance such as quench and scattering but the
method can be applied to a wide range of problems. In the case of quench,
for example, we found that the wave patterns predicted by the simulations
are characterized by the propapagation of bosonic waves all over the chain. 
For the case of scattering, 
on the other hand, we showed how boson packets get
strongly entangled after they collide and are then collected back on
the chain ends. As the maximum amount of entanglement grows 
logarithmically with the number of particles in the system, the
configuration proposed has potential application in experiments in
optical lattices with direct implications in quantum information
processing.

\chapter{The Bose-Hubbard model and time evolving block decimation}

\label{mede}

The BH model describes a collection of interacting bosons
that can hop among neighbouring sites and undergo on-site repulsion
when more than one boson occupies the same site simultaneously. The
BH model in one dimension is given by the Hamiltonian,

\begin{figure}
\begin{center}
\includegraphics[width=0.9\textwidth]{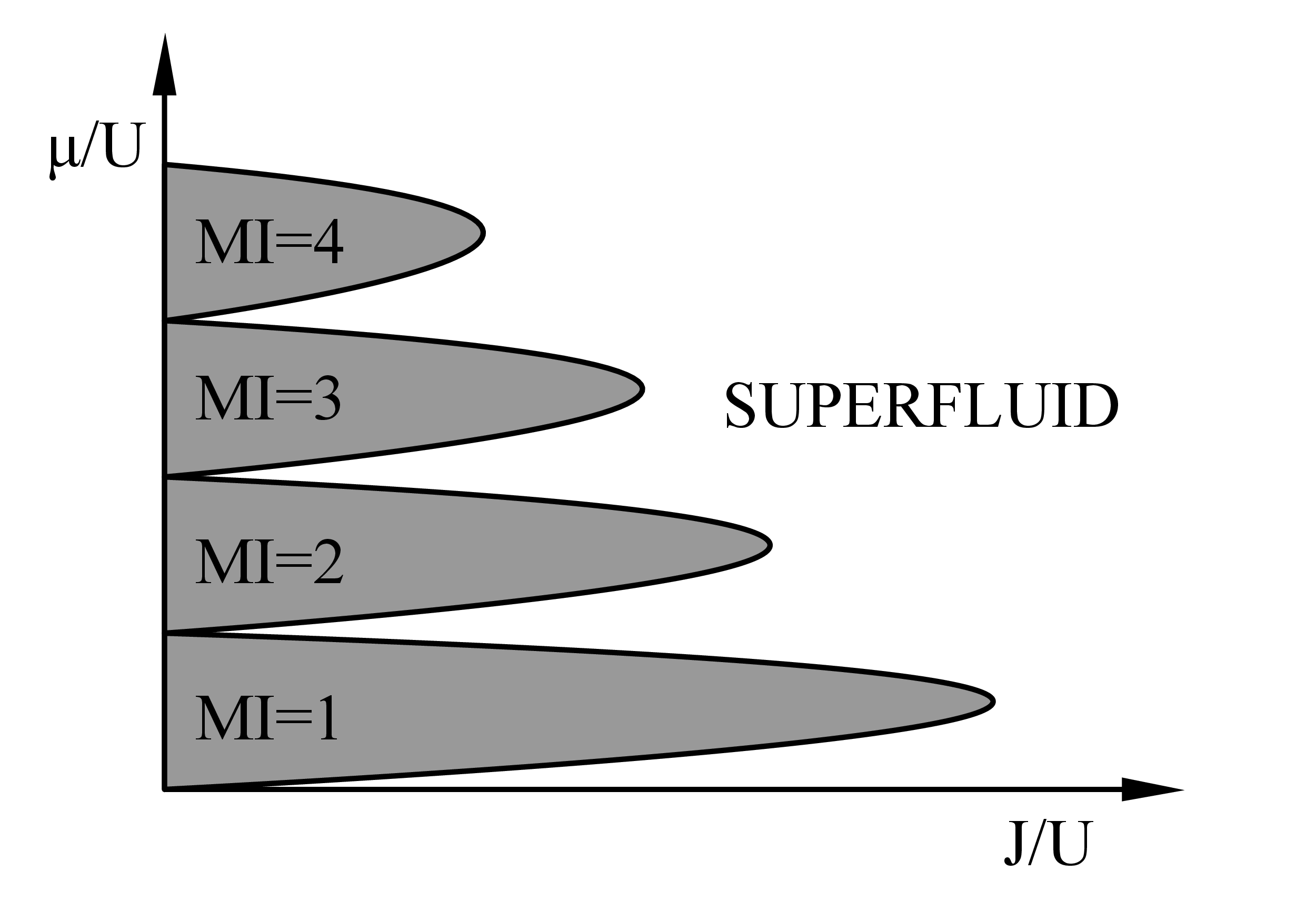}
\caption{Phase diagram of the BH model. Inside the Mott
insulator lobes the number of particles per site remains fixed.} 
\label{canty}
\end{center}
\end{figure}

\begin{equation}
\small{
\hat{H} = \sum_{k=1}^N {\frac{U_k}{2} \hat{a}^{\dagger}_k \hat{a}_k (\hat{a}^{\dagger}_k \hat{a}_k -1)} - \sum_{k=1}^{N-1} {J_k (\hat{a}^{\dagger}_{k+1} \hat{a}_k  + \hat{a}^{\dagger}_k \hat{a}_{k+1}  )}
} +  \sum_{k=1}^N \mu_k {\hat{a}_k^{\dagger} \hat{a}_k }.
\label{BH}
\end{equation}

As usual, the bosonic operators obey the commuting rules introduced before
in equation (\ref{comm}), while $N$ is number of sites in the
chain and the implicit integer $M$ is the number of bosons. The set of $U_k$ 
determine the 
intensity of the repulsion and the set of $J_k$ determine the intensity of the 
hopping across the chain. Finally, the set of $\mu_k$ account for the 
{\it chemical potential}. The BH model has been extensively studied 
since the discovery of superfluity in ${}^4He$ \cite{Fisher}. Although exact
analytical solutions are not available, a lot of insight can be obtained
out of careful inspection and clever mean field approaches. One important 
aspect of the study of the BH model is that such study is frequently carried 
out following a grand-canonical ensemble approach. This means it is
assumed that the boson chain is being constantly fed by a powerful supplier 
that continuously provides bosons as well as energy in order to maintain
the thermodynamic equilibrium. Nevertheless, it is important to mention
that here we do not follow such description. Instead, our chain is 
fully isolated from any outside disturbance and the system state is
given by the ground state of the Hamiltonian or by real-time unitary 
evolution. In figure \ref{canty} the phase diagram of the BH model
interacting with a supplier bath is presented. As can be seen, the
model is characterized by a superfluid phase and a Mott insulator phase.
The difference between the phases is quite clear. In the superfluid
state, bosons display non-vanishing
hopping-scope-length. We can say that this phase is driven by hopping,
since $J>>U$. On the other hand, in the Mott insulator state bosons 
remain tightly fixed on their hosting positions. In this phase hopping
is highly hindered due to the loss of energy balance when the Mott
insulator configuration is slightly disturbed, which results from 
$U>>J$. Formally, the Mott 
insulator phase is characterized by the following properties,

\begin{itemize}

\item Integer or commensurate density.

\item Existence of an energy gap.

\item Zero compressibility.

\item Exponentially decaying correlations 
$\langle \hat {a}_i^{\dagger} \hat{a}_j  \rangle$.

\end{itemize}

The first property is clear from the phase diagram of figure \ref{canty}.
The number of bosons per site is fixed inside the lobes that 
enclose the Mott phase. Commensurate densities correspond 
to configurations where the number of bosons is not an integer multiple 
of the number of sites, but a rational multiple. Over the y-axis and in 
between the lobes, the
Hubbard model describes a superfluid, since the total number of 
particles fluctuates between two fixed values. The second
property is of outstanding experimental importance, since
the energy gap allows us to identify the Mott phase in the laboratory 
by spectroscopically probing the sample \cite{Cirac}. Additionally, 
the gap can be used in EIT experiments, where 
very long storage times have been observed \cite{EIT}. 
The gap refers to the energy cost associated with 
making the bosons in the Mott state to jump out of their original 
positions. The third property means that the
density of the system does not change as the chemical potential is
varied, or equivalently, $\frac{\partial \rho}{\partial \mu} = 0$.
This also can be inferred from figure \ref{canty} by noticing that 
if we move
from inside one of the Mott lobes in the direction of the 
vertical axis the number of particles would remain fixed as long
as we stay in the Mott insulator state. 
We can think of this as a system in which the
number of particles remains fixed despite the channels that
define particle exchange between the system and its environment
remain open.

Conversely, the superfluid is characterized by particle tunnelling 
all across the chain. The main properties of the superfluid are,

\begin{itemize}

\item Can exist at any filling.

\item Eminently gapless.

\item Finite compressibility.

\item In one dimension, power law decay of correlations 
$\langle \hat {a}_i^{\dagger} \hat{a}_j  \rangle$.

\end{itemize}

The first three properties stand opposite to the
properties of the Mott insulator. The fourth property establishes
that the superfluid is liable to develop correlations at long scale.
This can be understood physically by noticing that as bosons jump
from one place to the other, they distribute information all over
the system. This property constitues the basis of the schemes that
will be proposed in subsequent sections in this work. Both the Mott 
insulator state 
as well as the transition to a superfluid have been verified using 
cold atoms in optical lattices \cite{Bloch,Fertig,Stoferle}. 
In a typical experiment, atoms are taken into a 
magnetic trap and then cooled down using an optical lattice of 
retroreflected diode lasers. This creates an 
arrangement of atoms where the resulting optical potential depths 
$V_{x,y,z}$ are proportional to the laser intensities and can be 
expressed in terms of the recoil energy, 

\begin{equation}
E_{R}=\frac{\hbar k^2}{2 m},
\end{equation}

with $m$ the atomic mass and $k$ the wavelength number. To prepare 
1D arrays, two 
lattice lasers are given high intensities in such a way that hopping can 
only efficiently take place across one axis \cite{Stoferle}. In terms 
of experimental parameters, hopping and repulsion coefficients are given 
by, 

\begin{equation}
J= A \left( \frac{V_{0}}{E_{R}} \right )^B e^{-C\sqrt{V_{0}/E_{R}}} E_{R}, 
\end{equation}

and, 

\begin{equation}
U=\frac{2 a_{s} E_{R}}{d} \sqrt{\frac{2 \pi V_{\perp}}{E_{R}}} \left( \frac{V_0}{E_{R}}  \right)^{\frac{1}{4}}, 
\end{equation}

where $V_0$ is the axial lattice 
depth, $V_{\perp}$ the depth of the lattice in the transverse directions, 
$a_s$ the $s$-wave scattering length, $d$ the lattice spacing. Capital
letters represent experimental constants established by the
geometry of the system. For example, for the experiment reported in
reference \cite{Fertig} such constants are \cite{Rey,Danshita1},

\begin{equation}
A=1.397, \hspace{0.5cm} B=1.051, \hspace{0.5cm} C=2.121. 
\end{equation}

Spatial variations in $U$ and $J$ 
can also be implemented using detuned lasers sent through specific sections of 
the lattice as shown in reference \cite{Chang}. 

Due to the vast number of quantum states necessary to span the Hilbert
space, we must utilize MPS in order to be able to handle chain sizes of 
physical relevance. In an isolated boson chain, the conservation of
the total number of bosons along with bosonic statistics determine the
size of the basis needed to write an arbitrary state, namely,

\begin{equation}
\text{basis size} = \frac{(M+N-1)!}{M!(N-1)!}.
\end{equation}

In practical terms, the amount of resources needed to simulate the
system grows exponentially with the number of sites and bosons.
Studies in boson chains using conventional bases have been done in
chains of up to 10 sites and 10 bosons \cite{Roth}.

In order to find the ground state of the chain, we apply imaginary 
time evolution to a given initial state written in MPS. Such an initial 
state must have some overlap with the system ground state. The imaginary 
evolution is explicitly given by the formula, 

\begin{equation}
|\psi_G \rangle = lim_{\tau \rightarrow \infty}
 \frac{e^{-\tau \hat{H}}|\psi_0 \rangle}{||  e^{-\tau \hat{H}}|\psi_0 \rangle ||}.
\label{sile}
\end{equation}

Additionally, we use second order Trotter expansion, which consists in 
splitting the evolution operator in a product of non-commuting operators as,

\begin{equation}
e^{-i \delta t \sum_{k=1}^N{\hat{A}_{k,k+1}}} = e^{-i \frac{ \delta t}{2} \sum_{k=1}^{N/2} {A_{2k-1,2k}}} e^{-i \delta t \sum_{k=1}^{N/2-1} {A_{2k,2k+1}}} e^{-i \frac{\delta t}{2} \sum_{k=1}^{N/2} {A_{2k-1,2k}}}.
\end{equation} 

\begin{figure}
\includegraphics[width=1.\textwidth]{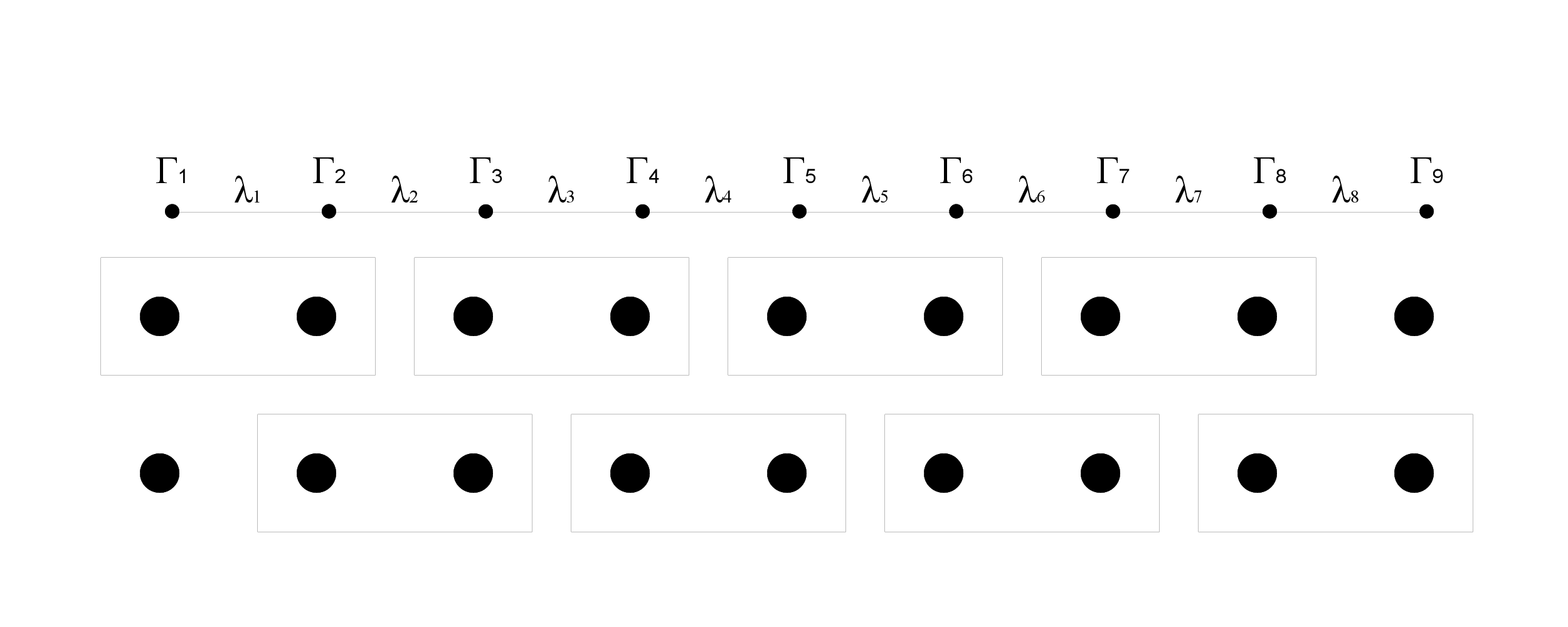}
\caption{Sketch of operator splitting.} 
\label{decom}
\end{figure}

Each operator in the above product can be split in a product of
several commuting operators, each corresponding to a specific pair of 
neighbour sites as shown in figure \ref{decom}. After every updating step, 
in which the state is evolved using the Trotter formula, we retain all the 
Schmidt coefficients greater than $\lambda_{1}^{k} \times 10^{-14}$, 
where $\lambda_{1}^{k}$ is the greatest coefficient at site $k$. 
Then we use the Schmidt coefficients to find out the canonical
representation. The maximum number of Schmidt coefficients in
the chain is denoted by the letter $\chi$. The larger $\chi$ the
more expensive it is for us to simulate the chain. In our program we 
use time slices in the range $10^{-5} < \delta \tau < 10^{-3}$ 
depending on the chain size.
Usually, simulations in long chains as well as very entangled chains 
require the smallest $\delta \tau$. Ground state convergence is evaluated 
through the criterion,

\begin{equation}
|1-\langle \psi(\tau) | \psi( \tau + \delta \tau)) \rangle |<10^{-14}.
\end{equation}

\begin{figure}
\includegraphics[width=0.35\textwidth,angle=-90]{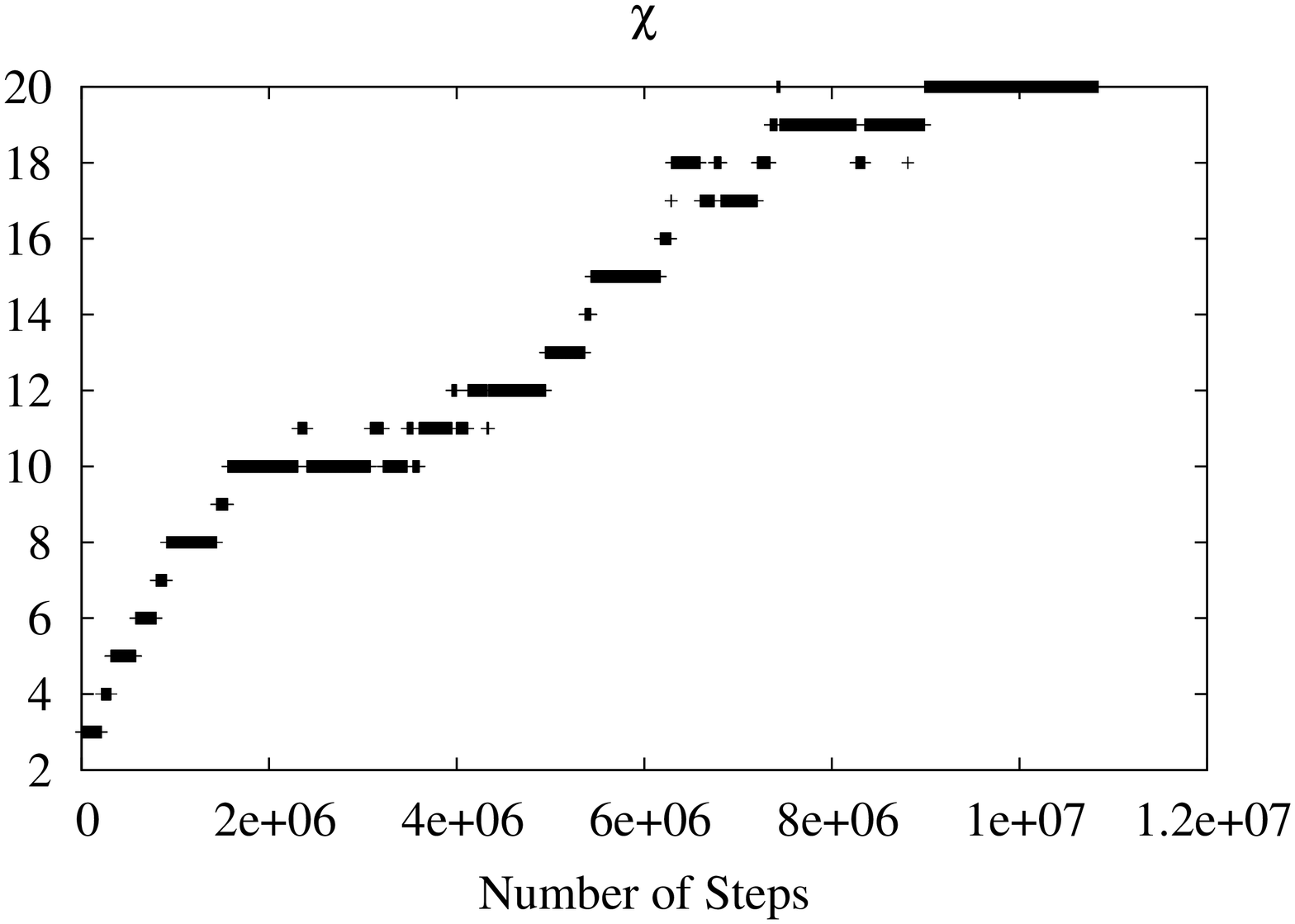}\includegraphics[width=0.35\textwidth,angle=-90]{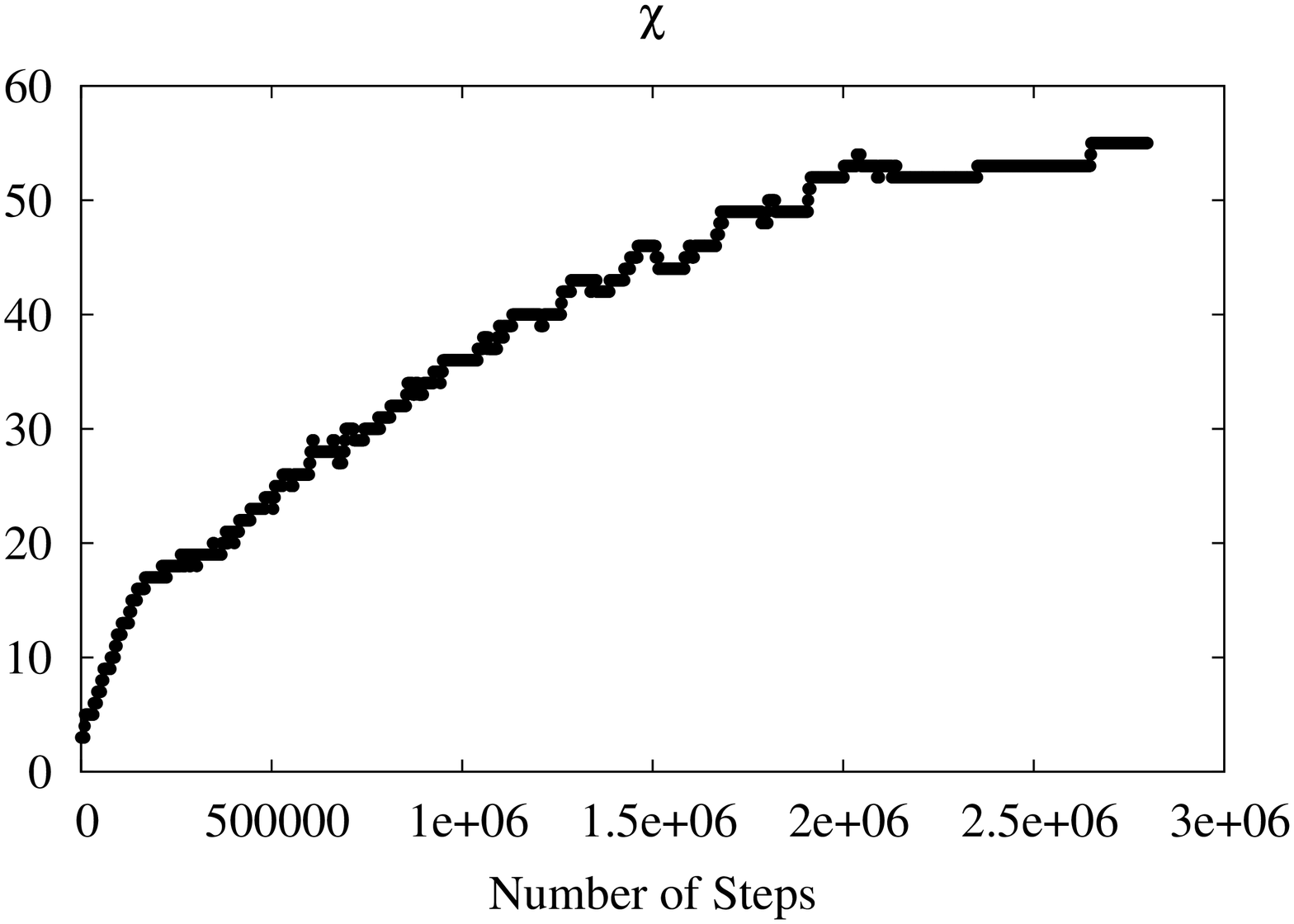}
\includegraphics[width=0.35\textwidth,angle=-90]{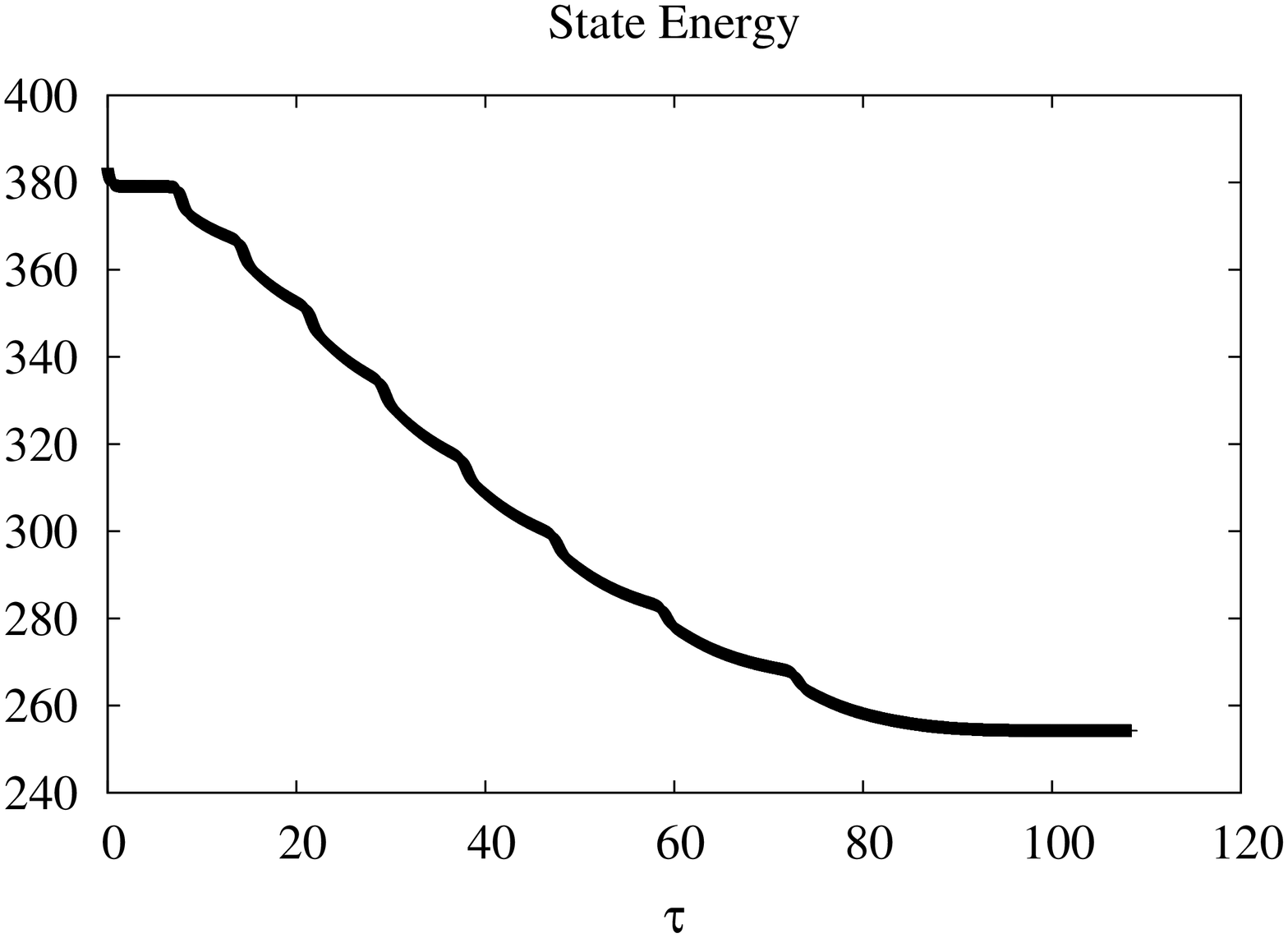}\includegraphics[width=0.35\textwidth,angle=-90]{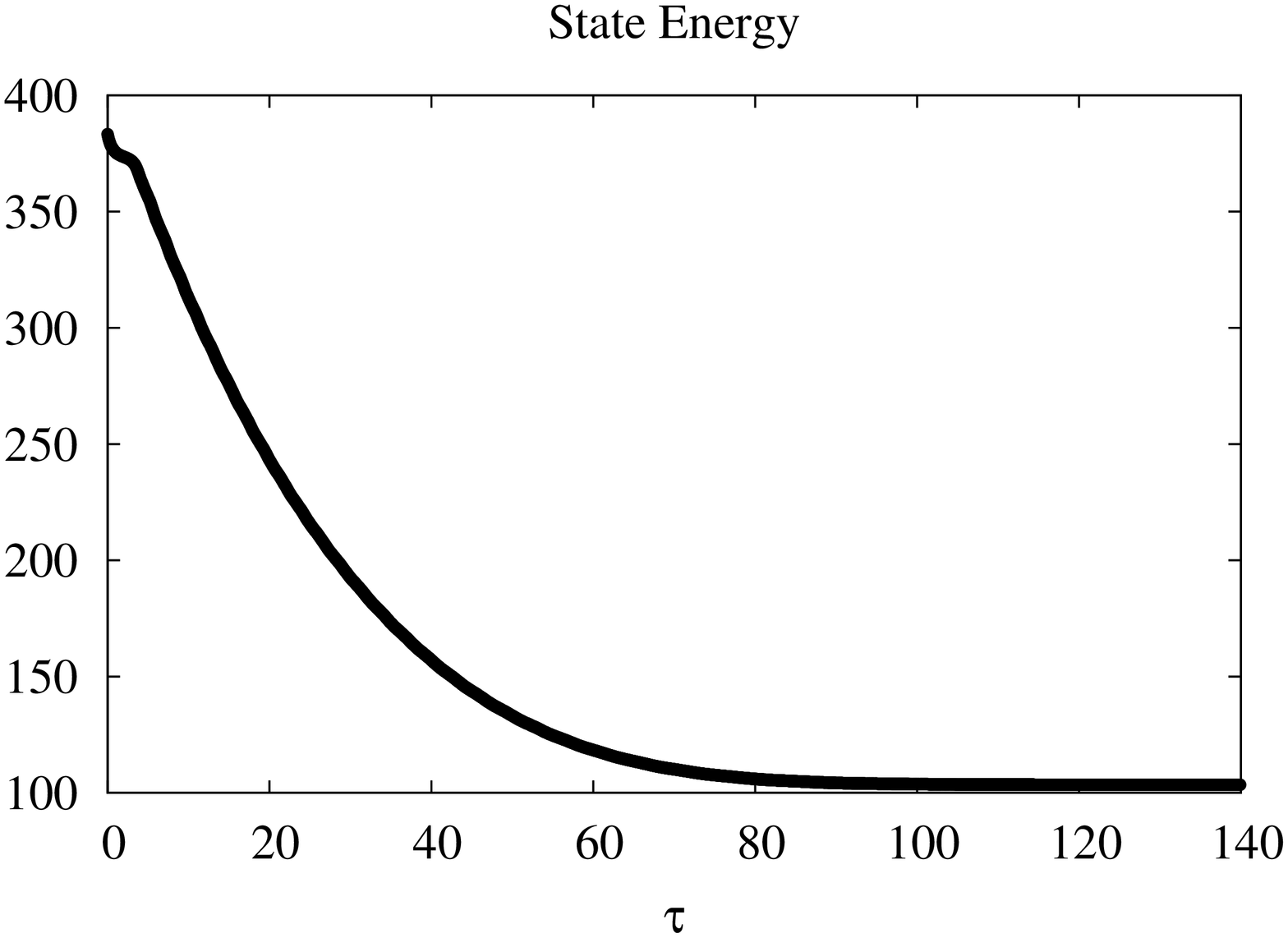}
\includegraphics[width=0.35\textwidth,angle=-90]{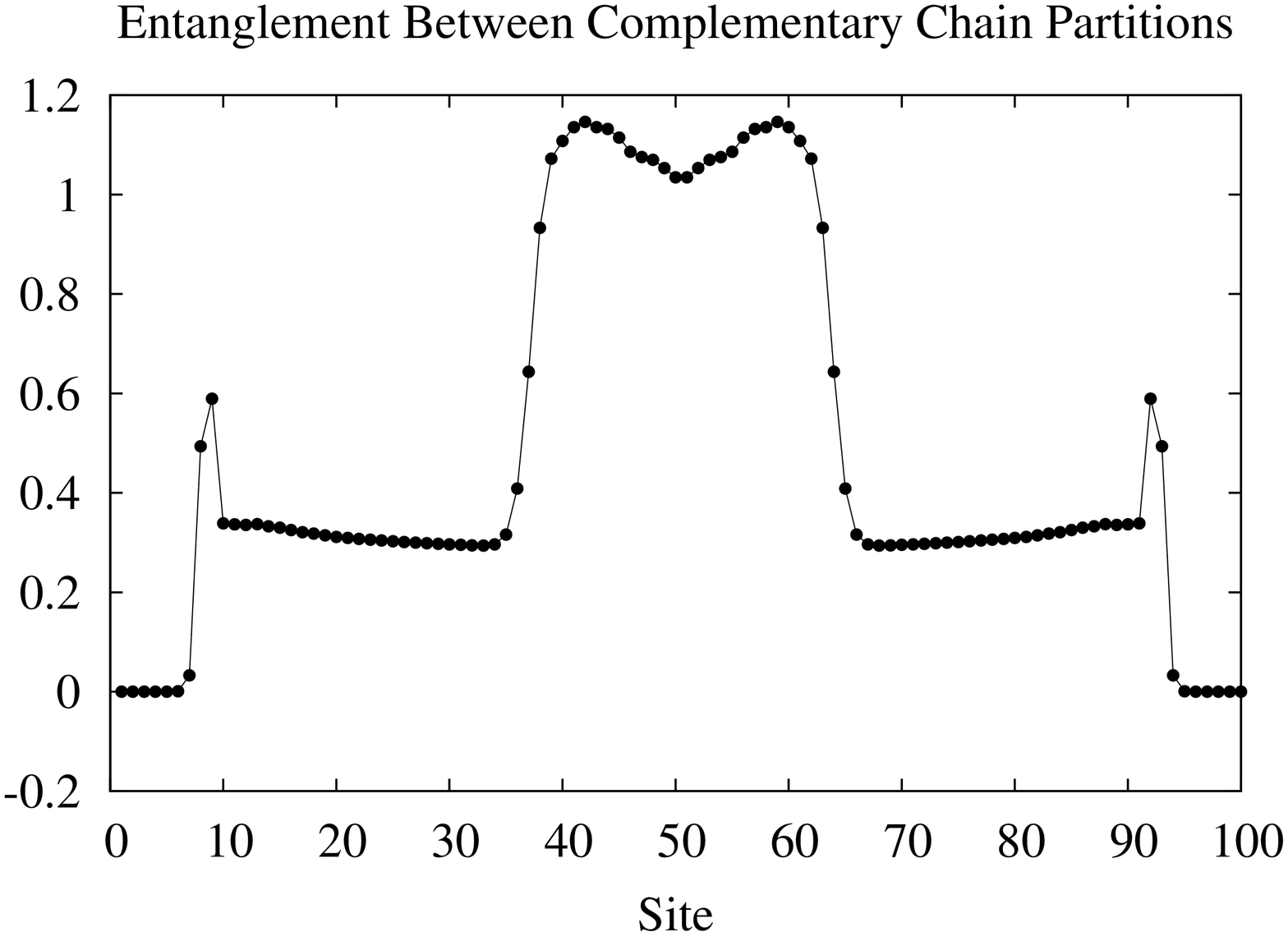}\includegraphics[width=0.35\textwidth,angle=-90]{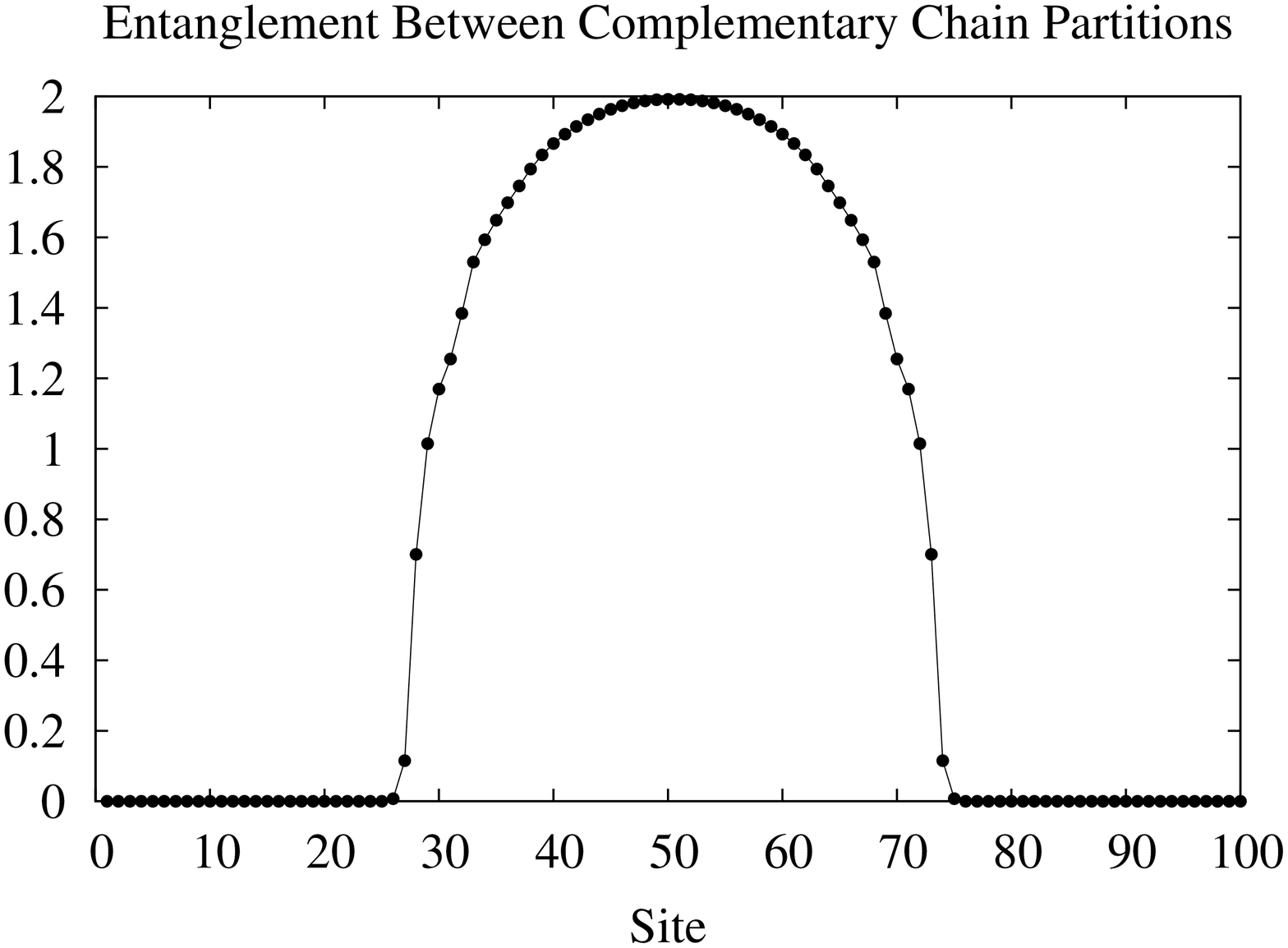}
\includegraphics[width=0.35\textwidth,angle=-90]{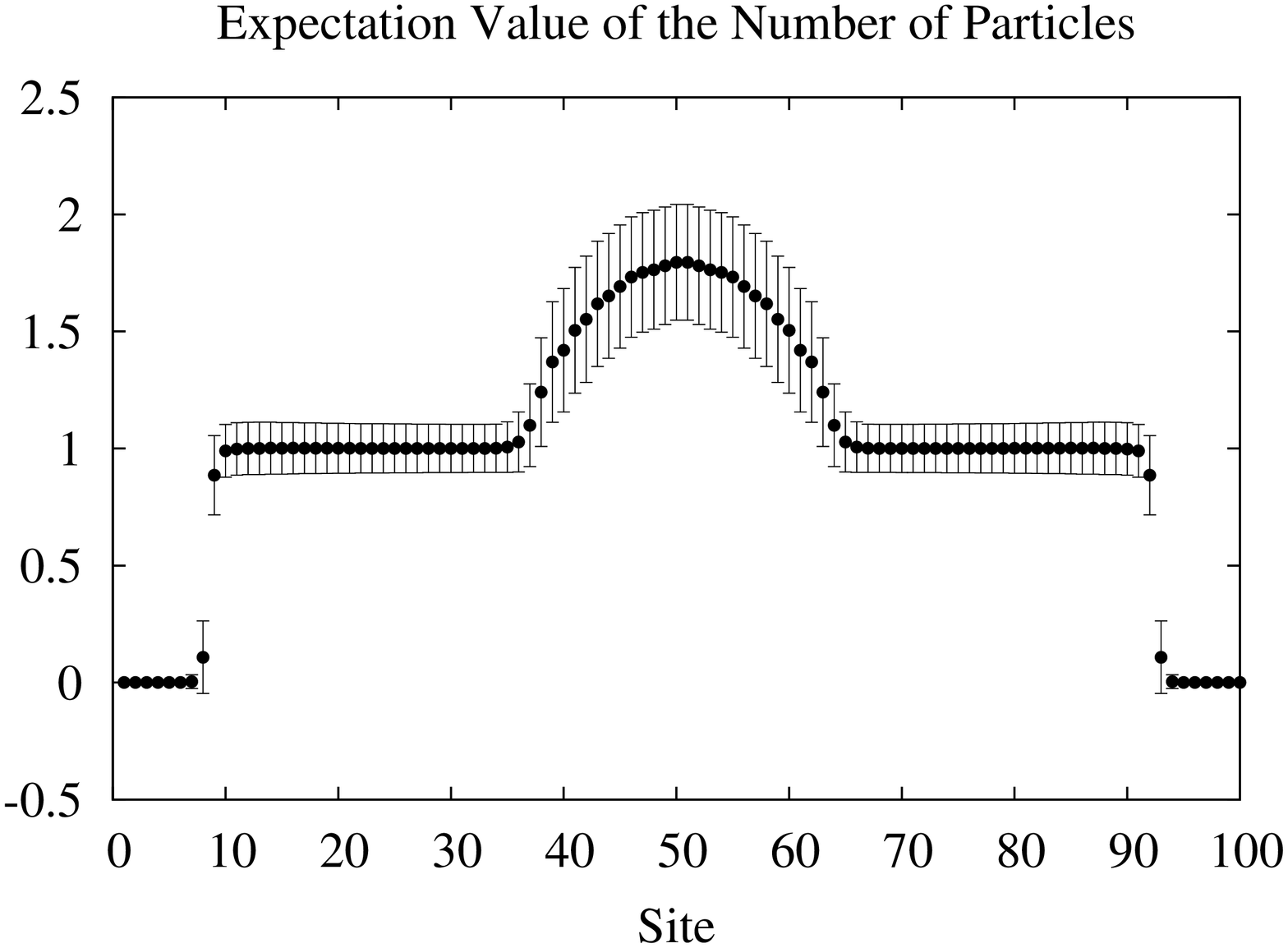}\includegraphics[width=0.35\textwidth,angle=-90]{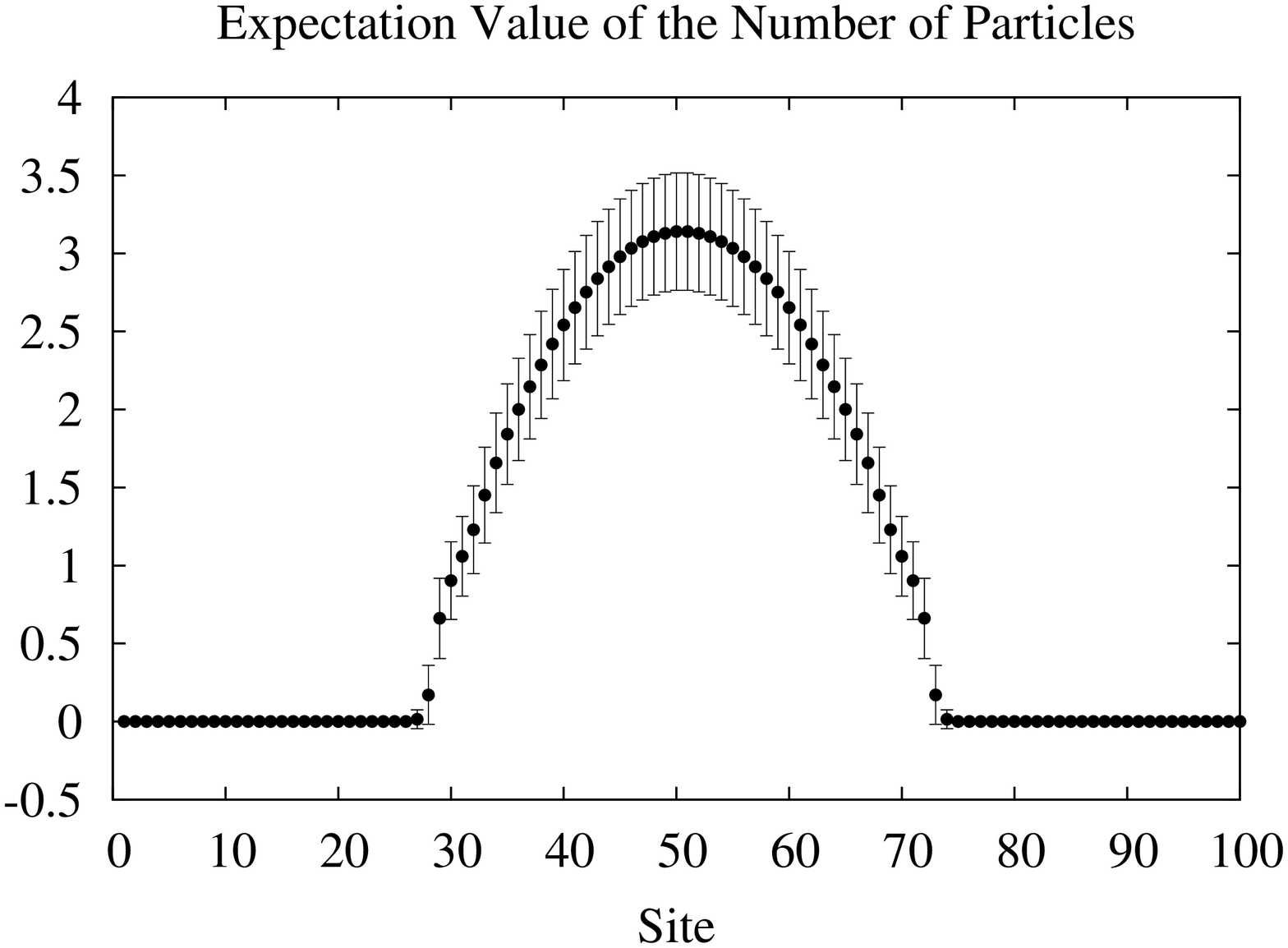}
\caption{Simulations in boson chains of 100 sites and 100 bosons using TEBD. 
The error bars in the lower graphs represent fluctuations. Left: $U_k=2$, 
$J_k=0.14$, $\mu_k = 0.0046(k-50)^2$ and $\delta \tau = 0.00001$. Right: $U_k=0.42$, $J_k=0.14$, $\mu_k = 0.0046(k-50)^2$ and $\delta \tau = 0.00005$. 
All these constants are given in arbitrary units.} 
\label{SFMT}
\end{figure}

The accuracy of the final state depends on factors such as 
the length of the time slice, the original state, and the parameters
in the Hamiltonian. In figure \ref{SFMT} we show simulations for trapped 
bosons in chains with different repulsion. The chosen regimes are
compatible with experimental observations of Mott insulator (driven by
repulsion) and superfluid (driven by hopping). The original state
is a separable state for which the canonical decomposition can be 
easily written, namely, a state with one boson in each site. In this
case the canonical decomposition can be written as,

\begin{eqnarray}
&\lambda_1^{[n]} = 1, \forall n & \nonumber \\
&\Gamma_{1,1}^{i[n]} = \delta_i^1, \forall n. &
\end{eqnarray}

\begin{figure}
\begin{center}
\includegraphics[width=0.6\textwidth,angle=-90]{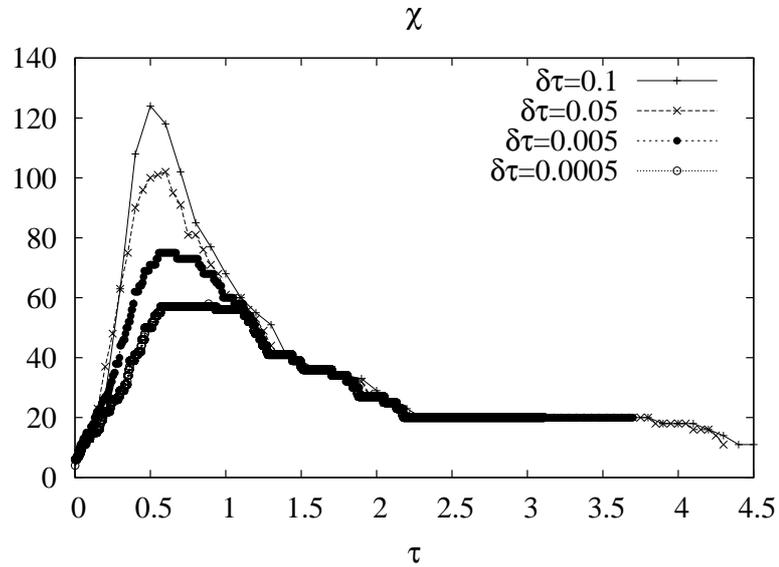}
\caption{Ground state convergence in PTH-repulsionless chains for 10 sites 
and 10 bosons. These simulations correspond to the energy values in 
table \ref{tab1}.} 
\label{drop}
\end{center}
\end{figure}

For this state $\chi=1$. In a standard simulation, the maximum
number of Schmidt coefficients $\chi$ gradually steps up as in 
figure \ref{SFMT}. In most
cases this variable saturates, but sometimes it instead peaks and
dips, then reaches an equilibrium value. This is particularly the case
for PTH chains with no repulsion. Figure \ref{drop} shows the 
behaviour of $\chi$ in imaginary time evolutions corresponding to
different time slices. From such graph we can see that simulations
with large $\delta \tau$ require large $\chi$. However, the
number of steps necessary to reach convergence is fairly low. On the
other hand, when the size of $\delta \tau$ is reduced, the maximum
$\chi$ goes down while the number of steps necessary to converge the
state goes up. We can actually establish the appropriate value of
$\chi$ by comparing the energy of the obtained ground state with
the exact analytical results. We know that the energy of a 
PTH chain with $M$ bosons is given by,

\begin{equation}
E_G = -\lambda \left(\frac{N-1}{2} \right) M.
\end{equation}

Similarly, we can estimate the energy $E_G$ of the state written in
tensor notation from the change in the state norm assuming that
the state itself is not hugely different from the ground state,

\begin{equation}
e^{-\delta \tau \hat{H}} | \psi_G \rangle = e^{-\delta \tau E_G}| \psi_G \rangle.
\end{equation}

Table \ref{tab1} shows the energy values obtained in this way. We 
can see that states obtained using relatively big $\delta \tau$ slightly 
underestimate the ground energy. In this case we can say that the
theoretical error involved in decomposing the evolution operator
in a product of small unitary transformations is more influential
than roundoff errors. Additionally, we can see there is
an optimal value of $\delta \tau$ for which the ground state energy 
contains the maximum number of significant correct figures without
it going below the actual ground energy. Similar analyses in
chains of different sizes show that in PTH chains optimal 
convergence is achieved using $\delta \tau = \frac{10^{-3}}{N}$. In 
the general case, the optimal size of $\delta \tau$ is inversely
proportional to the scope of the boson tunnelling in the 
ground state, which depends on the intensity of the hopping.
Hence, for arbitrary hopping and repulsion profiles we can
write,

\begin{equation}
\delta \tau = \frac{10^{-3}}{N_e},
\end{equation}

where $N_e$ is the effective tunnelling length of bosons in
the state, which must be estimated out of physical insight. 

\begin{table}
\begin{center}
\begin{tabular}{|c|c|} \hline
$\delta \tau$ & $E_G$ \\ \hline
0.1     & -.902854035422649E+02 \\ \hline
0.05    & -.900740540911757E+02 \\ \hline
0.01    & -.900029984667920E+02 \\ \hline
0.005   & -.900007499039339E+02 \\ \hline
0.001   & -.900000299962281E+02 \\ \hline
0.0005  & -.900000074855496E+02 \\ \hline
{\bf 0.0001}  & {\bf -.899999999159858E+02} \\ \hline
0.00005 & -.899999988644993E+02 \\ \hline
\end{tabular}
\end{center}
\caption{Ground state energy in PTH-repulsionless chains with no repulsion for 
10 sites and 10 bosons. These values correspond to the energy graphs in figure 
\ref{drop}. The value closest to $90$, the actual energy, is written in boldface.}
\label{tab1}
\end{table}

As it has been pointed out in several studies \cite{Perales,Kollath}, it is
the variable $\chi$ that ultimately determines the simulation efficiency.
This fact is quite reasonable, on the one hand $\chi$ defines how much
memory is necessary to store the canonical decomposition, and on the
other hand it determines the number of numerical operations that must
be performed in order to upgrade the state \cite{vidal2,Kollath}. 
Similarly, this number is
directly linked to the entanglement profile displayed by the state.
As can be seen in figure \ref{SFMT}, states showing more entanglement
also develop higher $\chi$. Therefore, practical simulations can be 
done in systems where analytical correlations are not too strong. 
In the BH model particularly, we know that close to the Mott 
insulator regime quantum correlations are quite moderate. Additionally,
the superfluid state is characterized by certain regularity which 
somehow keeps $\chi$ from growing indefinitely. However, it is the
state in between the phases what brings more challenge. In this case,
a kind of strongly enhanced entanglement embraces the state and
$\chi$ grows well above manageable limits. According to mean field 
theory, the transition from Mott insulator to superfluid in the
thermodynamic limit takes place at $\frac{U}{J} = 11.6$ \cite{Fisher,Cirac} 
for a chain with unit filling. Figure \ref{SFMT} shows simulations
on each side of such transition point. These graphs
show the characteristics of the Mott insulator and the superfluid.
As can be seen, the expectation value of the number of particles
in the Mott state shows integer density over a wide range of positions 
as well as small fluctuations. The superfluid features large fluctuations,
which anticipates the existence of long range correlations. In general,
the simulations of the superfluid require more computational resources
than the Mott state. As discussed before, the Mott 
state is characterized by the existence of a gap as well as 
integer density. The superfluid on the other hand is
associated with intense boson hopping and long range correlations.

\begin{figure}[t]
\begin{center}
\includegraphics[width=0.35\textwidth,angle=-90]{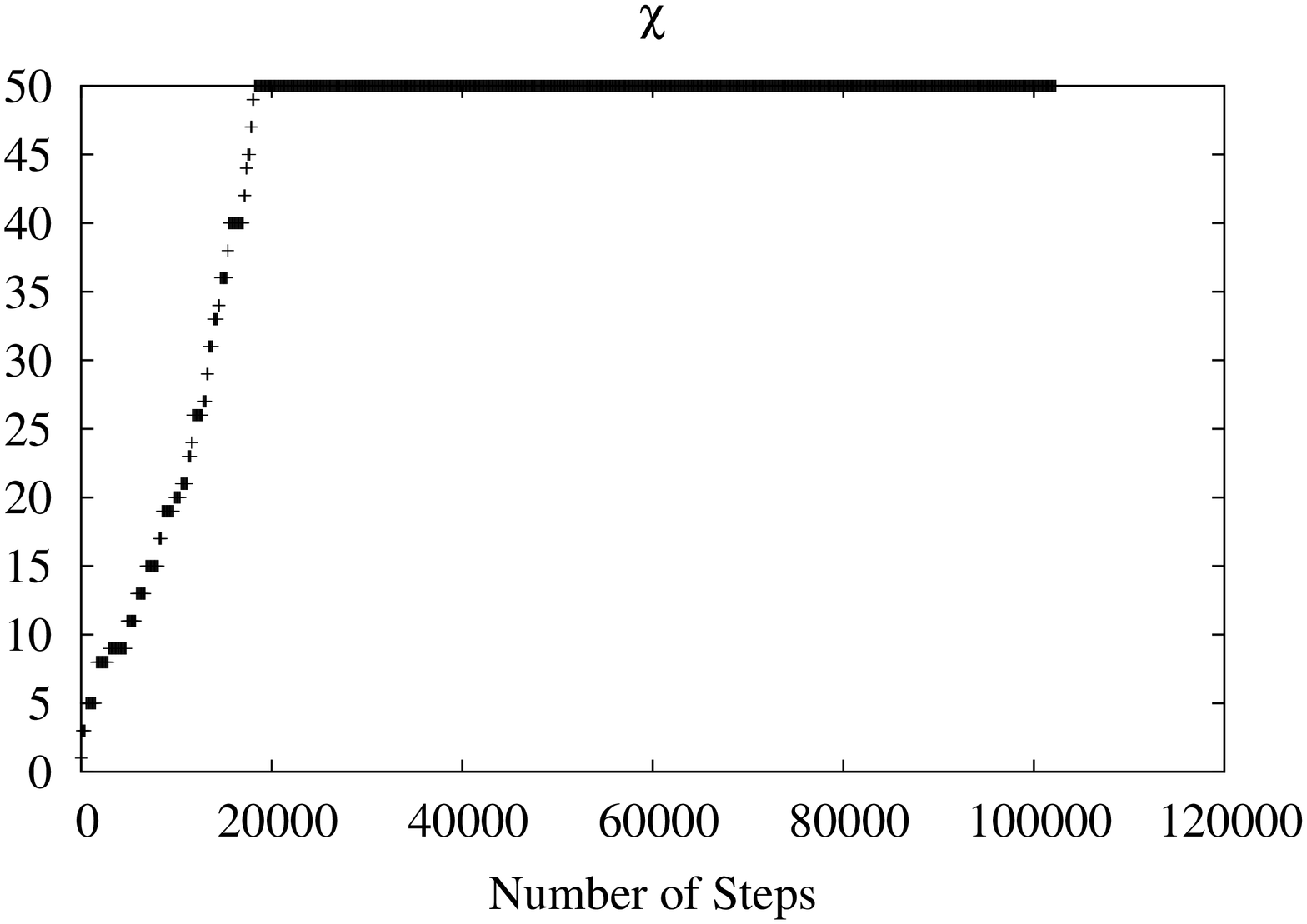}\includegraphics[width=0.35\textwidth,angle=-90]{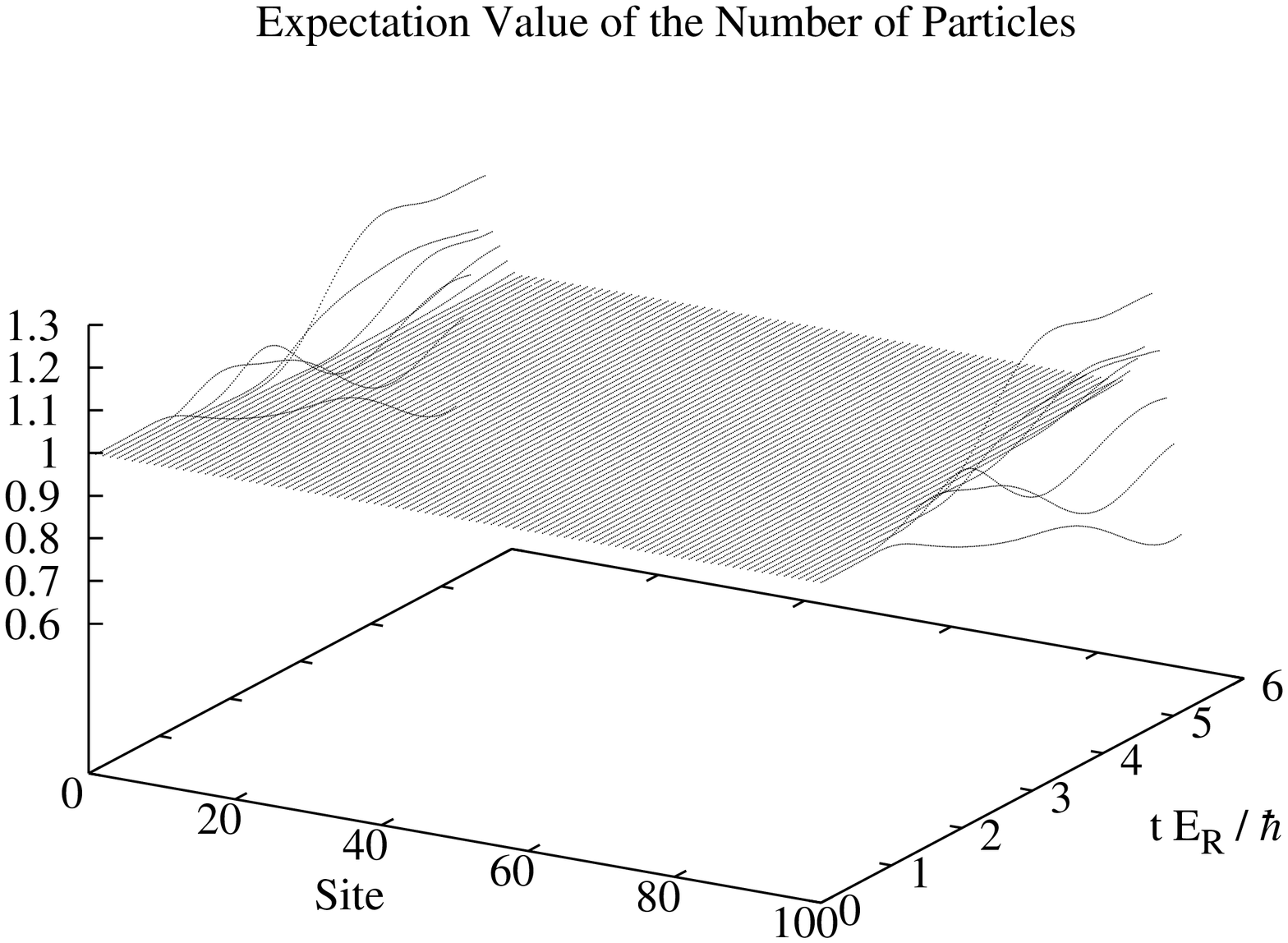}
\includegraphics[width=0.35\textwidth,angle=-90]{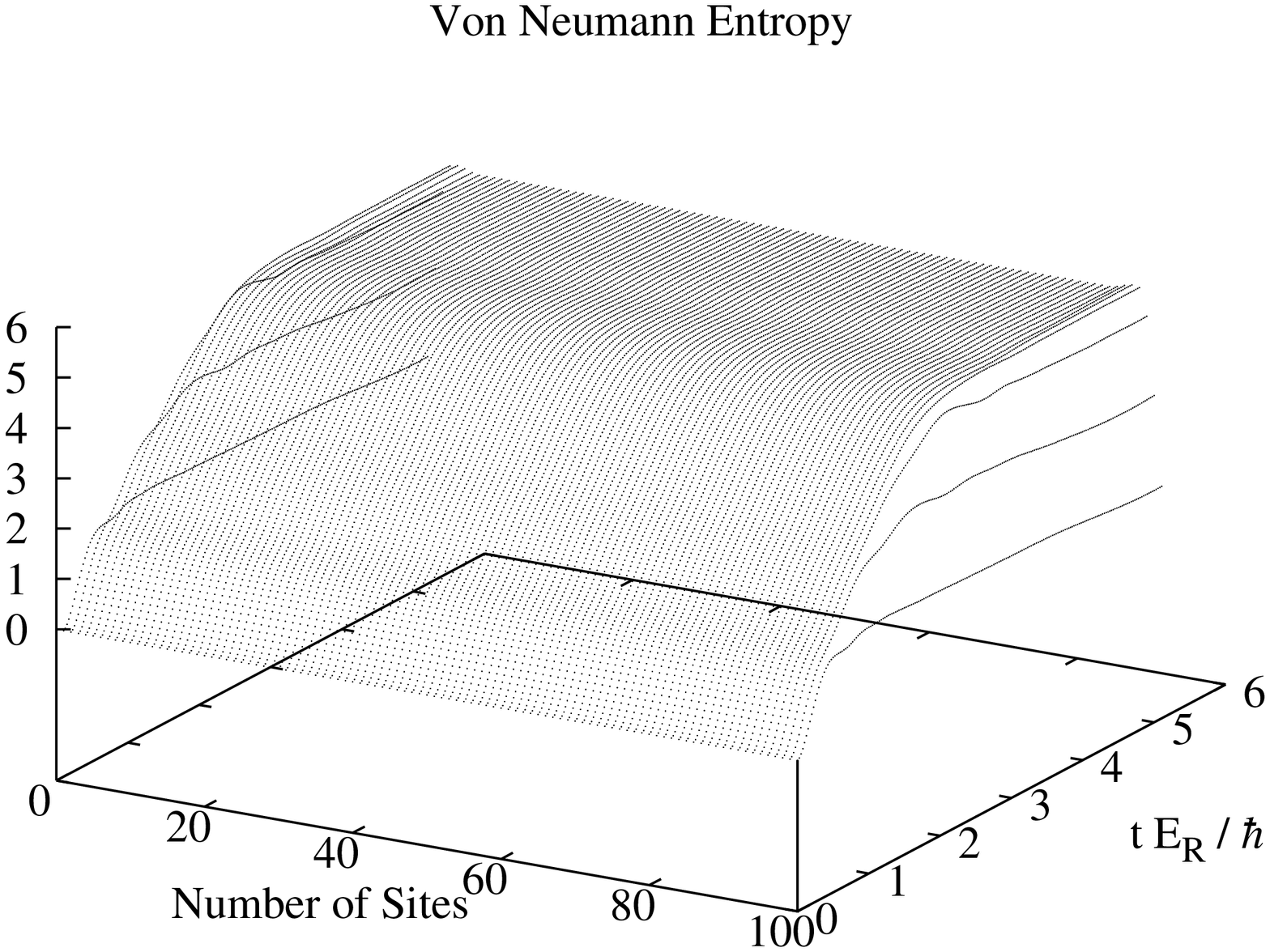}\includegraphics[width=0.35\textwidth,angle=-90]{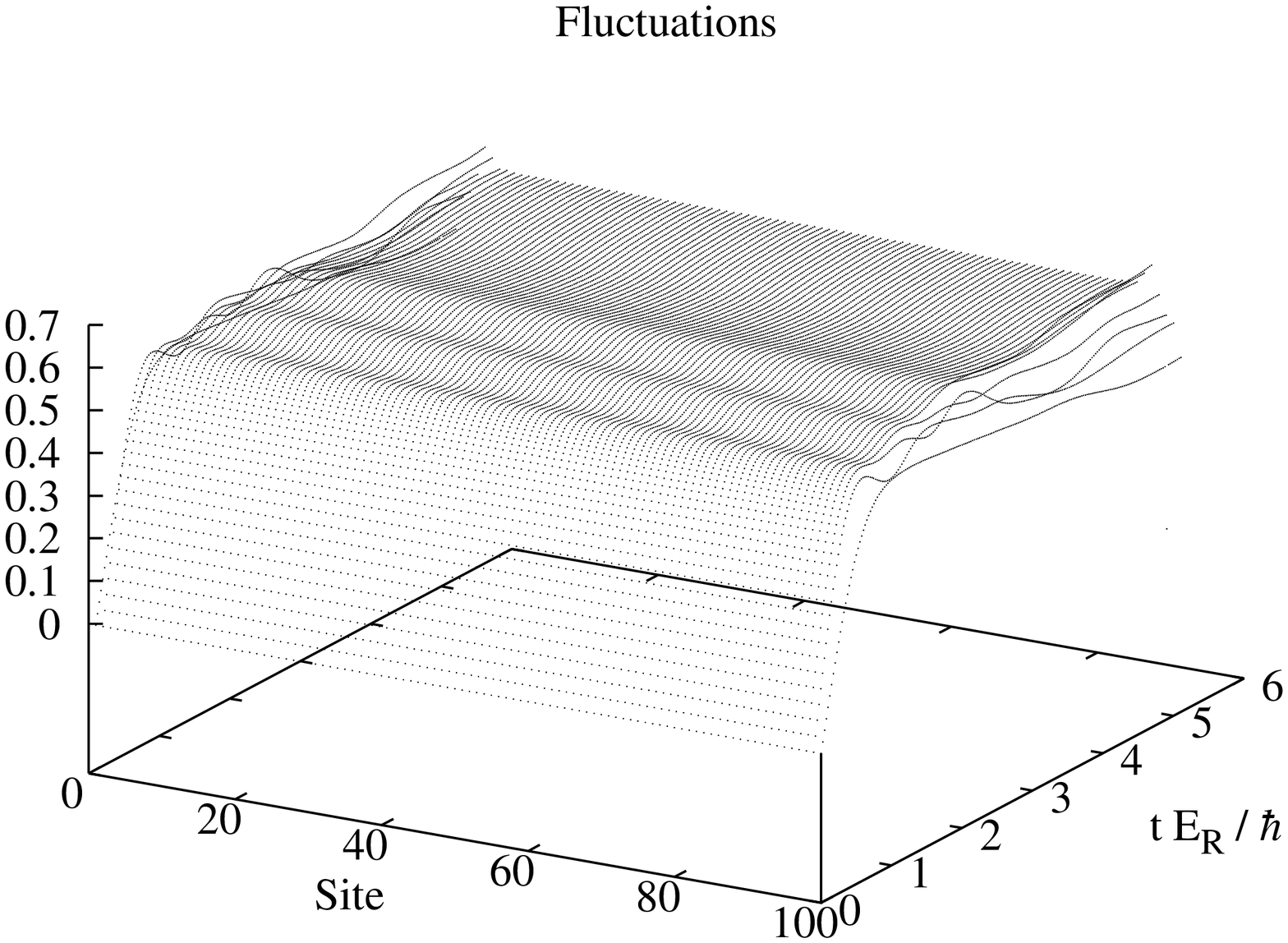}
\caption{Dynamics of a boson chain using TEBD. The system is prepared
in a Mott insulator state with one boson per site. The chain parameters
are characteristic of the superfluid phase, namely, $U_k=0.4E_R$,
$J_k=E_R$ and $\mu_k=0$. In order to get reliable results, we use a very
thin time slice $\delta t = 0.00005 E_R/\hbar$. As a way of keeping the size
of the canonical decomposition at a manageable level, we {\it deliberately} 
fix $\chi_{MAX} = 50$. It has been verified that the same
simulations with $\chi_{MAX} = 60$ generate identical results.}  
\label{zig}
\end{center}
\end{figure}

Similarly to ground state simulations, the performance of TEBD in the 
real time context is highly 
dependent on the variable $\chi$. However, the development of
this variable in dynamical problems is different from the one
observed in imaginary-time evolution. While for the latter case
$\chi$ saturates to a reasonable value, for the real time
case it grows steady, the same that the number of resources necessary to
describe the state \cite{Osborne}. As a consequence, numerical dynamics using 
TEBD is usually restricted to short times regimes, where the 
size of the canonical decomposition can still be handled. Alternatively,
we can set a maximum value for $\chi$. This
allows us to obtain results for extended time intervals in the
range where the system behaviour remains consistent. Figure \ref{zig}
presents several dynamical quantities obtained from simulating a
boson chain prepared in a Mott insulator state with one boson
per site. The apparent lack of dynamics in the graph of the expectation value
of the number of bosons 
can be explained by noticing that in a regularly filled chain 
with one particle in every site the boson hopping 
generates equitable mass exchange, specially around the central part. 
Therefore, the expectation value remains relatively stable 
through very long times. However, the system develops an underlying
dynamics, as can be inferred from the graph of fluctuations,
given by the formula,

\begin{equation}   
f_i = \sqrt{ \left \langle \hat{n}_i^2 \right \rangle - \langle \hat{n}_i \rangle^2},
\end{equation}

where $\hat{n}_i = \hat{a}_i^{\dagger}\hat{a}_i$. As can be seen, 
fluctuations saturate faster than the von Neumann entropy between 
complementary blocks. This suggests that the incidence of high order 
effects in the chain is very important, not least in the short run. 
It also can be seen that the system dynamics close to the chain
ends is notoriously different from the flat patterns displayed
over the central part. These border effects are characteristic
of chains with open boundary conditions \cite{Lauchli}. In the long run
we expect that these border fluctuations propagate all across the
chain and the system acquires a wavy profile. Nevertheless,
long time simulations are likely to contain flawed data as
a result of the truncation of MPS. 

In this section we have given a short review of the BH model, 
emphasising in the aspects that are more relevant to our investigation. 
We have described the use of imaginary-time TEBD in boson chains and its 
relation with the phases associated with the model. Similarly, we have 
discussed the main characteristics of real-time numerics and the way
our simulations are affected by them. In chains with strong correlations, 
which are the focus of subsequent sections, the presence of entanglement 
considerably hinders the simulation efficiency. As a result, we can estimate 
the ground state as well as the dynamics of highly entangled chains in 
systems of moderate size. Further analysis and complementary
results are to be presented in subsequent sections.

\chapter{Entanglement and its relation with physical processes in 
engineered Bose-Hubbard chains}

\label{cart}

As the increasing development of cooling techniques in optical lattices has 
led to the experimental implementation of engineered quantum systems, it has
become important to explore how the quantum nature of such configurations can
be used in quantum computation processing and other interesting fields of
research. Consequently, given the connection between entanglement and quantum
computation, we have concentrated on studying the entanglement in systems that 
are now the object of intensive experimental exploration. In doing so we
have focused on one kind of entanglement that, to our understanding, is most 
useful for developing quantum computation tasks, this is, the entanglement
between distant places of the chain. Likewise, given the volume of technical 
complexities associated with a multi-disciplinary investigation, we have 
narrowed this research to the study of entanglement between the ends of
the chain and the entanglement register provided by the coefficients
of the canonical decomposition. In addition to the characterization of 
entanglement, in this section we look at the relation of such entanglement
with the physical mechanisms associated with the model under scrutiny.

\section{End-to-end entanglement and generalities}

\begin{figure}[t]
\includegraphics[width=1.0\textwidth, angle=-0]{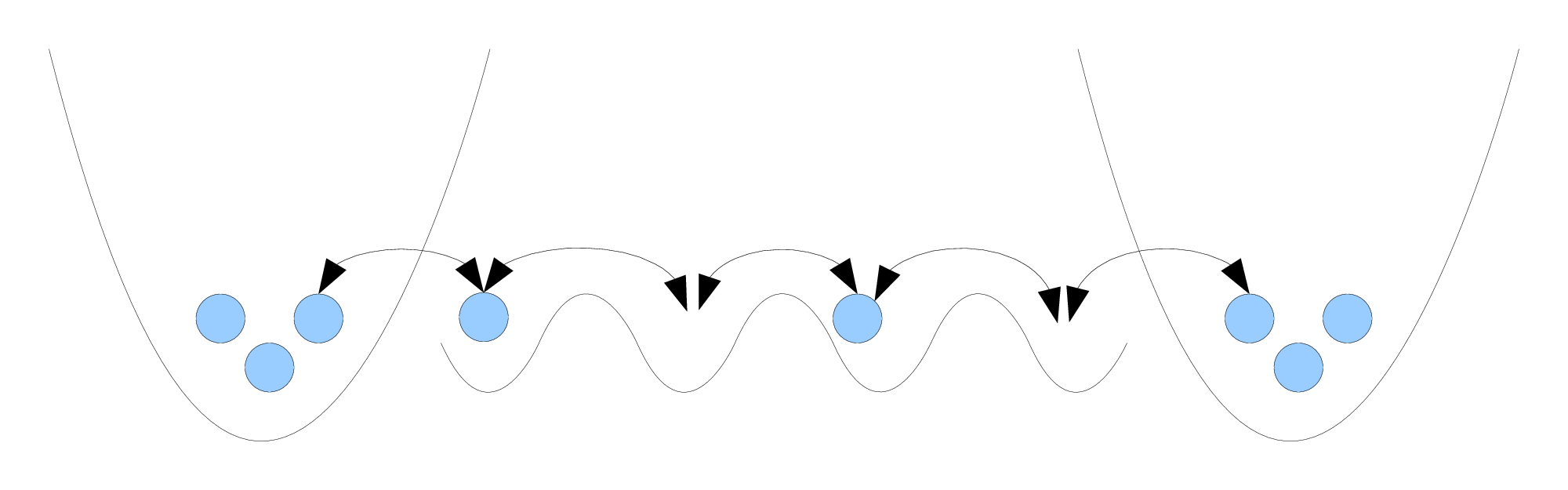}
\caption{Potential realization of BH model with spatially dependent repulsion in an optical lattice.} 
\label{fig:zero}
\end{figure}

In a boson chain with no chemical potential or constant chemical potential
on each site, the Hamiltonian is given by,

\begin{equation}
\small{
\hat{H} = \sum_{k=1}^N {\frac{U_k}{2} \hat{a}^{\dagger}_k \hat{a}_k (\hat{a}^{\dagger}_k \hat{a}_k -1)} - \sum_{k=1}^{N-1} {J_k (\hat{a}^{\dagger}_{k+1} \hat{a}_k  + \hat{a}^{\dagger}_k \hat{a}_{k+1}  )}
}.
\label{BH1}
\end{equation}

The reason for not including additional terms is because this slightly
simplified form allows us to focus exclusively on repulsion and hopping, so that
the effects of chemical potential can be considered on the light
of the results obtained in this simpler scenario. 

\begin{figure}[t]
\includegraphics[width=0.35\textwidth,angle=-90]{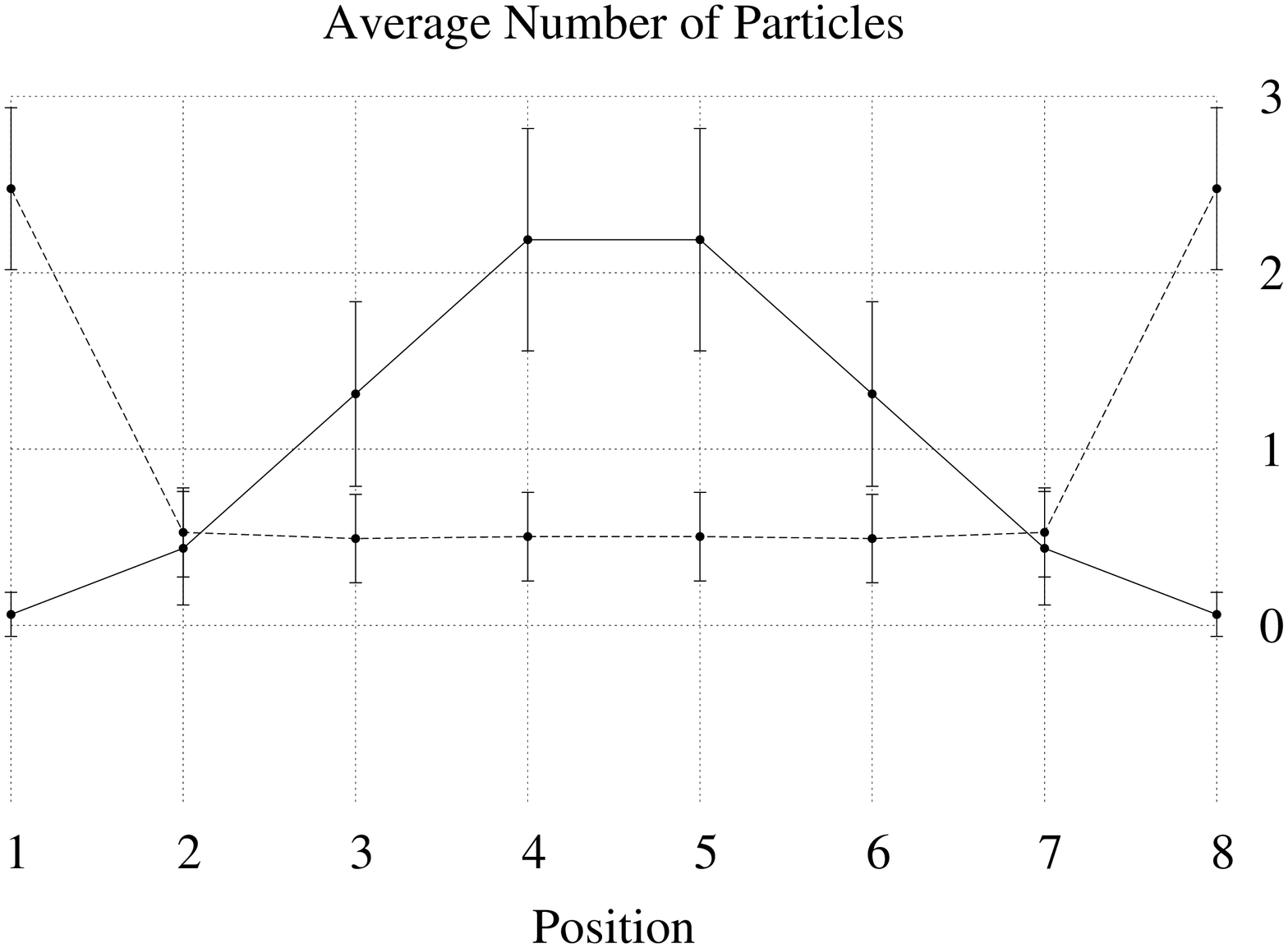}\includegraphics[width=0.35\textwidth,angle=-90]{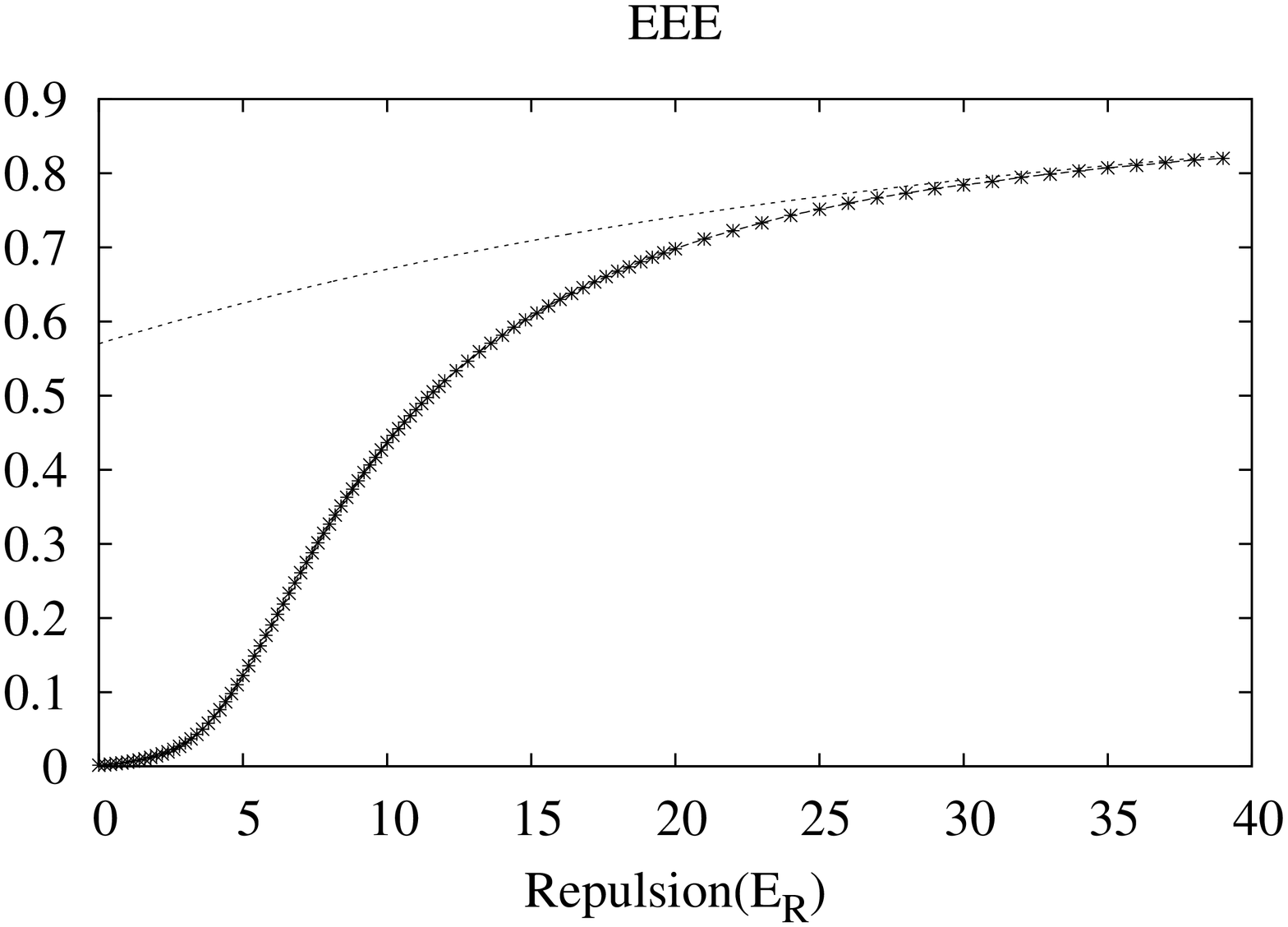}
\caption{Left. Particle distribution profile and fluctuations for the ground state in
PTH chains. (---) No repulsion at all. ($--$) $U_{2-7}$ = 100. Right. Ground state 
entanglement between the ends of a chain of eight places and eight bosons against 
repulsion. PTH and repulsionless ends are assumed . Fitting 
$y = 0.91 -0.34 e^{-0.035 x}$.} 
\label{fig:one}
\end{figure}

From our initial simulations we observe that ground state
entanglement between the ends is highly sensitive to the particle
distribution profile along the chain. This is due to the effect of
tracing out intermediate sites in order to get the reduced density
matrix of the terminals. When population is accumulated in the middle of the chain
we must trace out states with a considerable number of particles in
order to obtain the mixed state of two distant sites at the borders.
Lower population at the ends implies a lower number of possibilities
for distinct particle number states available at the ends and
consequently a lower ability to correlate particle numbers between
these sites and entangle them. Thus for the repulsionless case and PTH,
where population accumulates primarily in the middle of the chain
(which is shown in figure \ref{fig:one}), the
entanglement is vanishingly small. The choice of PTH, as opposed to
a uniform $J_k$, thus clearly illustrates the negative effect of
population concentration in the middle of the chain. To get a higher
entanglement, we then proceed to deliberately frustrate boson
accumulation in the central part of the chain by evenly increasing
the repulsion in every site but the ends which in turn are left
repulsionless (figure \ref{fig:zero}). This approach allows us to look
closely at the physical mechanisms underlying the interplay between
hopping and repulsion and its effects on quantum correlations at
long distances. In figure \ref{fig:one} we can see how
the boson population in intermediate positions concentrate on the
lowest levels of occupation creating a fine arrangement of
superfluid particles that connects the ends, which in turn become
highly populated. Figure \ref{fig:one} also depicts the behaviour of
Log-negativity as a function of repulsion. It can be seen that
entanglement grows monotonically, undergoing saturation for large
repulsion, suggesting that maximum entanglement is achieved when
particles in intermediate places behave as hardcore bosons. In this
limit, bosons find it difficult to stay at the intermediate
positions as the hopping of particles from neighbour sites would
give rise to multi-occupied states with strong repulsion. Instead,
bosons organize in superposition states that on average contain only
half a boson per site for the intermediate sites, creating a
globally delocalized profile that favours long-range quantum
correlations. Similarly, analogous plots corresponding to expectation
values of number of particles also display saturation for high repulsion,
establishing a clear signature of rescaling behaviour, that is,
bosons reorganize under the influence of larger repulsion constants
in such a way that the phenomenology of the system is altered,
roughly, just by a proportionality factor. Similar conclusions can
be drawn for chains with constant coefficients $J_k$ and
repulsionless ends, although PTH chains are more efficient at
enhancing entanglement. For instance, in a chain of ten sites, 
ten bosons, PTH and high intermediate repulsion $U_{2-9} = 100$, 
Log-negativity was found to be $0.83$ against $0.60$ of the same 
chain with uniform unity hopping. In the study ahead, we further 
explore this remarkable behaviour.

\section{Entanglement in the ground state}

\begin{figure}[t]
\includegraphics[width=0.35\textwidth,angle=-90]{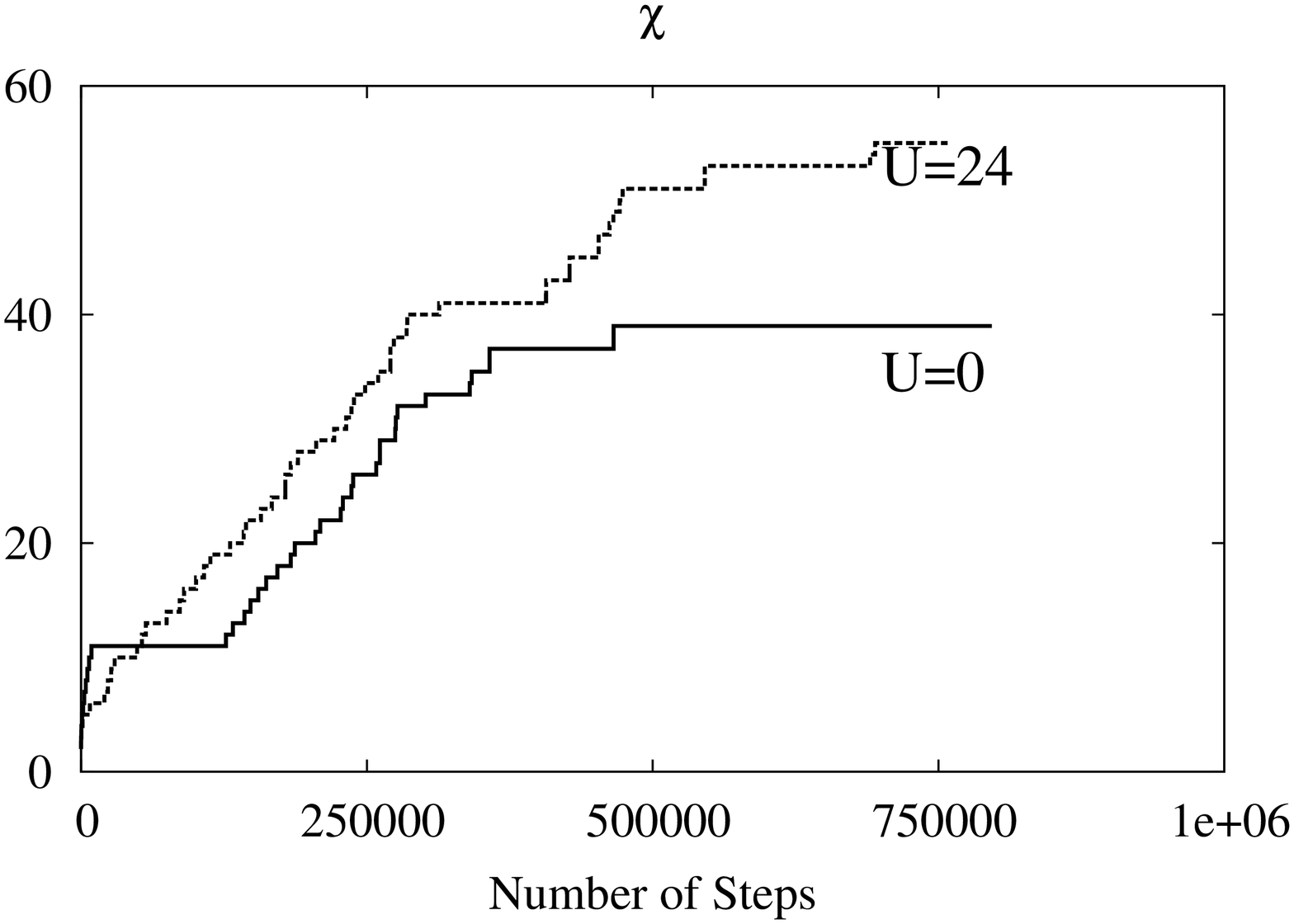}\includegraphics[width=0.35\textwidth,angle=-90]{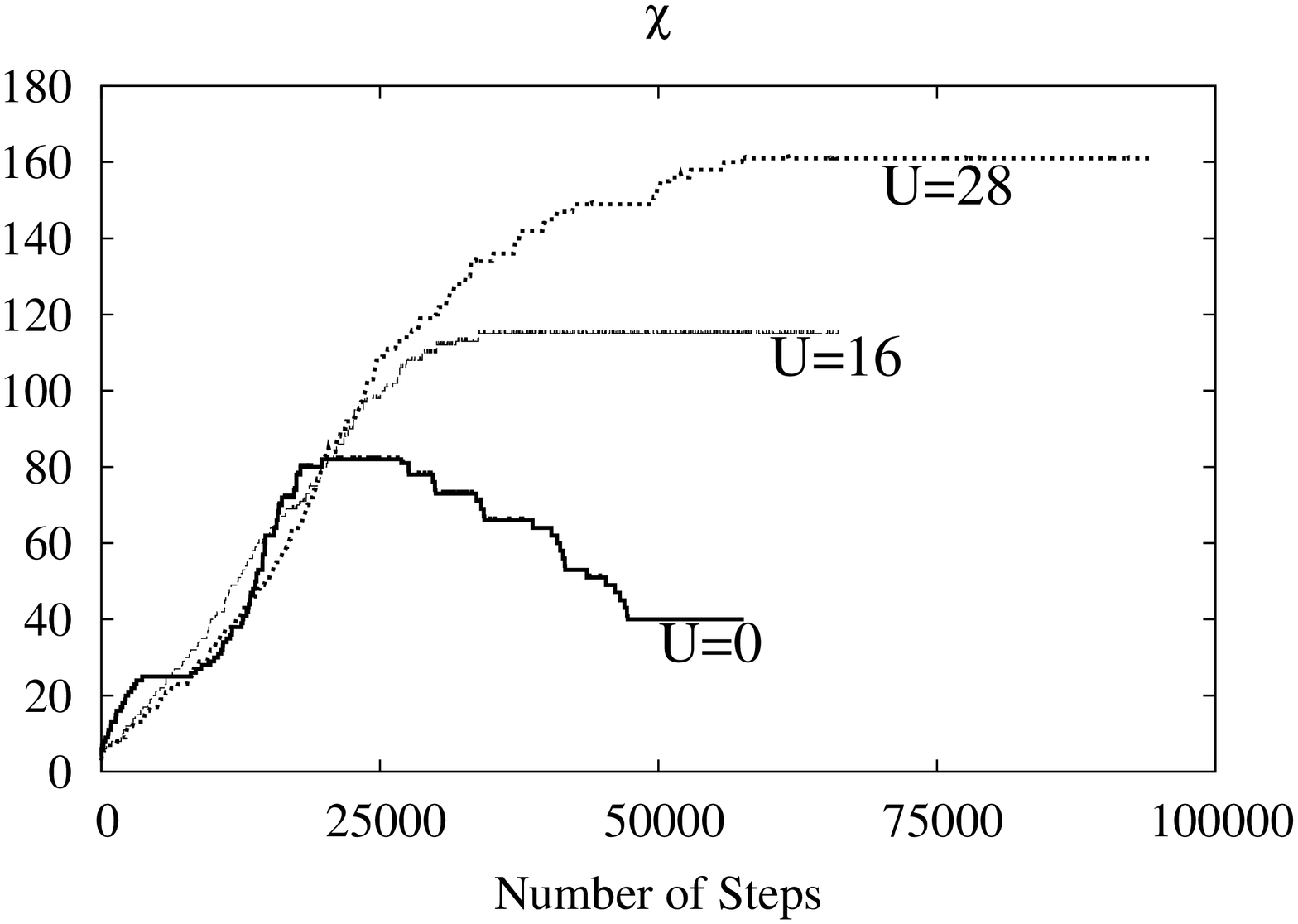}
\caption{Numerical cost of ground state simulations of Hamiltonian (\ref{BH1}) for
$N=M=20$. These computations correspond to some of the points in figures 
\ref{G_lo} and \ref{hern}. Left. CH. Right. PTH.} 
\label{higui} 
\end{figure}

The analysis in the previous section suggest that the relative population at the
ends is in some manner linked to EEE. Indeed, after some test computations we 
found that as a consequence of the universality characteristics of the problem it
is appropriate to plot our results using the fraction of particles on the ends
(dimensionless), 

\begin{equation}
\zeta= \frac{\langle \hat{a}^{\dagger}_1 \hat{a}_1 \rangle + \langle \hat{a}^{\dagger}_N \hat{a}_N \rangle }{M},
\end{equation}

which allows a comparative perspective of the system behaviour for
different scales. $\zeta$ varies according to the intensity of 
the repulsion in intermediate sites. As can be seen in figure 
\ref{fig:one}-left, when repulsion is small few particles remain 
on the ends and $\zeta \approx 0$. From the same graph it follows 
that when repulsion is strong the amount of particles on the ends 
is maximum and, 

\begin{equation}
\zeta = \frac{1}{2} + \frac{1}{M}.
\end{equation}

Notice that $\zeta$ keeps a close relation with
the repulsion constant $U_k$. This variable $\zeta$ enables us to
depict our results for different chain sizes in a comparative manner.
Before presenting our results and the corresponding analysis, we
want to briefly comment on the computational cost associated with
the simulations shown on this section. In order to make
our analysis as robust as possible, we performed computations in
chains as large as our resources allowed them to be. In a strongly
entangled chain, nevertheless, TEBD is not very efficient, and the 
biggest system we worked on corresponds to $20$ bosons in $20$
sites. These computations, however, are considerably demanding
(using normal desktop computers, finding the ground state can
take up to two weeks). As a comparison, note that the maximum $\chi$ 
in the simulations presented in figure \ref{SFMT} for systems of 
$100$ bosons in $100$ places is always below $60$ while the same 
cost-measuring parameter goes well above $140$ in figure \ref{higui} 
which corresponds to the results to be shown in this section.

\begin{figure}
\includegraphics[width=0.35\textwidth,angle=-90]{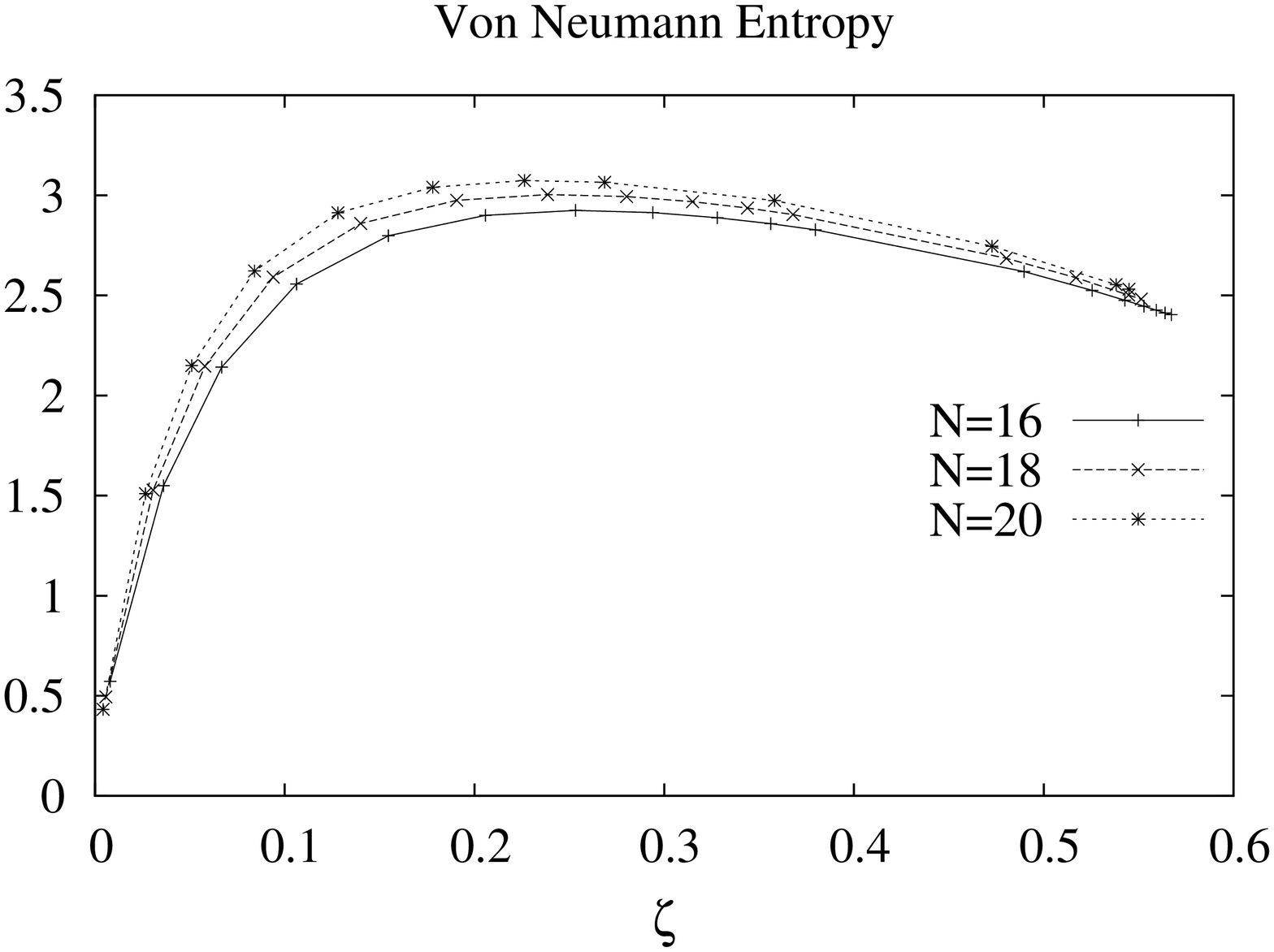}\includegraphics[width=0.35\textwidth,angle=-90]{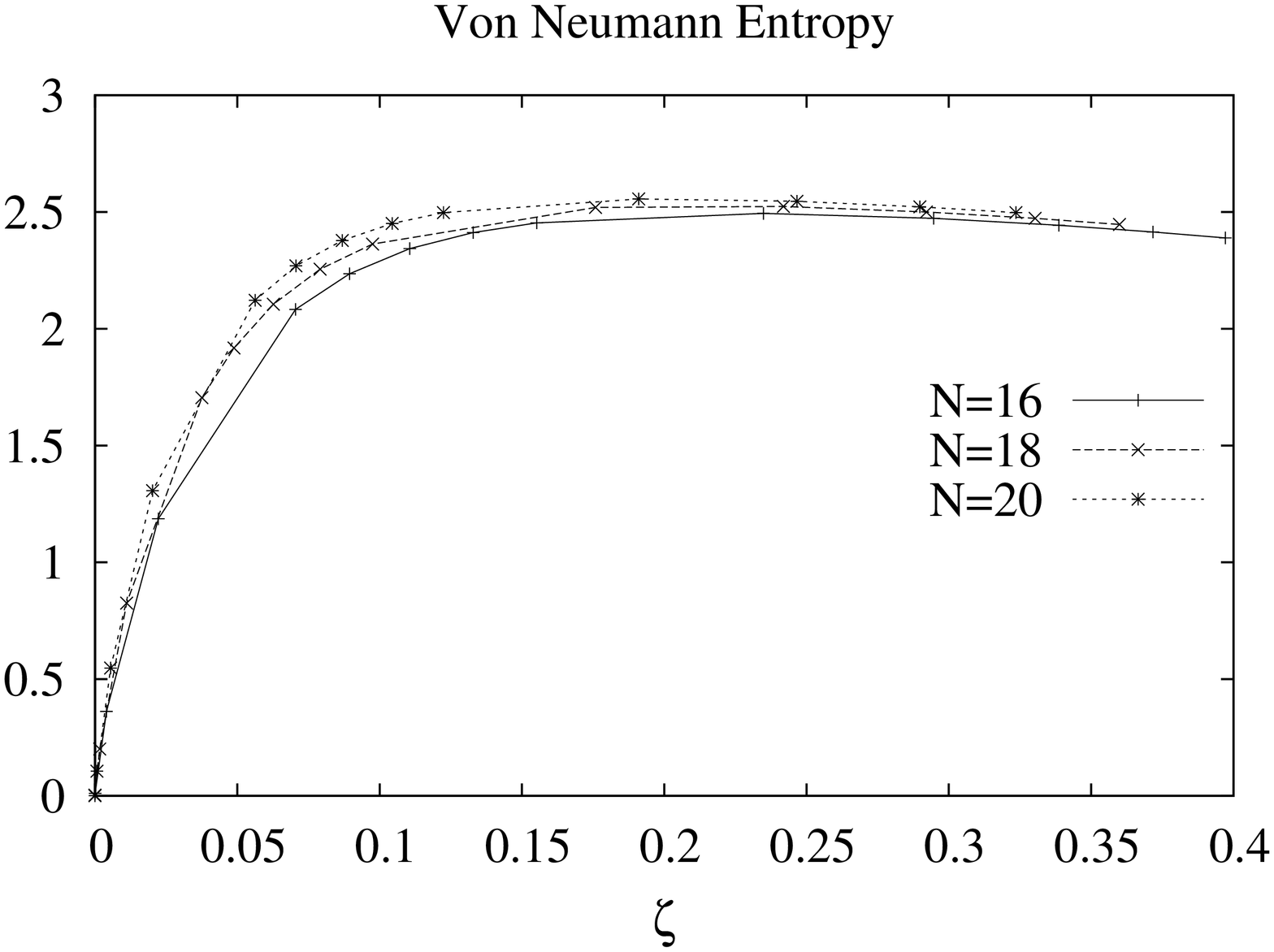}
\caption{Entanglement between the terminals and the rest of the system in a
boson chain governed by Hamiltonian (\ref{BH1}) and with no repulsion on the 
ends. Left. CH. Right. PTH. As usual, the number of boson in each chain
is equal to the indicated number of sites.}  
\label{barr}
\end{figure}

In our physical model, independently of the particularities of the 
hopping profile, 
the ground state of repulsionless chains is eminently superfluid,
although with most of the tunnelling taking place in the central 
part, leaving the terminal positions nearly empty
(figure \ref{fig:one}-left). In this case entanglement is at the 
same time strong and localized. Indeed, in these circumstances 
entanglement between half-chain-blocks is strongly enhanced. 
Similarly, particles are forced to tunnel through longer 
distances and correlations develop at longer scales as repulsion 
in intermediate positions climbs up. Additionally, bosons
increasingly migrate towards the terminal positions since 
the repulsion prevent particles from staying in
intermediate sites. Outstandingly, on account of the repulsionless 
ends, boson fluidity stands no matter how strong the 
repulsion in intermediate sites is. As repulsion  
slowly grows away from zero, entanglement between the
chain halves goes down while entanglement between the ends
and the rest of the system goes up, an indication that 
the originally localized correlations are being spread all 
over the chain following the particle distribution profile. 
When repulsion is increased even more, the terminals start getting 
macroscopically occupied so that the expectation value of the number of 
bosons and the fluctuations in intermediate sites both go down asymptotically 
towards $\frac{1}{2}$, which indicates that bosons get highly
delocalized. In this case correlations among places near the ends 
and in opposite sides of the chain are strongly enhanced, in 
contrast to the correlations in the centre of the chain. This can 
be seen in the graphs of figure \ref{barr} where, interestingly, 
von Neumann entropy between both ends and the rest of the system 
slightly comes down after the original redistribution of entanglement 
mentioned above occurs. This effect is more evident for the 
CH case. The decrease in the entanglement between the ends and
the rest of the system can be interpreted as a reduction of the
correlations in intermediate sites which takes away local combination
of states with substantial entangling potential. On the other
hand, this reduction of the correlations between both ends and
its complement is convenient for the emergence of EEE, since the
reduced state of the terminals becomes closer to a pure state.
This fact alone, however, is not sufficient for the arising of
EEE. Even if the state of the terminals is completely pure, it
is possible that the quantum state of such reduced system is
separable. In this context it is interesting to explore the 
mechanisms that ultimately define the onset of long-range 
entanglement. 

\begin{figure}[t]
\begin{center}
\includegraphics[width=0.7\textwidth, angle=-90]{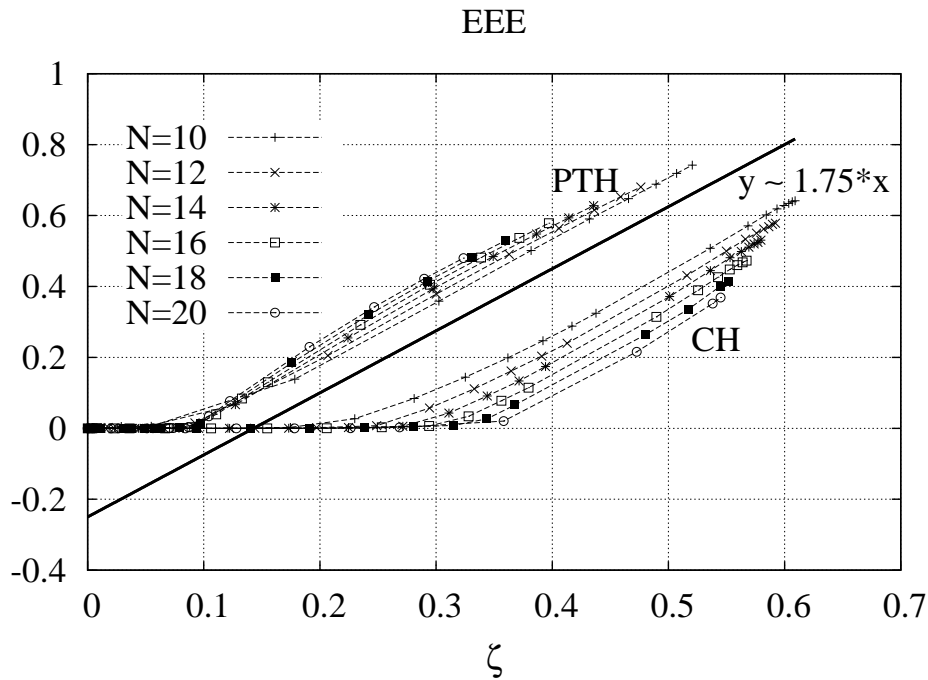}
\caption{Logarithmic negativity of the terminals. In all cases the
number of bosons in the chains is equal to the indicated number of sites.} 
\label{G_lo}
\end{center}
\end{figure}

The issues discussed so far are independent of either hopping 
profile we feed into Hamiltonian (\ref{BH1}). We now want to
focus on those characteristics associated exclusively
with the values given to the hopping coefficients. 
Figure \ref{G_lo} summarizes the results we obtained from
computing the logarithmic negativity in both CH and PTH chains.
Notice that the curves in figure \ref{G_lo} are analogous to
the results in figure \ref{fig:one} which are depicted in a
slightly different way.

As a matter of fact, EEE has a more interesting behaviour in PTH
chains than in CH chains. The most notorious advantage of PTH
is the persistence of entanglement, even when the number of sites
and bosons augment. In contrast, the amount of entanglement in
CH chains does not stand against increasing chain size. Similarly,
when the entanglement values are seen at any given fixed $\zeta$,
it is clear that PTH chains display more EEE than their CH equivalent.
This is straightforward since in figure \ref{G_lo} PTH entanglement 
arises well before CH entanglement. Given the increasing cost of
the simulations, specially for the PTH case, not all the possible
values of $\zeta$ are considered, but further information can be
extrapolated from the data at hand. Indeed, PTH curves fit nicely
into a single line with slope $1.75$. As expected, the cost of the 
simulation 
is proportional to the amount of entanglement, but we can get
a reliable scaling characterization from the information
contained in figure \ref{G_lo}. Correspondingly, numerical fitting 
yields

\begin{eqnarray}
EEE^{PTH}_{\zeta=0.5} \sim log(N^{0.1}), \text{and} \nonumber \\
EEE^{CH}_{\zeta=0.5} \sim log(N^{-0.2}). 
\end{eqnarray}

Indeed, the logarithmic behaviour of EEE has been inherited from 
the measure we use to quantify EEE, but the fitting formally 
establishes that the PTH profile provides a platform that is 
highly convenient to entanglement.

The fact that entanglement is better for some configuration
of parameters than for others makes us wonder what is so special
about a hopping profile that delivers substantial entanglement
between the terminals. Curiously, the particle distribution
profile in general looks like the one in figure \ref{fig:one}-left
for chains with very high repulsion in intermediate sites, no
matter which hopping profile, PTH or CH, we choose. This means
that the mechanisms giving rise to entanglement must be determined
by high order terms of the imaginary-time-evolution operator \cite{Muller}.
Such high order terms induce essentially two kinds of physical  
processes, namely,

\begin{itemize}

\item Exchange of more than one particle among places of the chain.

\item Exchange of particles among distant places of the chain.

\end{itemize}

\begin{figure}
\includegraphics[width=0.7\textwidth, angle=-90]{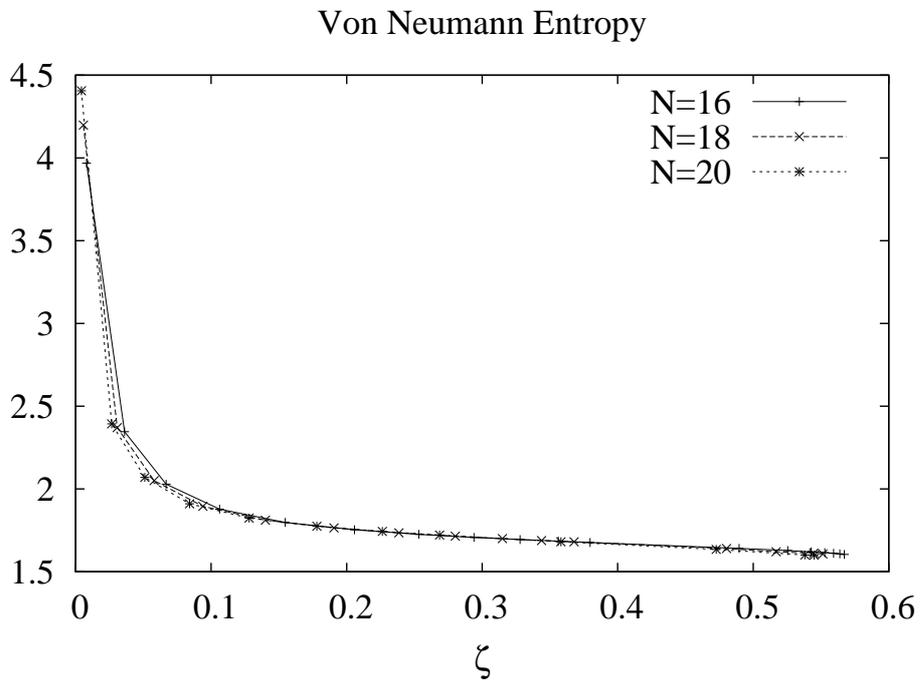}
\includegraphics[width=0.7\textwidth, angle=-90]{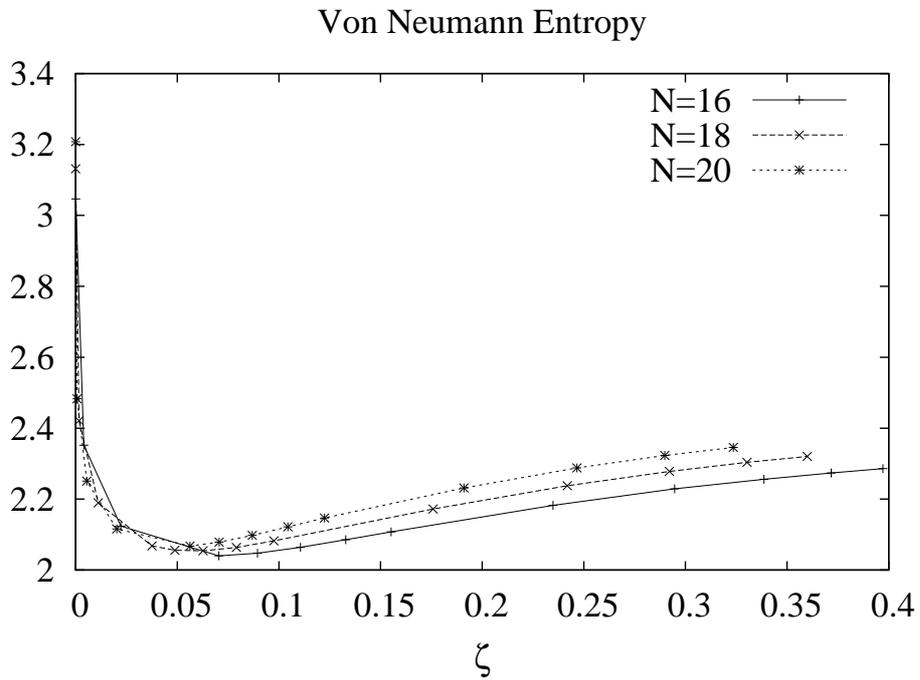}
\caption{Entanglement between the halves of the chain. In all the 
graphs the number of bosons in equal to the indicated number of sites. 
Top. CH. Bottom. PTH. } 
\label{hern}
\end{figure}

These mechanisms can take place simultaneously, and as a result,
correlations develop at a macroscopic scale. When the conditions are
optimal, as for instance with PTH, correlations are massively 
enhanced and this leads to long range entanglement. This 
interpretation coincides with further results, especially with 
the logarithmic negativity graphs shown in figure \ref{hern}. 
As can be seen, entanglement between the halves of the chain
decreases monotonically as the repulsion in intermediate sites
is turned up. Such behaviour is in certain way not surprising
since as repulsion becomes more intense, more particles are 
forced away from the centre and into the terminals, which
reduces the amount of ``quantum channels'' that can be used
by every half-chain block to speak directly to each other.   
Nevertheless, the behaviour of logarithmic negativity in
PTH chains is quite different. In such a case the graphs 
display a clear minimum instead of a constantly falling curve.
Consequently, in spite of the reduction in communication
resources, there is a critical repulsion value for which
entanglement stops falling and slowly turns its way up. 
The bigger $N$, the sharper the effect, which suggests this
would lead to a second order phase transition in the thermodynamic
limit. This sudden reinforcement of entanglement can be 
understood as coming from an enhancement in the particle exchange
between the chain terminals. Certainly, when the delocalization
length of bosons is equal to the distance between the chain
ends, particles held on the terminals can be exchanged directly 
between the ends. This induces a transition of the system state, since
those particles originally squeezed against the chain borders are 
suddenly unleashed and, instead of being highly localized around the
system boundaries, their wave functions now envelop the chain to its 
whole extent, keeping on average most of the wave function weight on 
the terminals. This interesting effect can also be intuitively 
identified by assuming a particle-like behaviour of bosons. Indeed, 
we can think that as repulsion in the middle is continuously turned on, 
bosons are obliged to hop through longer distances and thereby become 
more and more delocalized. The enhancement of entanglement is
therefore determined by the {\it tunnelling scope} rather than
by plain particle accumulation, although the latter is a necessary
condition for EEE emergence. In CH chains, the form of von Neumann 
entropy between both halves in consistent with a regime in which 
entanglement is being continuously redistributed across the system 
as a result of increasing tunnelling scope, 
but the fact that there is no turning point in figure \ref{hern}-top
indicates that such hopping never takes place across the complete 
length of the chain. However, such hopping is enough to 
induce EEE at finite $N$, but not in the thermodynamic limit
since EEE dies down against increasing chain size. It is
as if such finite $N$ entanglement results from the overlapping
of neighbouring wave functions. The transition to a strongly entangled 
state takes place when such local wave functions flatten and successively 
combine to create a strongly global description of the state. 
Once the previously mentioned long scale tunnelling takes over, 
increasing the repulsion in intermediate sites reinforces end-to-end 
exchange  and induces an increase in the amount of entanglement 
shared by the terminals, as can be seen in figure \ref{hern}-bottom. 
This can also be understood in terms of the so
called {\it healing length}, which defines the distance that takes
the condensate to form its bulk. We can say that the healing length
determines the space over which the superfluidity of the system is
suppressed. In terms of physical constants, this length scales as
$L_{healing} \sim \frac{1}{\sqrt{U}}$. In ${}^4 He$, the healing
length is around $0.1 nm$. The phenomenon that we report in this
section can be seen as a suppression of the healing 
length, in such a way that superfluid features are present all over 
the chain, and then these features are reinforced by the PTH hopping
profile.

\section{Entanglement and dynamics}

\begin{figure}[h]
\includegraphics[width=0.7\textwidth, angle=-90]{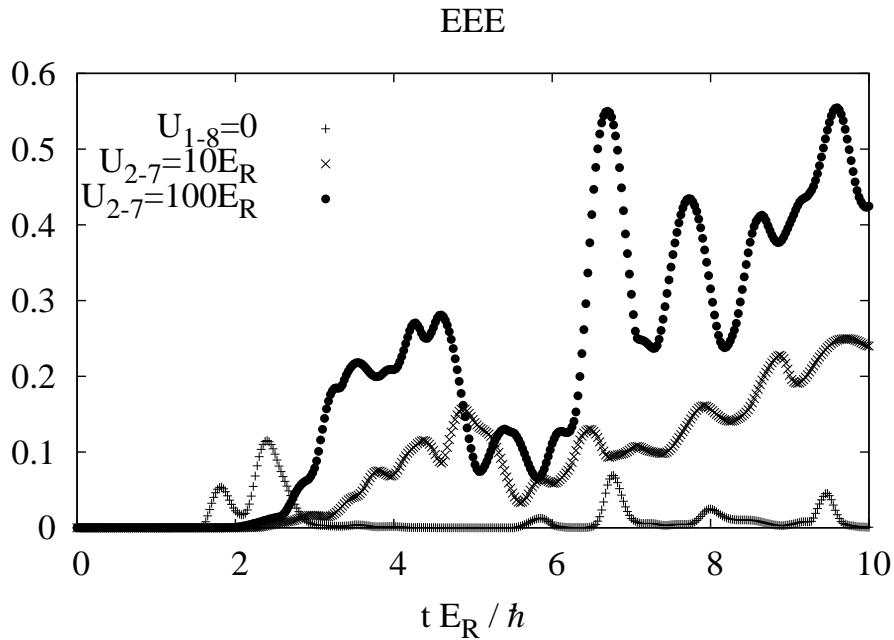}
\caption{Logarithmic negativity for $N=M=8$ and CH. The system is 
prepared in a Mott insulator state with one boson in each site. As 
can be seen, strong repulsion
constants in intermediate places are necessary to induce substantial EEE.} 
\label{pere}
\end{figure}

Here we study how entanglement arises dynamically in boson
chains with superfluid characteristics. This is the most relevant
situation since any form of entanglement is null in the Mott insulator 
state. In the discussion that follows we assume that our 
chains are initially prepared in a Mott insulator state and 
the parameters are suddenly changed to generate the dynamics.
The new set of parameters correspond to a given hopping profile,
CH or PTH, and a given repulsion profile which always maintains
the ends repulsionless. This is the most natural scenario one can 
expect in a BH chain where one first prepares the Mott 
state with one boson per site and then turns all the interactions 
globally, the same as in \cite{Moeckel,Cramer,Flesch}, but opposite 
to the approach in \cite{Corinna,Lauchli} where the quench goes 
from the superfluid to the Mott insulator. As a generic example 
we present in figure \ref{pere} the evolution 
of the logarithmic negativity for different values of intermediate 
repulsion. In a chain with all the repulsion constants set
to zero, EEE shows very little development in the course of evolution 
as can be seen in the indicated figure. Dynamics in
repulsionless chains is, in fact, known to lead to a progressive
thermalization of reduced density matrices of each site
\cite{Cramer,Flesch}. Two site entanglement is thus very low in this
repulsionless scenario. High repulsive evolution, on
the other hand, shows a less regular profile, product of interference
among many unsynchronized phases accumulated locally. From 
figure \ref{pere} we infer that as in the case of ground state 
entanglement, dynamic entanglement does not build up
unless a significant accumulation of particles assists the enhancement
of fluctuations at the ends. As can be observed in the evolution patterns 
of chains with high repulsion in intermediate sites, the amount of 
entanglement contained in the system after reasonable times can be 
significant, yet smaller than what can be obtained from the ground state 
in optimal configurations (compare with figure \ref{fig:one}-right).
Here, we assume uniform hopping but our conclusions are valid also for PTH.

\begin{figure}
\includegraphics[width=0.35\textwidth,angle=-90]{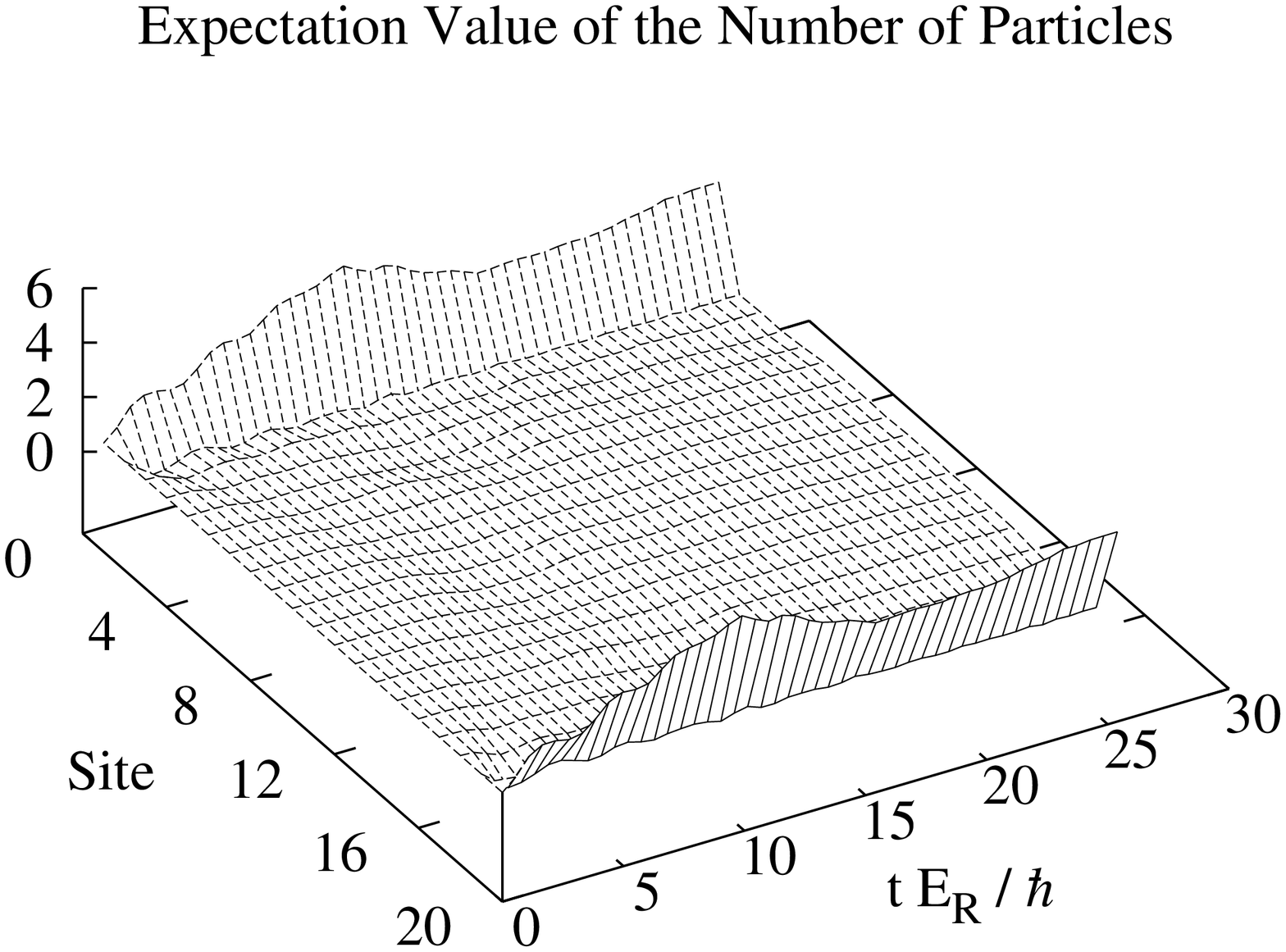}\includegraphics[width=0.35\textwidth,angle=-90]{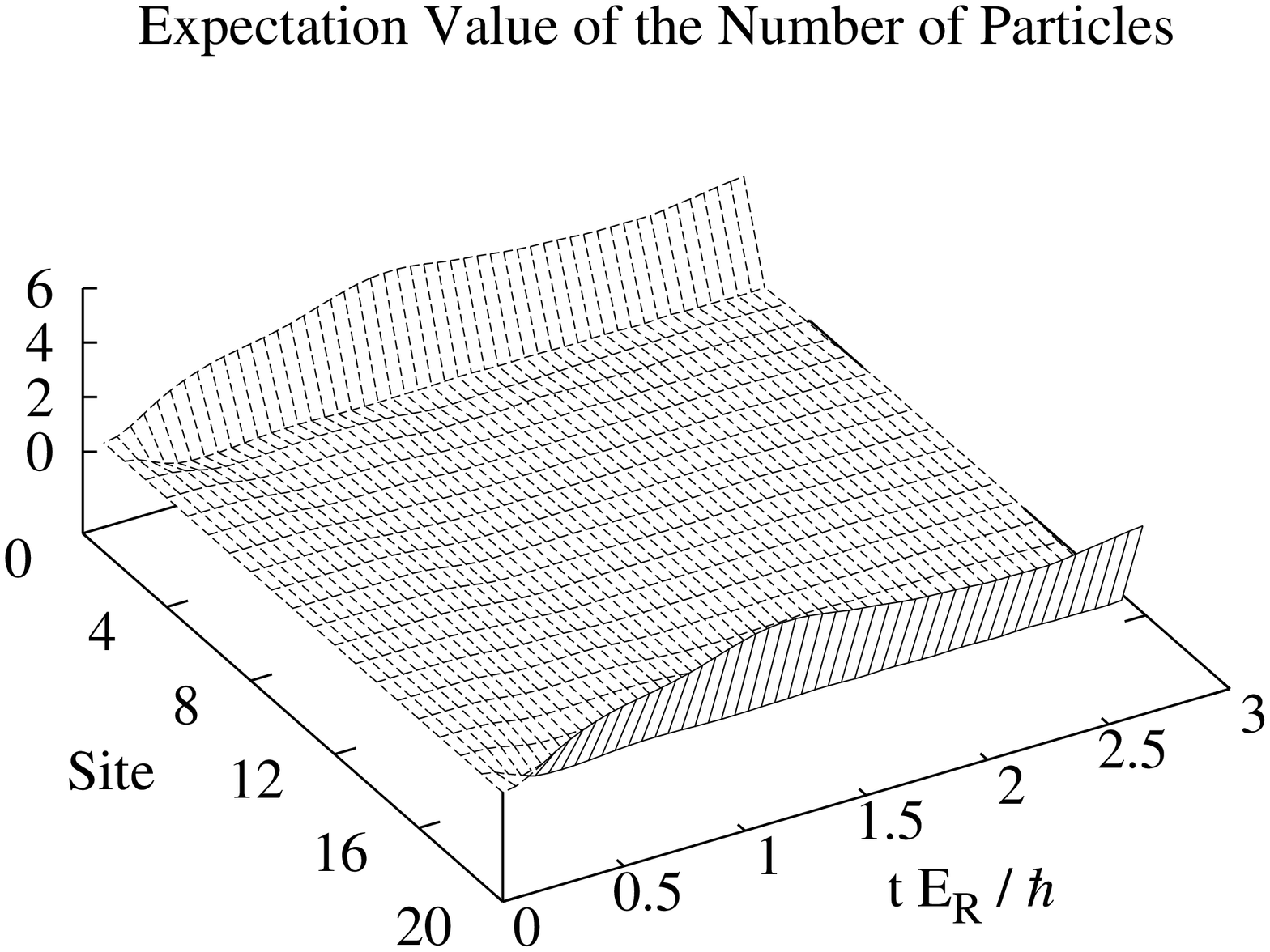}
\caption{Particle distribution in bosonic chains for $N=M=20$. 
The system is 
prepared in a Mott insulator state with one boson per site. Strong 
repulsion constants in intermediate sites forces the bosons into 
the ends. Left. CH and $U_{2-19}=100E_R$. Right. PTH and $U_{2-19}=200E_R$.} 
\label{leon}
\end{figure}

\begin{figure}
\includegraphics[width=0.35\textwidth,angle=-90]{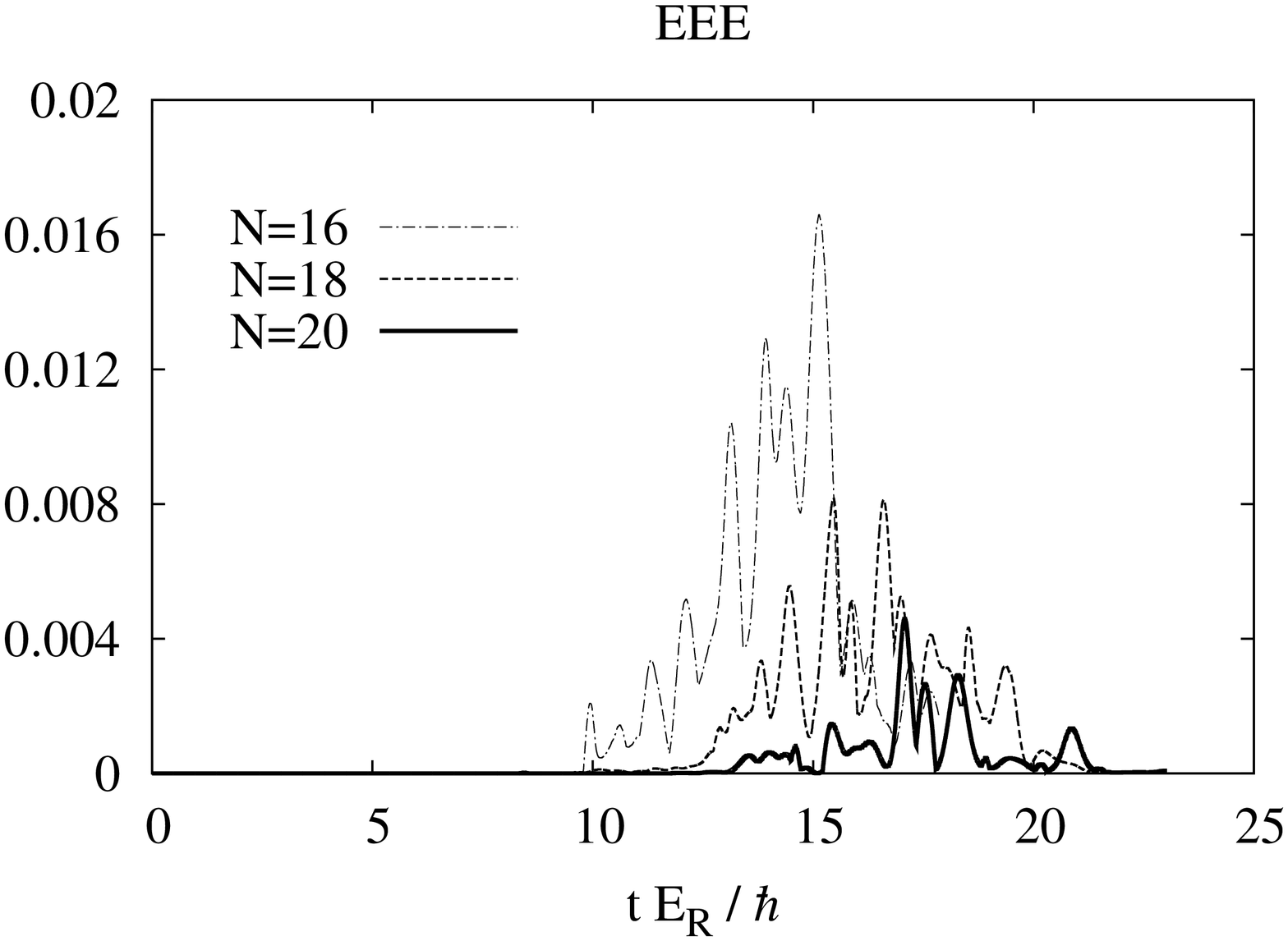}\includegraphics[width=0.35\textwidth,angle=-90]{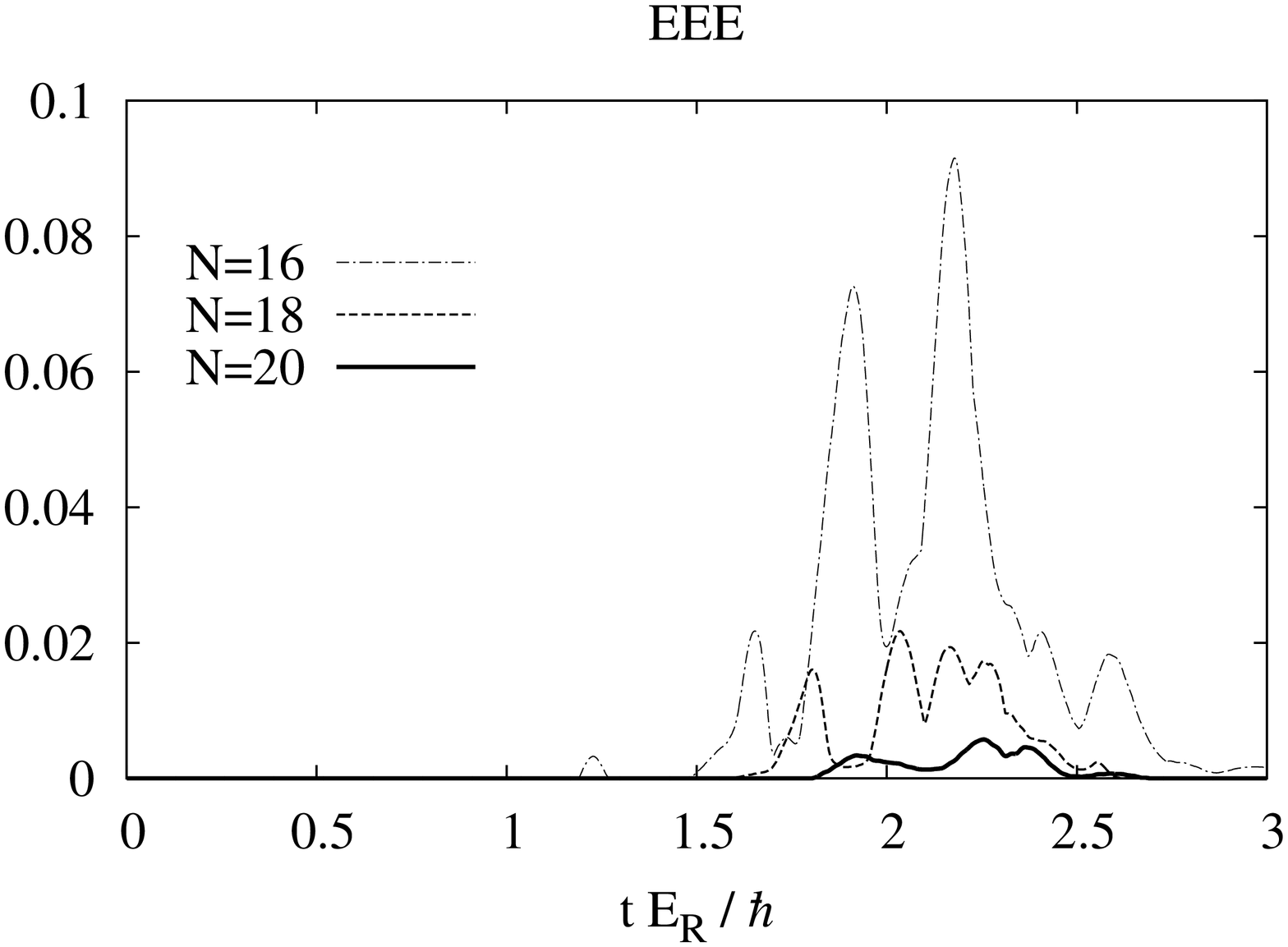}
\caption{The system is 
prepared in a Mott insulator state with one boson in each site. These 
graphs correspond to equivalent systems shown in figure \ref{leon}. 
Left. CH and $U_{2-19}=100E_R$. Right. PTH and $U_{2-19}=200E_R$.} 
\label{chon}
\end{figure}

In order to identify relevant differences between the entanglement
behaviour in chains with different hopping profiles, we found it
necessary to perform simulations in chains of up to $20$ places with 
$20$ bosons. As we  already know that strong repulsion is necessary 
to induce EEE, we just set high repulsion constants on every site except 
the terminals and simulate the evolution using the TEBD algorithm.
The resulting dynamics can be seen in figure \ref{leon}, where 
the expectation value of the number of particles is plotted as a 
function of time.
The ends get macroscopically occupied and tunnelling in intermediate 
places intensifies. EEE arises some time after the ends have been occupied, 
as can be appreciated in the graphs of figure \ref{chon}. In general, 
the bigger the chain the longer it takes for EEE to emerge.
For PTH, a natural time scale is determined by the transmission period, 
$T=\pi$, on account of our particular choice of $\lambda=2$.
Entanglement shows up more or less at half of the period, 
$t\approx \frac{\pi}{2}$, when some particles have travelled from 
one end of the chain to the other. It could be, however, that EEE in
longer chains behave differently, since repulsion in intermediate
places hinders the boson mobility. This negative effect is likely
to be more pronounced when transport takes place at long distances.
For longer times entanglement might grow above the values reported in
figure \ref{leon}, independently of the hopping profile, but this 
effect is hard to follow as long time simulations require additional 
computational resources \cite{Osborne,Perales}.

\section{Entanglement and perturbative dynamics}

We have seen how boson chains can be made to contain substantial
entanglement in the ground state as well as in the course of dynamics.
Ideally, if both approaches are combined we then could 
multiply the amount of entanglement in the chain, which would 
result in a highly beneficial use of the several quantum degrees
of freedom that come from the multiple possibilities of arranging
a given number of particles on the system. Hence, we now assume that 
a chain initially prepared in an entangled ground state
evolves by the action of a perturbation added to the original
Hamiltonian. The reason for choosing a perturbation rather than
an abrupt change of parameters in the Hamiltonian is because a
perturbation is expected not to modify the particle distribution
profile substantially, so that the terminals of the chain remain
macroscopically populated which has been already identified as 
a necessary condition for entanglement generation. In formal terms, 
the state evolution is given by,

\begin{equation}
|\psi (t) \rangle =  e^{-i t \hat{H}_1} |\psi_g \rangle,
\label{eq:twelve}
\end{equation}

where,

\begin{equation}
\hat{H}_1 =  \hat{H}_0 + \epsilon \sum_{j=1}^N\hat{h}_j,
\label{panorama}
\end{equation}

in such a way that $|\psi_g \rangle$ is the ground state of
Hamiltonian $\hat{H}_0$, from equation (\ref{BH1}), and 
$\epsilon$ is a small real number which determines the intensity 
of the perturbation. $\hat{h}_j$ represents a local operator acting 
on site $j$. The explicit form of operators $\hat{h}_j$ is to be 
worked out according to our convenience. For consistency, the 
first condition to be satisfied is,

\begin{equation}
[\hat{H}_0,\sum_{j=1}^N \hat{h}_j]\ne 0, 
\end{equation}

otherwise dynamics would be trivial. In order to simplify our analysis 
we assume that the local operators can be written as functions of the 
corresponding local number operator,

\begin{equation}
\hat{h}_j = h_j(\hat{n}_j)
\end{equation}

where,

\begin{equation}
\hat{n}_j = \hat{a}_j^\dagger \hat{a}_j, 
\end{equation}

so that we can now focus on the functional forms $h_j(x)$ rather than
in quantum operators. We can establish a non-vanishing commutator between 
the perturbation and the Hamiltonian by defining, 

\begin{equation}
h_j(x) = k x,
\label{rochi}
\end{equation}

\begin{figure}
\includegraphics[width=1.\textwidth,angle=-0]{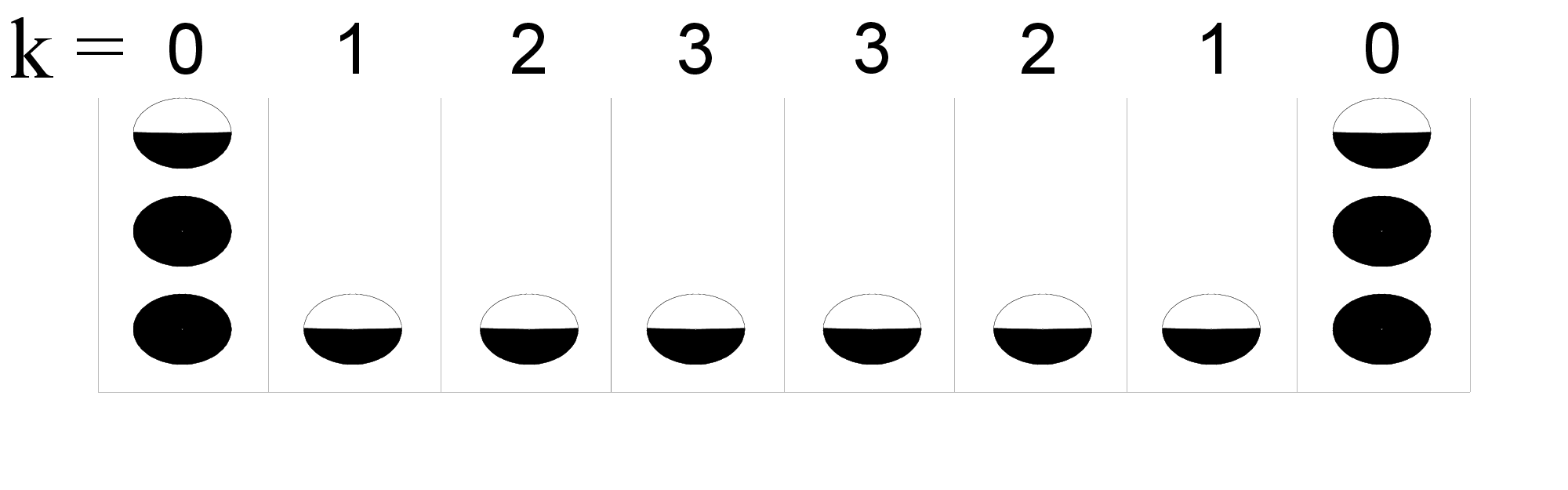}
\caption{Sketch of bosons spread across a chain of repulsionless ends.} 
\label{bolitas}
\end{figure}

where $k$ is the integer distance between site $j$ and the closest end,
as indicated in figure \ref{bolitas}. This form holds only for intermediate 
places. Similarly, to determine the 
form of the operators on the terminals, $h_{1,N}$, the primary factor to 
take into account is the optimization of boson-transfer from intermediate sites 
towards the ends. If the expectation value of the number of particles remains
stable during time evolution we can carry out a classical analysis in
the basis of the number of bosons. Let us think that in the configuration
shown in figure \ref{bolitas}, which sketches one of the particle distribution
shown in figure \ref{fig:one}-left, we take half a boson from the centre of the 
chain and put it in the closest terminal.
We can assume that this repositioning does not cause much change in
the average value of $\hat{H}_0$, since there is no contribution
from the repulsion terms for bosons in intermediate positions, because
repulsion is zero for one or zero bosons, or for bosons on the chain ends,
because repulsion is off on the terminals. As a consequence, we would 
expect that the change in the average values of the system could be 
calculated focusing only on the averages of the perturbative potential.
Additionally, because we want particles to go from the middle of the
chain to the terminals and vice versa, we need that the energy cost of 
extracting particles from intermediate positions balances the energy cost
of putting the same particles on the chain ends. For example, starting
from the state depicted in figure \ref{bolitas}, if all the particles 
go consecutively from the middle of the chain to the ends, the first
process to take place is that bosons in positions with $k=3$ disappear 
from their respective positions and pop up on the terminals. The
corresponding energy balance can be written as,

\begin{equation}
h \left( n_0 + \frac{1}{2} \right ) - h(n_0) = \frac{3 \epsilon}{2}.
\end{equation}

In this expression $n_0$ represents the expectation value of the number 
of particles
on one of the terminals, which can be simply taken as $n_0 = 2.5$ 
for this specific case. Once this process is over, there are
fewer particles in intermediate sites and more on the ends. Hence,
taking first bosons with $k=2$ and putting them on the ends and 
then doing the same for bosons with $k=1$ we obtain,

\begin{eqnarray}
h \left( n_0 + 1 \right ) - h \left( n_0 + \frac{1}{2} \right ) = \epsilon, \nonumber \\
h \left( n_0 + \frac{3}{2} \right ) - h \left( n_0 + 1 \right ) = \frac{\epsilon}{2}. 
\end{eqnarray}

In a chain of size $N$ these equations lead to one single expression, 
namely,

\begin{equation}
h \left ( n_0 + \frac{1}{2} \left( \frac{N}{2}-k \right)+\frac{1}{2} \right )-h \left ( n_0 + \frac{1}{2} \left( \frac{N}{2}-k \right) \right )=\frac{\epsilon k}{2}
\label{eq:13}.
\end{equation}
 
After some preliminary tests we find that the optimal choice for
the unknown function is,

\begin{equation}
h(x) = c_2 x^2 + c_1 x.
\label{nazi}
\end{equation}

Constants $c_2$ and $c_1$ are used to balance equation (\ref{eq:13}).
Combining these equations and after some simplification we get,

\begin{eqnarray}
& c_2 \left ( n_0 + \frac{1}{2} \left ( \frac{N}{2} -k \right) \right) + \frac{c_2}{4} + \frac{c_1}{2} = \frac{\epsilon k }{2}   & \nonumber \\
& \Rightarrow c_2 n_0 + \frac{c_2 N }{4} + \frac{c_2}{4} + \frac{c_1 }{2}   \mbox {\boldmath $-\frac{c_2 k}{2}$} = \mbox{\boldmath $\frac{\epsilon k }{2}$  } & 
\end{eqnarray}

In order to work out a set of constants valid for every $k$ we 
must define,

\begin{eqnarray}
&& c_2 = -\epsilon, \nonumber \\
&& c_1 = \left( \frac{N+1}{2} + 2 n_0 \right) \epsilon.
\end{eqnarray}

\begin{figure}
\includegraphics[width=0.35\textwidth,angle=-90]{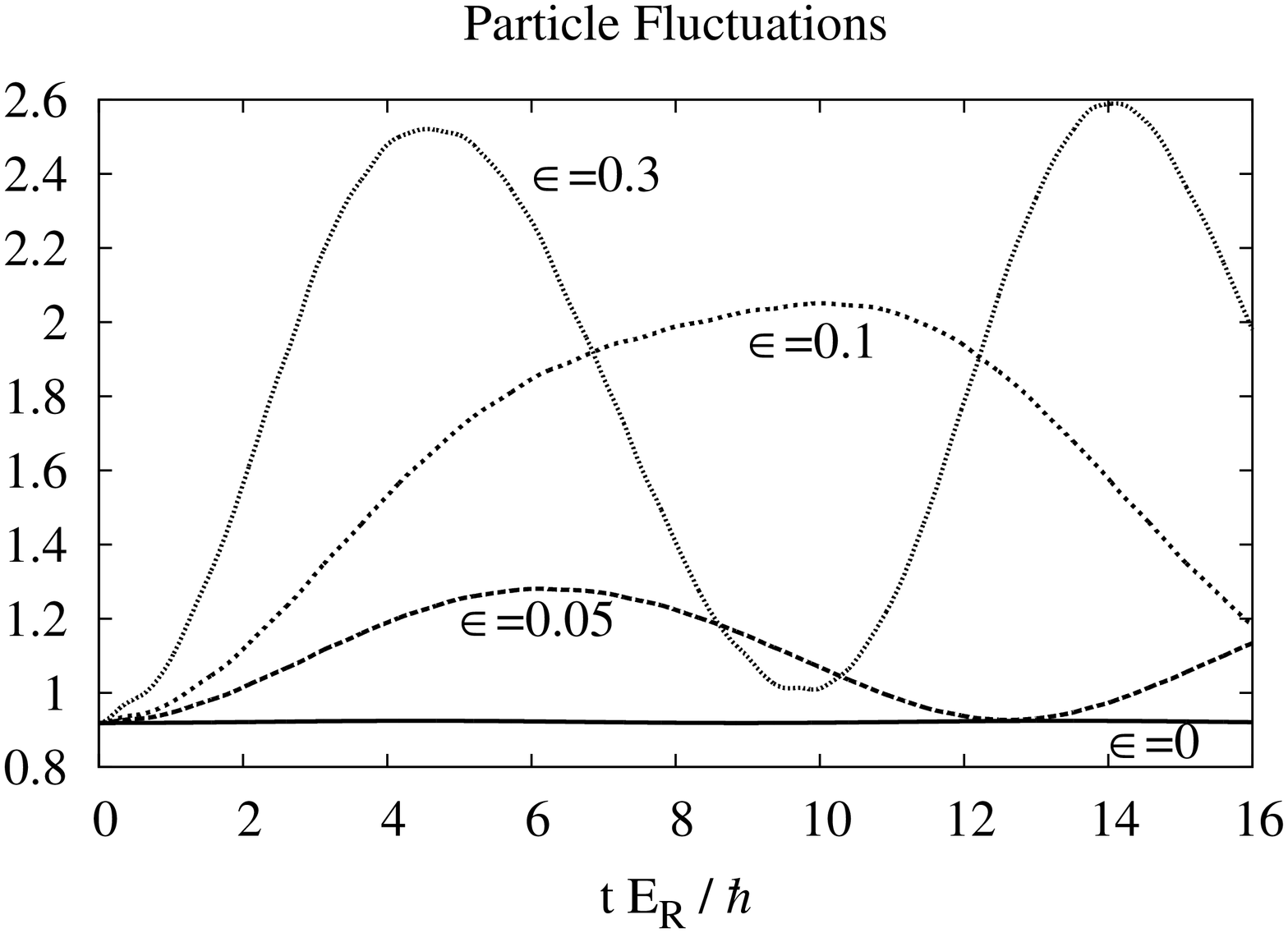}\includegraphics[width=0.35\textwidth,angle=-90]{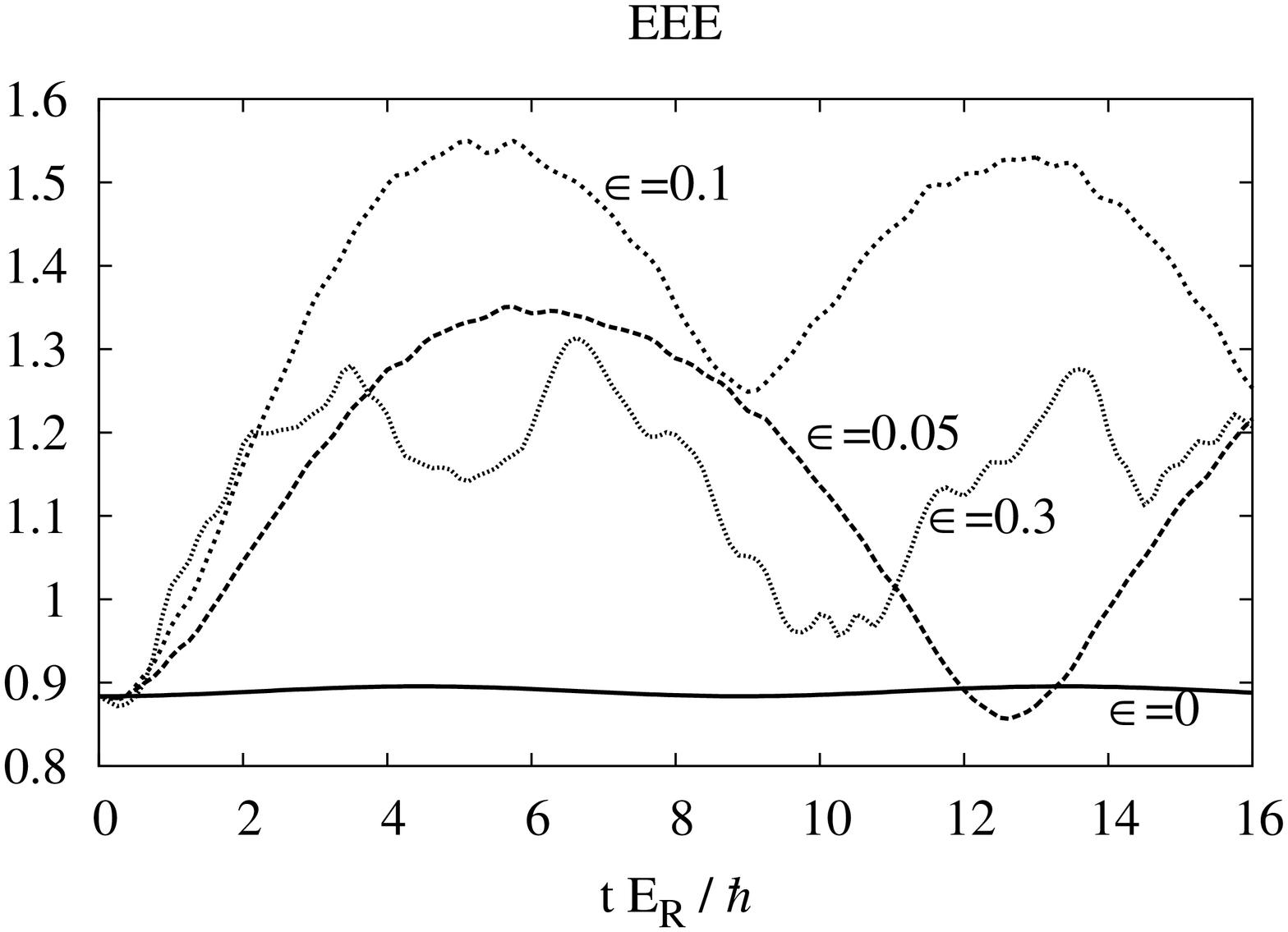}
\caption{ 
Evolution in chains of $N=M=8$ for PTH and $U_{2-7}=100E_R$. The system 
responds 
according to the intensity of the perturbation. Equivalent graphs corresponding 
to population averages deviate less than 2\% from their initial values.
Left. Fluctuations in the number of bosons allocated in one end of a chain. 
Right. Dynamical entanglement.
} 
\label{osor}
\end{figure}

Moreover, for a perturbation written as above, the mean number of 
bosons does not fluctuate much since the dynamics is still governed by 
a Hamiltonian with strong repulsion coefficients in intermediate sites 
and bosons are forced to stay on the ends. In this situation one
would expect that the evolution is dominated by high order dynamics
such as the exchange of particles between the terminals and the
rest of the system. Evidence of this can be seen in the graph of
fluctuations shown in figure \ref{osor}-left. As can be expected from our 
particular choice of function $h(x)$, the transit of bosons between 
the ends and the rest of the chain is enhanced and therefore the 
fluctuations on the terminals are developed according to the
strength of the perturbation. Certainly, such enhancement of fluctuations 
leads to an improvement of communication between distant places
which results in entanglement. Figure \ref{osor}-right shows Log-negativity as
a function of time for different perturbation intensities.
Roundoff errors accumulated in the ground state during
the imaginary time evolution cause slow oscillations in the dynamics
at $\epsilon=0$, but these are tiny compared with the more complex
evolution seen at finite $\epsilon$. In the initial stages of
evolution the dynamics is characterized by a sharp increase of
quantum fluctuations on the ends of the chain accompanied by little
change in the expectation value of the number of bosons. Hence, entanglement
development is enhanced while avoiding massive migration of bosons
towards the centre of the chain. This effect is maximal at around
$\epsilon = 0.1$ and becomes less important for higher values of
$\epsilon$, where perturbative terms induce a less predictable dynamics.
Significantly, entanglement generation is improved not by adding
interaction terms to the Hamiltonian but by evolving an already
entangled state using a locally-tuned perturbative potential.

\begin{figure}[t]
\begin{center}
\includegraphics[width=0.7\textwidth, angle=-90]{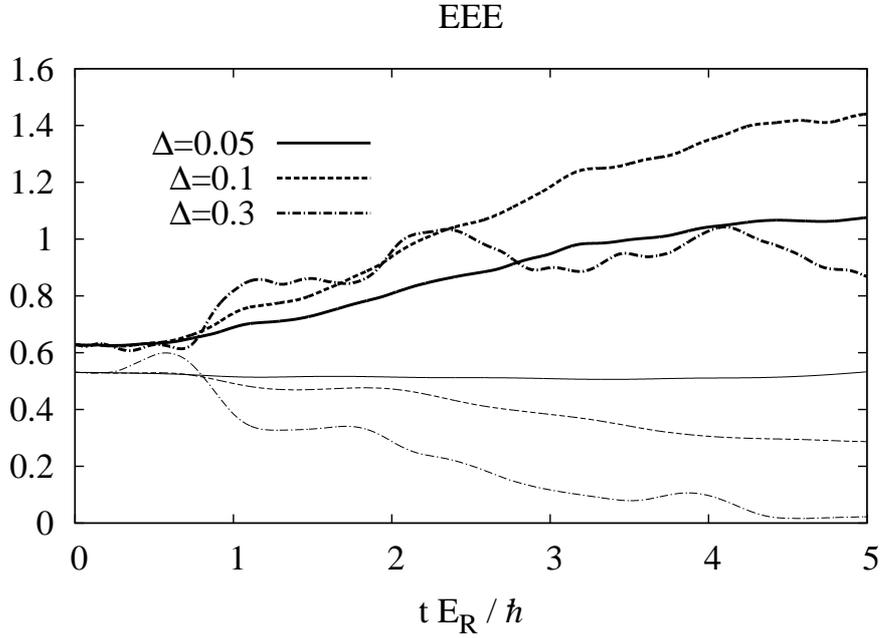}
\caption{
Logarithmic negativity in chains of $N=M=14$
for both CH and PTH. PTH lines are thicker than their CH
equivalents. EEE in the ground state can be improved as a
result of dynamics in PTH chains. $\chi=200$ for PTH and
$\chi=100$ for CH.
} 
\end{center}
\label{tren}
\end{figure}

In order to see if this effect lingers on when the size of the
chain is increased, we applied the same ideas to chains of
$N=M=14$. Interestingly, EEE behaves in a very similar
way than in the previous case, with almost the same values at
the same times, as indicated in figure \ref{tren}. In this
figure we also include CH results, which make it clear that
the enhancement effect is characteristic of PTH chains. This is
a curious matter, since the perturbative analysis does not
include any reference to any specific hopping profile, it just
balances the energy costs coming from different terms in
an optimal trajectory. In principle, one can think that the
lack of entanglement development may be due to either,

\begin{itemize}

\item Lack of boson mobility.

\item Dynamics with deficient entangling capability.

\end{itemize}

In general, entanglement seems not to team up with CH, as
opposite to PTH. Whatever the specific reason why this 
compatibility between entanglement and some specific 
profiles occurs, our results clearly suggest that particle
transport plays a crucial role in the development of
correlations in the quantum state.

\section{Entanglement detection}

\begin{figure}[t]
\begin{center}
\includegraphics[width=0.7\textwidth, angle=-90]{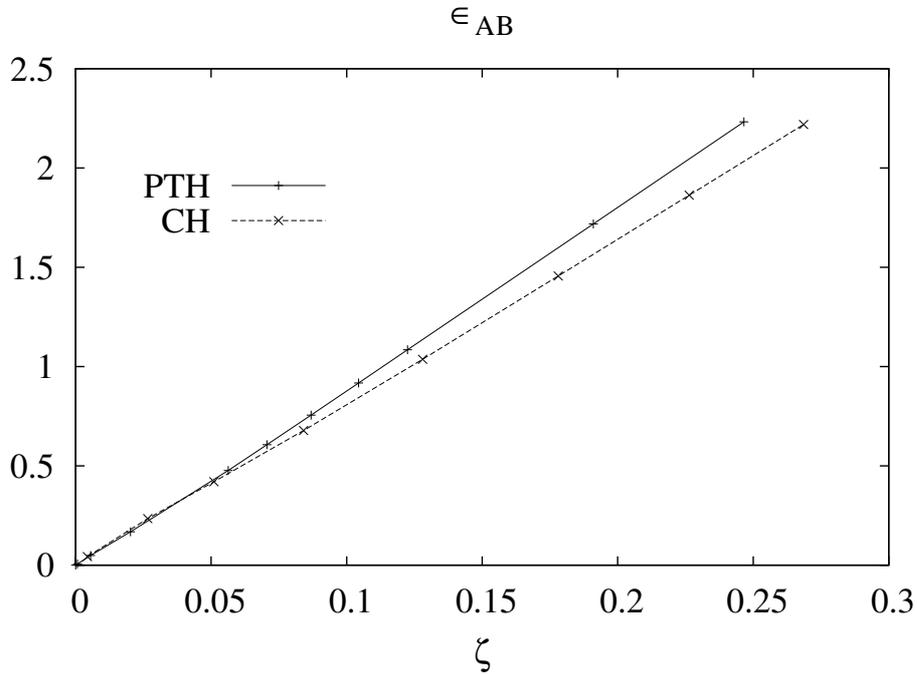}
\caption{Entanglement detection} 
\end{center}
\label{arno}
\end{figure}

In this short section, we want to explore how this entanglement
could be measured in the laboratory. Once bosons are cooled down
in the ground state or after evolution has taken place and the
interactions are turned off,
the detection protocol studied in \cite{Vedral,Simon} could be used. 
The idea behind most entanglement-verification procedures
is to uncover the non-local behaviour of the entangled state. This
non-locality can manifest itself in a spatial basis, such as it
happens, for instance, while probing the non-locality of a single
particle \cite{Jacob1}, but it can also arise in the basis of
the number of particles. The protocol proposed in \cite{Vedral}
exploits the idea presented in \cite{Simon}, that there cannot
be a pure separable state with fixed particle number and full
interference among spatial modes. Let us suppose that after the 
intermediate degrees of freedom in the chain has been reduced,
we are left with a density matrix $\hat{\rho}$, which represents our
knowledge of the resulting two-side mixed state, or equivalently, 
the state of the chain ends. 
Following the protocol of reference \cite{Vedral}, bosons on the 
terminals are sent through a
50:50 splitter and then the number of particles on the
outputs is measured, for example following the technique proposed
in \cite{Feld}.

The effect of the 50:50 splitter is to combine the
modes of the original system, $\hat{a}_1^{\dagger}$ and 
$\hat{a}_N^{\dagger}$, according to the following transformation, 

\begin{eqnarray}
\hat{a}_1^{\dagger} = \frac{1}{\sqrt{2}}(\hat{a}_c^{\dagger} + \hat{a}_d^{\dagger}), \nonumber\\
\hat{a}_N^{\dagger} = \frac{1}{\sqrt{2}}(\hat{a}_c^{\dagger} - \hat{a}_d^{\dagger}),
\end{eqnarray}

where $\hat{a}_c^{\dagger}$ and $\hat{a}_d^{\dagger}$ are the 
modes of the system after the 50:50 splitter. Such splitter can
be considered as the physical place where interference between
the bosonic bulks takes place.
In this way, as $\hat{a}_c^{\dagger}$ and $\hat{a}_c$
are the operators describing the particle occupation on one 
of the outputs, EEE can then be detected from the experimentally 
measurable variable,

\begin{equation}
\epsilon_{AB} = tr(\hat{a}^{\dagger}_c \hat{a}_c \hat{\rho} )  
- \frac{N}{2} = tr(\hat{a}^{\dagger}_1 \hat{a}_N \hat{\rho} ),
\end{equation}

where, 

\begin{equation}
tr(\hat{a}^{\dagger}_c \hat{a}_c \hat{\rho} ),
\end{equation}

is the number of particles in one output and $\hat{\rho}$ is the reduced 
density matrix of the ends. As it is characteristic of a good 
entanglement measure,

\begin{equation}
\epsilon_{AB}=0, 
\end{equation}

for separable states and, 

\begin{equation}
\epsilon_{AB}>0,
\end{equation}

for entangled states. In this way, entanglement can be detected
by just comparing the amount of bosons in one output, which can
be done with relative efficiency. In order to show how this method 
can be applied to our model, we present in figure \ref{arno} a 
graph of $\epsilon_{AB}$ obtained from the ground state of chains
with PTH and CH for some of the parameters in figure \ref{G_lo}. 
We conclude that $\epsilon_{AB}$ correctly
determines whether the state is entangled or not, but it does not
faithfully depicts the actual amount of entanglement in the system
as it is shown in figure \ref{G_lo}. 
This might be due to the fact
that $\epsilon_{AB}$ accounts for the entanglement contained in the
degrees of freedom corresponding to first order physics, that is,
those given in terms of averages and mean values of measurable
observables, while most of the entanglement in the system comes
from high order physics, as for instance long range particle
exchange. In any case it would be of great interest to investigate
this entanglement measure in the laboratory, since this would 
definitely tell whether long range entanglement can be achieved 
in bosonic chains.

\chapter{Kicked dynamics in systems of ultracold atoms}

\label{cucu}

\section{Dynamics of a double-kicked bosonic condensate}

So far we have studied diverse aspects of bosonic systems described by
fully quantized Hamiltonians. Such an approach is necessary when we want
to deal with quantities such as entanglement, since one needs to treat 
sections of the system as individual entities. For instance, in order 
to find EEE we need to get the reduced density matrix of the chain ends, 
which describes the statistics of such two-site system.  As we have
shown, this description of the state bears some technical complications.
For this reason, alternative approaches are of interest. In cases when
the object of study depends on the system's behaviour as a whole and 
not on the combination of mutually interacting contributions, a simpler 
approach provides valuable insight. It is in this context that we want to
address the problem of the kicked condensate \cite{Rancon,Raizen1,Raizen2}. 
Here we want to study the stability of a condensate of bosons when
it is subject to a periodic kicking. The way we manoeuvre the problem
is the following: we first study the evolution of the non-linear 
Schrodinger equation and find the condensate wave function. Then,
we treat this solution as the ground state of the system, and
analyse how the condensate population migrates towards excited
levels. It is worth mentioning that this latter part 
is carried out following the procedure introduced by Castin and
Dum in reference \cite{CasDum}.
This procedure is in many aspects similar to the
Bogoliubov method to study interacting Bose gases. However, in the
latter case the state of the system is described by a {\it coherent}
state and therefore the number of particles is not well defined. The
Castin and Dum approach overcomes this issue by expanding the atomic
field operator in a sum of condensed and non-condensed parts and 
assuming that the non-condensed contribution can be considered as
a perturbation with appropriate scaling behaviour with respect to
the condensed part. In this way, the total number of particles
is always fixed and the proportion of particles in the condensate
can be calculated as the expectation value of the condensate field
operator.
Suppose we have a bosonic system which can be described by means of the
Gross-Pitaevskii equation, which corresponds to a mean field approximation
of the Hamiltonian (\ref{BH}), plus a periodic kicking. The particles
are confined in a ring-shaped trap of radius $R$. We assume that the
lateral dimension $r$ of the trap is much smaller that its circumference,
and thus the system can be effectively considered as one-dimensional.
In this model, the one-particle wave-function, $\psi$, is given by the 
following second order differential equation,

\begin{equation}
i\frac{\partial \psi}{\partial t} = -\frac{1}{2}\frac{\partial^2 \psi}{\partial \theta^2} + g |\psi|^2\psi + k \cos \theta \left ( \sum_{n=0}^\infty {\delta(t-n\beta)-\delta(t-n\beta-\epsilon) } \right )\psi. 
\label{eZq:1}
\end{equation}

This model can be physically associated with a Bose Einstein Condensate (BEC) 
that is being double-kicked with periodicity $\beta$. Two consecutive kicks 
have opposite sign so that they nearly cancel each other and depletion is
contained. The time interval $\epsilon$ is a small number and can be considered 
as a perturbative parameter. The intensity of the external 
potential is $k$ and the interaction strength is $g$, which is described
by a mean-field term with scaled strength \cite{Raizen1,Raizen2}, 

\begin{equation}
g=\frac{8 M a_s R}{r^2},
\end{equation}

where $a_s$ is the $s$-wave scattering length, and $M$ is the total number
of bosons. The length is measured in units of $R$ and energy in units of,

\begin{equation}
E_R = \frac{\hbar}{2 m R^2},
\label{sontren}
\end{equation}

with $m$ the atomic mass. In addition, we assume periodic boundary 
conditions $\psi(0)=\psi(2 \pi)$. It is worth mentioning that in
writing equation (\ref{eZq:1}) we have discontinued the notation of previous 
sections in order to make our formulae compatible with standard notation 
in the field. 
Equation (\ref{eZq:1}) is non-linear, hence conventional quantum-mechanics 
methods cannot be applied. In simple terms, our problem consists in getting
an analytical solution of the GP equation for the initial condition, 

\begin{equation}
\psi_0(\theta) = \frac{1}{\sqrt{2 \pi}}. 
\label{knik}
\end{equation}

Let us point out that because this initial wave function is symmetric 
with respect to $\theta$, the wave function must reflect the same symmetry 
for all times, since such symmetry is explicit in the GP equation as well. 
Equation (\ref{eZq:1}) along with the initial condition 
(\ref{knik}) can be solved straightforwardly for $k=0$ or $\epsilon=0$. 
The case $g=0$ is a double-kicked quantum rotor 
\cite{Creffield3,Creffield2,Stocklin}. The time evolution of 
the initial wave function is dictated by the Floquet operator,

\begin{equation}
\hat F (\beta,\epsilon) = \hat T_1(\beta) e^{i k \cos (\theta)} \hat T_2 (\epsilon) e^{-i k \cos (\theta)},
\label{eq:2}
\end{equation}

where $\hat T_1(\beta)$ and $\hat T_2(\epsilon)$ are the evolution operators 
for times $\beta$ and $\epsilon$ respectively. As $\hat T_2(\epsilon)$ induces
evolution for a small interval of time, the wave function can be approximated
by its average value during this short period so that we can write, 

\begin{equation}
e^{i k \cos (\theta)} \hat T_2(\epsilon) e^{-i k \cos (\theta)}  \approx e^{i k \cos (\theta)} e^{i\epsilon(\frac{1}{2}\frac{\partial ^2}{\partial \theta^2} - g |\psi_{fixed}|^2  )}    e^{-i k \cos (\theta)}. 
\label{eq:3}
\end{equation}
 
This expression is exact to first order in $\epsilon$. As can be seen, the
right-hand side of the equation is reminiscent of a unitary transformation. 
It is not difficult to see that such a transformation induces the following 
changes in the quantum operators,

\begin{equation}
e^{i k \cos (\theta)} \frac{\partial}{\partial \theta}  e^{-i k \cos (\theta)} = \frac{\partial}{\partial \theta} + i k \sin (\theta),
\label{eq:4}
\end{equation}
\begin{equation}
e^{i k \cos (\theta)} |\psi_{fixed}(\theta)|^2  e^{-i k \cos (\theta)} = |\psi_{fixed}(\theta)|^2.
\label{eq:5}
\end{equation}

Inserting these results in equation (\ref{eq:2}) we obtain,

\begin{equation}
\hat F (\beta,\epsilon) \approx \hat T_1(\beta) e^{i\epsilon \left ( \frac {1}{2}\left( \frac{\partial}{\partial \theta} + i k \sin \hat \theta \right)^2 - g |\psi_{fixed}|^2  \right )}.
\label{eq:6}
\end{equation}

When written in this way, the Floquet operator allows us to identify the 
effective perturbation due to two consecutive kicks. It is worth recalling 
that $\epsilon k$ must be small so that the net effect of two consecutive 
kicks can be considered as a perturbation and the wave function evolves 
slowly as seen stroboscopically after pairs of kicks. Similarly, it should 
be noticed that equation (\ref{eq:6}) is exact for $g=0$. We would like to 
find a way of introducing a perturbative potential in equation 
(\ref{eZq:1}) from the information contained in equation (\ref{eq:6}). 
Curiously, should the expression on the exponential be exclusively 
$\theta$-dependent, it would be easy to carry out such a procedure. We then could 
simply introduce a perturbative potential like,

\begin{equation}
\delta V(\theta, t) = f(\theta) \sum_{n=0}^\infty {\delta(t-\beta n)}.
\end{equation}

However, the operator on the exponential in equation (\ref{eq:6}) depends 
explicitly on $\frac{\partial}{\partial \theta}$ and therefore dynamical
effects take part in the perturbation process as well. To see this more
explicitly, let us put 
the term on the exponential in equation (\ref{eq:6}) in the following form 
(up to an $i$ factor),

\begin{equation}
\left \{ \frac{\epsilon}{2} \frac{\partial^2}{\partial \theta^2} - \epsilon g |\psi_{fixed}|^2 \right \}  + 2 i \frac{\epsilon}{2} k \sin \hat \theta \frac{\partial}{\partial \theta} + i \frac{\epsilon}{2} k \cos \hat \theta -  \frac{\epsilon}{2} k^2 \sin^2 \hat \theta. 
\label{eq:7}
\end{equation}

The term in brackets represents free evolution while the other terms account 
for the perturbation. We would like to disentangle these contributions in 
order to simplify the dynamics. As the commutator between terms is second 
order in $\epsilon$, {\it splitting the exponential preserves first order 
accuracy}. Similarly, it can be shown that as long as we stay in 
the small perturbation regime, the term that goes with the first derivative
actually depends on {$\epsilon^2$}, hence we drop it. After these 
simplification the Floquet operator can be written as,

\begin{equation}
\hat F (\beta,\epsilon) \approx \hat T_1(\beta) e^{i\epsilon \left ( \frac{1}{2} \frac{\partial^2}{\partial \theta^2} - g |\psi_{fixed}|^2 \right )} e^{i\left ( i \frac{\epsilon}{2} k \cos \hat \theta -  \frac{\epsilon}{2} k^2 \sin^2 \hat \theta \right )}.
\label{eq:211}
\end{equation}

Furthermore, putting together terms representing time evolution and obtain,

\begin{equation}
\hat T_1(\beta) e^{i\epsilon \left ( \frac{1}{2} \frac{\partial^2}{\partial \theta^2} - g |\psi_{fixed}|^2 \right )} = \hat{T}(\beta + \epsilon).
\label{eq:221}
\end{equation}

This step amounts to sticking the short time slice between kicks to the 
longer time interval between pairs of kicks. This sticking is mathematically 
correct and does not introduce additional errors to our analysis except by the 
fact that $\psi$ is considered time-independent in the short time
interval. 

Introducing expression (\ref{eq:221}) in (\ref{eq:211}) we find that the 
Floquet operator can be written as the kicking term followed by free 
dynamics. Moreover, from the Floquet operator we can go back to equation 
(\ref{eZq:1}) and rewrite it as,

\begin{equation}
i\frac{\partial \psi}{\partial t} = -\frac{1}{2}\frac{\partial^2 \psi}{\partial \theta^2} + g |\psi|^2\psi + \left ( - i \frac{\epsilon k}{2}  \cos (\theta) + \frac{\epsilon k^2}{2} \sin^2 (\theta)  \right )   \left ( \sum_{n=0}^\infty {\delta(t-n(T))} \right )\psi,
\label{eq:8}
\end{equation}
 
where $T=\beta + \epsilon$. Now the label $n$ in the sum runs over pairs of 
kicks rather than on single kicks. Correspondingly, the original problem has
been re-formulated in terms of better-understood single-kick dynamics 
\cite{Fishman,Grempel,Shepe,Creffield2}. The 
fact that the term proportional to $\cos (\theta)$ is not hermitian does not 
lead to any inconsistency because the whole term accompanying the delta 
function does not show up in the energy formula (equation (\ref{eq:33}) 
ahead). This complex contribution in the new GP equation is related to 
kinetic effects coming from differentiating equation (\ref{eq:6}). 

In what follows, we apply the Castin and Dum linearisation-approach presented
in references \cite{Castin,CasDum}. Briefly, the formalism consists in 
linearising the time-dependent GP equation for a small time-dependent 
perturbation. Such approach leads to a set of two-variable differential 
equations driven by inhomogeneous terms. The condensate evolution is then 
determined by the form of the small time-dependent perturbation or source. 
In order to review the Castin and Dum's method \cite{Castin}, let us write 
the solution of equation (\ref{eq:8}) as a sum of two terms,

\begin{equation}
\psi(\theta,t) = \frac{1}{\sqrt{2\pi}} + \delta \psi(\theta,t),
\label{zuza}
\end{equation}

so that replacing (\ref{zuza}) in (\ref{eq:8}) we get,

\begin{equation}
i \frac{\partial \delta \psi }{\partial t} = -\frac{1}{2}\frac{\partial^2 \delta \psi}{\partial \theta^2} + \frac{2 g \delta \psi}{2\pi} + \frac{g \delta \psi^*}{2\pi} + \frac{\delta U}{\sqrt{2\pi}},
\label{pepo}
\end{equation}

where,

\begin{equation}
\delta U(\theta,t) = \left ( - i \frac{\epsilon k}{2}  \cos \theta + \frac{\epsilon k^2}{2} \sin^2 \theta \right )   \left ( \sum_{n=0}^\infty {\delta(t-n(T ))} \right ).
\end{equation}

In addition, we have neglected a term proportional to $|\delta \psi|^2$
in equation (\ref{pepo}), which means we have removed the non-linearity.   
As the wave function is complex, the real and imaginary part can be 
treated as independent real functions. Nevertheless, it is more elegant to
keep the original function $\delta \psi$ and introduce its complex
conjugate as the other unknown. It can be shown that in doing so equation
(\ref{pepo}) leads to,

\begin{equation}
i\frac{\partial}{\partial t}
\left(
\begin{array}{c}
\delta \psi  \\
\delta \psi^*
\end{array}
\right) =
\left(
\begin{array}{cc} 
-\frac{1}{2} \frac{\partial^2}{\partial \theta^2} + \frac{g}{2\pi} & \frac{g}{2\pi}\\
-\frac{g}{2\pi}  & - \left( -\frac{1}{2} \frac{\partial^2}{\partial \theta^2} + \frac{g}{2\pi} \right)
\end{array}
\right)
\left(
\begin{array}{c}
\delta \psi  \\
\delta \psi^*
\end{array}
\right) 
+
\left(
\begin{array}{c}
S(\theta,t) \\
S(\theta,t)^*
\end{array}
\right). 
\end{equation}

In our case, the source term is given by,

\begin{equation}
S(\theta,t) = \delta U(\theta,t) \psi_0 = \left ( - i \frac{\epsilon k}{2}  \cos \theta + \frac{\epsilon k^2}{2} \sin^2 \theta \right )   \left ( \sum_{n=0}^\infty {\delta(t-n(T ))} \right )\frac{1}{\sqrt{2 \pi}}.
\label{eq:9}
\end{equation}

This is essentially a linear equation and as such it can be solved
using standard methods. Some care must be taken, for instance,
in order to make sure that function $\delta \psi$  has a physical
meaning so that it remains orthogonal to the homogeneous
solution $\psi_0 = \frac{1}{\sqrt{2 \pi}}$. Issues like this are
extensively discussed in reference \cite{Castin}. In what follows,
we limit ourselves to using the results of this reference to get 
a solution to our problem. In synthesis, the wave function is given
by,

\begin{eqnarray}
&& \psi(\theta,t) =  \\
&& \frac{1}{\sqrt{2 \pi}} \left( 1 + U_1 b_1 (t) e^{i \theta} + V_1 b_1^* (t) e^{-i \theta} + U_{-1} b_{-1} (t) e^{-i \theta} + V_{-1} b_{-1}^* (t) e^{i \theta} \right. \nonumber \\
&& \left.  + U_2 b_2 (t) e^{2 i \theta} + V_2 b_2^* (t) e^{-2 i \theta} + U_{-2} b_{-2} (t) e^{-2 i \theta} + V_{-2} b_{-2}^* (t) e^{2 i \theta} \right). \nonumber
\label{eq:23}
\end{eqnarray}

The unknown coefficients can be obtain from time integration,

\begin{equation}
b_j(t) = \int_0^t \frac{d \tau}{i} s_j (t - \tau) e^{-i \epsilon_j \tau},
\label{eZq:12}
\end{equation}

where,

\begin{equation}
s_j (t) = \int_0^{2 \pi} d \theta \left( U_j \left( \frac{e^{-i j \theta}}{\sqrt{2 \pi}}  \right)S(\theta,t) + V_j \left( \frac{e^{-i j \theta}}{\sqrt{2 \pi}}  \right)S^*(\theta,t)    \right),
\label{eq:90}
\end{equation}

and finally, $U_j$ and $V_j$ are given by \cite{Castin},

\begin{eqnarray}
U_j + V_j = \zeta_j, \nonumber \\ 
U_j - V_j = \frac{1}{\zeta_j},
\label{eZq:17}
\end{eqnarray}

with,

\begin{equation}
\zeta_j = \left( \frac{\frac{j^2}{2}}{\frac{j^2}{2} + \frac{g}{\pi} }\right)^{\frac{1}{4}},
\end{equation}

and,

\begin{equation}
\epsilon_j = \sqrt{\frac{j^2}{2}\left( \frac{j^2}{2} + \frac{g}{\pi}  \right)}.
\label{eZq:19}
\end{equation}

In principle, $j$ can be any integer number. Fortunately, only four modes 
contribute, namely, $j=-2,-1,1,2$. Performing the integrations we find,

\begin{equation}
s_1(t) = \frac{i \epsilon k}{4} (-U_1 + V_1)  \left ( \sum_{n=0}^\infty {\delta(t-n(T ))} \right ),
\label{eZq:10}
\end{equation}

\begin{equation}
s_2(t) = -\frac{\epsilon k^2}{8} (U_2 + V_2)  \left ( \sum_{n=0}^\infty {\delta(t-n(T ))} \right ),
\label{eZq:11}
\end{equation}

and $s_{-1}=s_1$, $s_{-2}=s_2$. Now we must calculate the time-dependent 
coefficients (\ref{eZq:12}), which are found to be,

\begin{eqnarray}
& b_1(t) = \frac{\epsilon k}{4} (-U_1 + V_1) \sum_{n=0}^{N(t)-1} e^{-i \epsilon_1 (t - n T)}, & \label{eZq:13} \\
& b_2(t) = \frac{i \epsilon k^2}{8} (U_2 + V_2) \sum_{n=0}^{N(t)-1} e^{-i \epsilon_2 (t - n T)}, & \label{eZq:14} \\
& b_{-1}(t) = b_1(t), & \\ 
& b_{-2}(t) = b_2(t). \label{eZq:15} &
\end{eqnarray}

In equations (\ref{eZq:13}) and (\ref{eZq:14}) $N(t)$ is the number of pairs of 
kicks ($N(t)=1,2,...$) and $\epsilon_{1,2}$ are the eigenenergies associated 
to each mode. As can be seen these coefficients are written in terms of time, 
nevertheless, it 
is more convenient to put them in terms of the number of kicks in order to 
compare with numerical results, which lead us to replace $t$ by $(N-1) T$. 
After some rearrangement coefficients $b_1$ and $b_2$ adopt the
following form,

\begin{equation}
b_1 (N) = -\frac{\epsilon k}{4 \zeta_1} e^{-i\epsilon_1 (N-1) T} \sum_{n=0}^{N-1} e^{i \epsilon_1 T n} = -\frac{\epsilon k}{4 \zeta_1 \sin \frac{\epsilon_1 T}{2}} e^{-i \frac{\epsilon_1 T}{2}(N-1)}\sin {\frac{\epsilon_1 T}{2} N},
\label{eq:21}
\end{equation}

\begin{equation}
b_2 (N) = i \frac{\epsilon k^2 \zeta_2}{8} e^{-i\epsilon_2 (N-1) T} \sum_{n=0}^{N-1} e^{i \epsilon_2 T n} = i \frac{\epsilon k^2 \zeta_2}{8 \sin \frac{\epsilon_2 T}{2}} e^{-i \frac{\epsilon_2 T}{2}(N-1)}\sin {\frac{\epsilon_2 T}{2} N}.
\label{eq:22}
\end{equation}

On the other hand, the wave function symmetry can be made explicit by 
factorizing the exponentials in equation (\ref{eq:23}). This leads to,

\begin{equation}
\psi(\theta,N) = \left( 1 + 2 (U_1 b_1 (N) + V_1 b_1^*(N))\cos \theta + 2 (U_2 b_2 (N) + V_2 b_2^*(N))\cos {2 \theta} \right).
\end{equation}

Furthermore, inserting $b_1$ and $b_2$ in $\psi(\theta,N)$ and after 
some direct simplifications we obtain, 

\begin{eqnarray}
&& \psi(\theta,N) = \nonumber \\
&& \frac{1}{\sqrt{2 \pi}} \left( 1 -\frac{\epsilon k}{2 \sin \omega_1}  \sin {\omega_1 N } \left( \cos {\omega_1 (N-1)} - \frac{i}{\zeta_1^2} \sin{\omega_1 (N-1)}    \right)\cos \theta  \right. \nonumber \\
&& \left. + \frac{\epsilon k^2}{4 \sin \omega_2}  \sin {\omega_2 N } \left( \zeta_2^2 \sin{\omega_2 (N-1)} + i \cos {\omega_2 (N-1)} \right) \cos {2 \theta}  \right),\label{eq:24}
\end{eqnarray}

\begin{figure}[t]
\begin{center}
\includegraphics[width=0.4\textwidth,angle=-0]{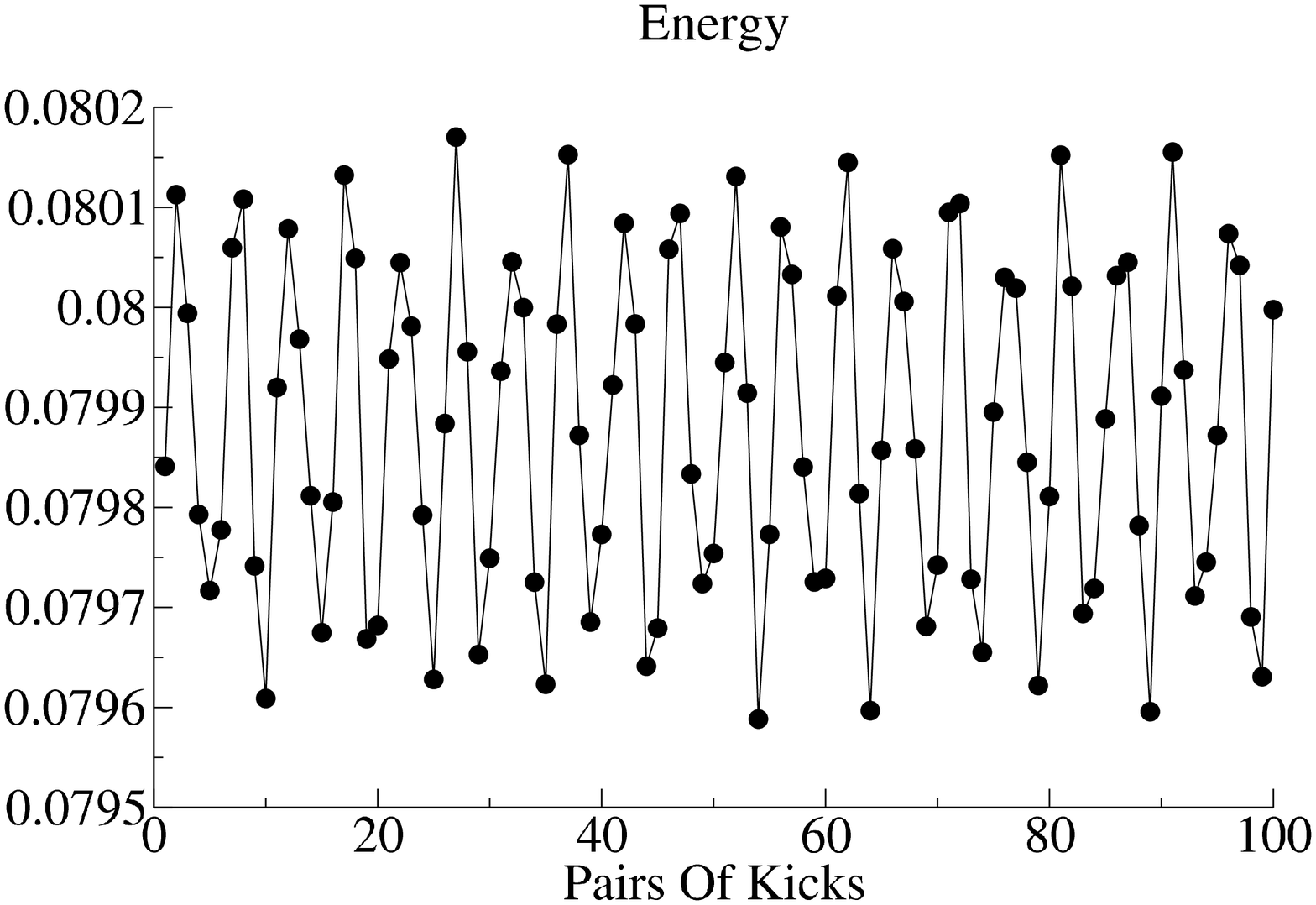}\includegraphics[width=0.4\textwidth,angle=-0]{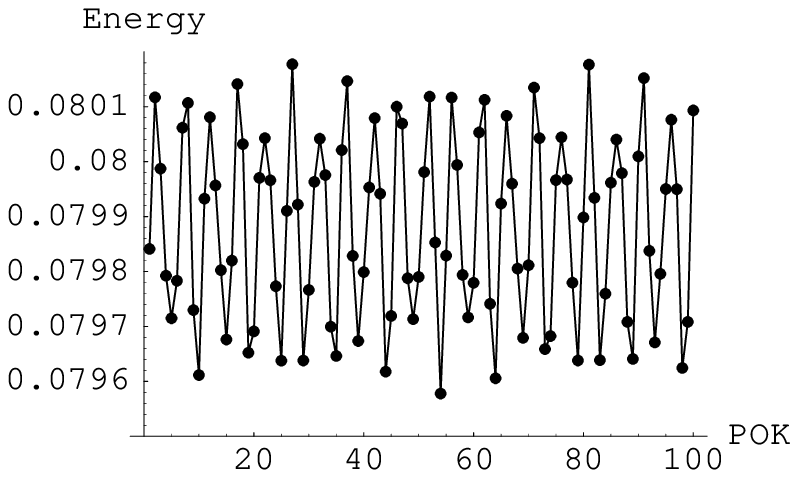}
\includegraphics[width=0.4\textwidth,angle=-0]{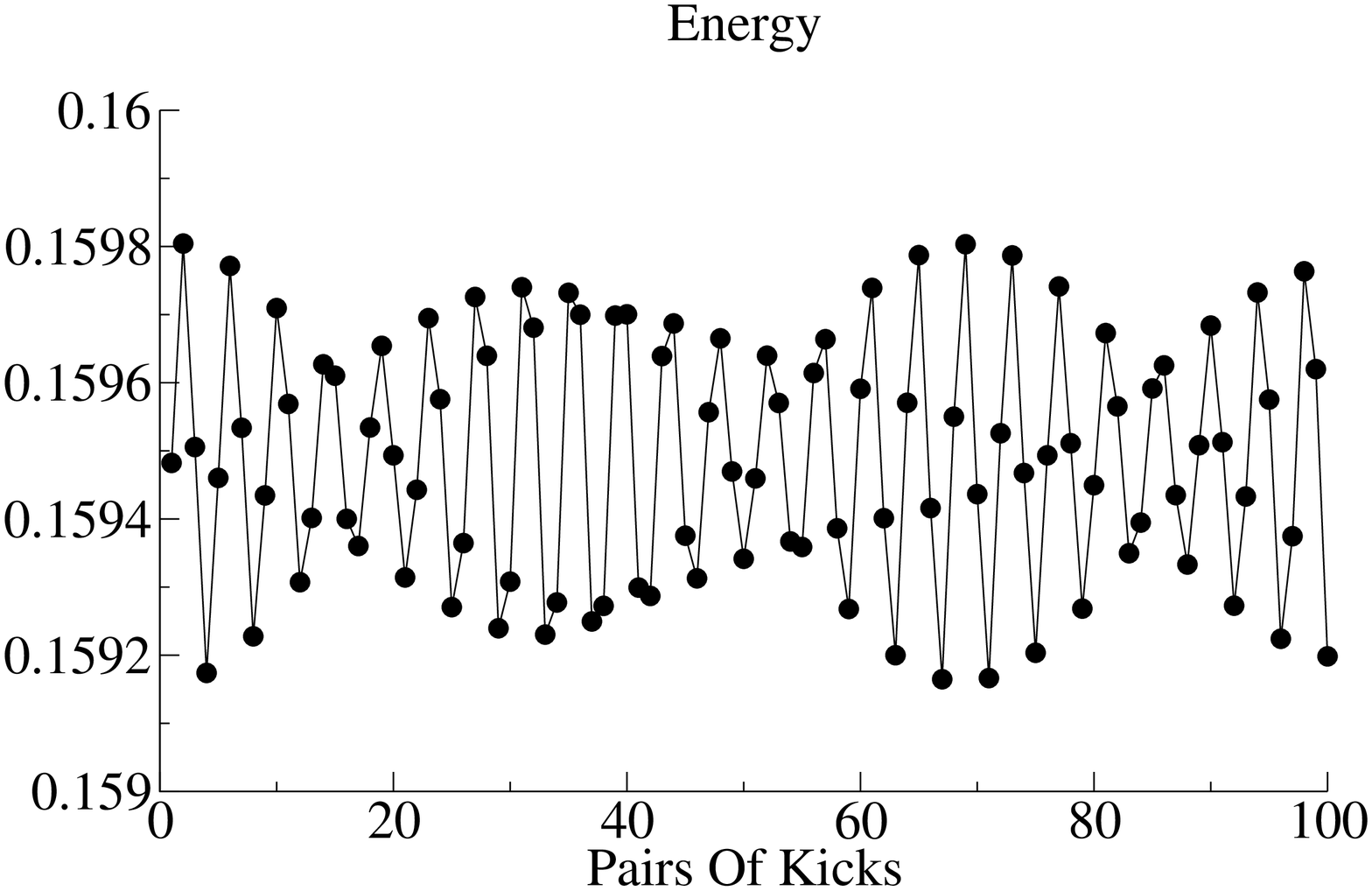}\includegraphics[width=0.4\textwidth,angle=-0]{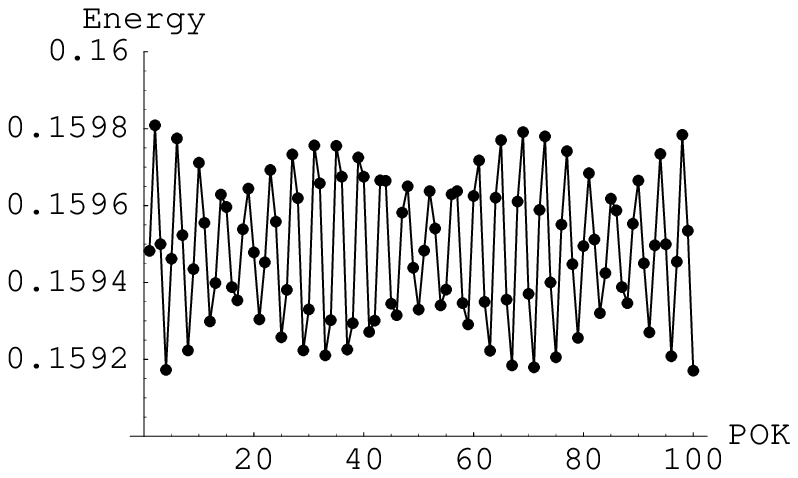}
\includegraphics[width=0.4\textwidth,angle=-0]{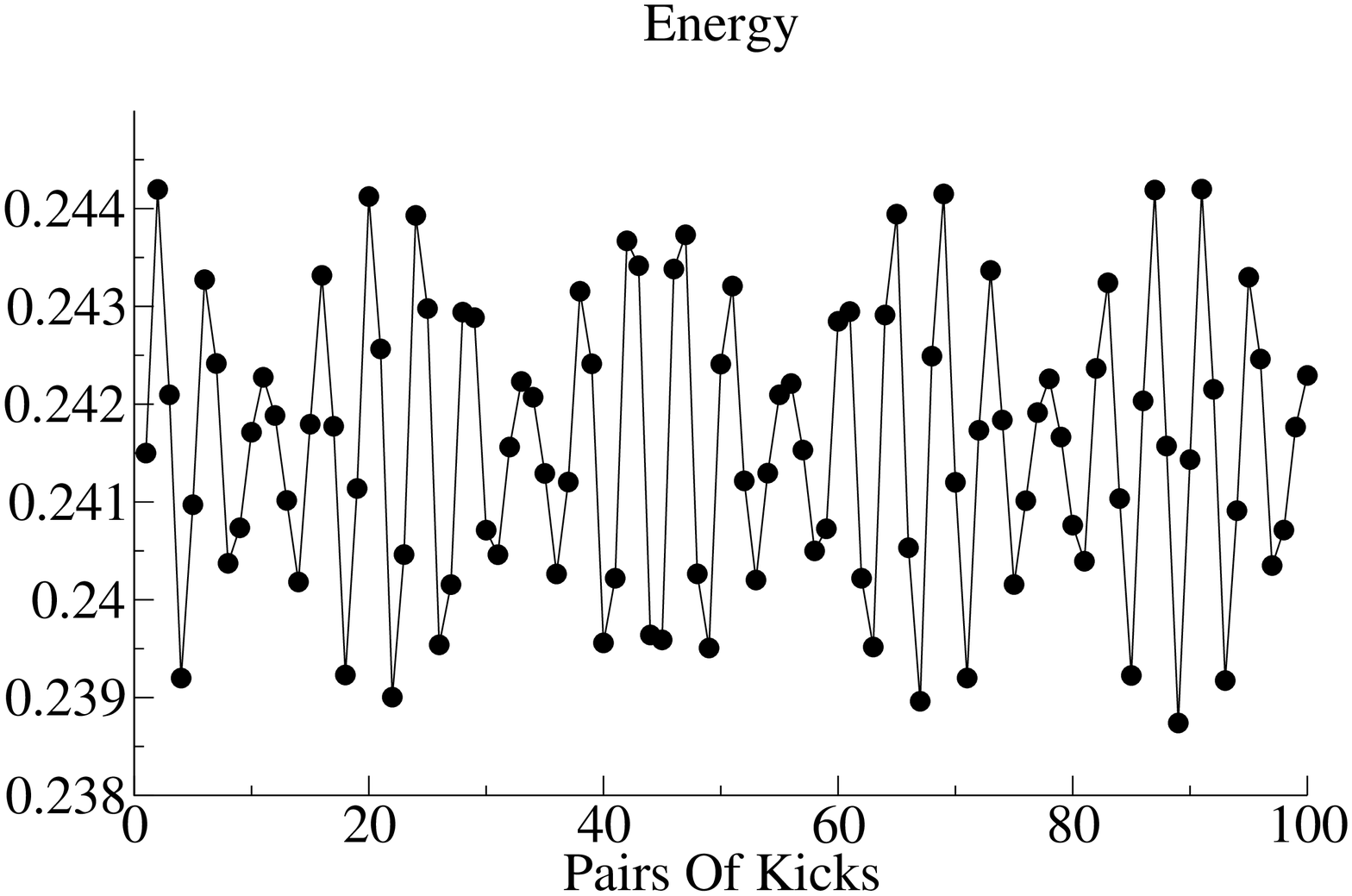}\includegraphics[width=0.4\textwidth,angle=-0]{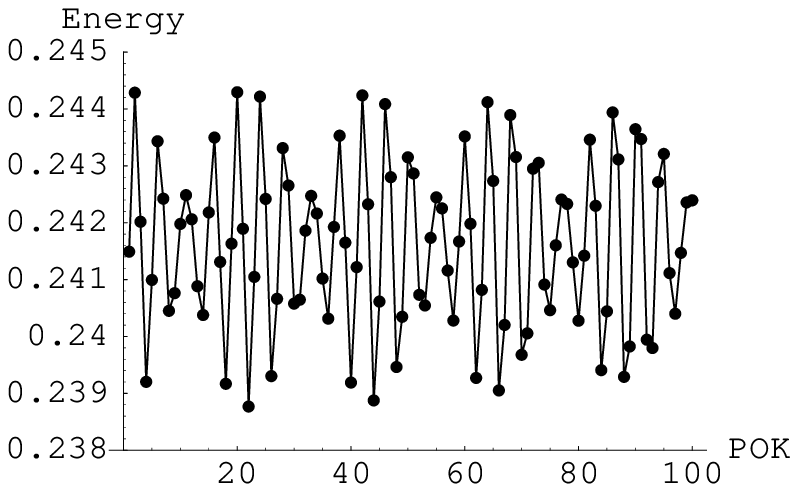}
\caption{Right graphs show results obtained from equation (\ref{eq:24}) and 
left graphs show the same results obtained numerically. Top. $k=1,g=1$. Medium position. $k=1$,$g=2$. Bottom. $k=2$,$g=2$. Energy is measured in
terms of the recoil energy given in equation (\ref{sontren}).}  
\label{caribe}
\end{center}
\end{figure}

where $\omega_j = \frac{\epsilon_j T}{2}$. This wave function reads 
$\frac{1}{\sqrt{2 \pi}}$ for $\epsilon=0$ or $k=0$ as expected. In order 
to check the accuracy of equation (\ref{eq:24}) we calculate the state 
energy by the formula,

\begin{equation}
E(N) = \int_0^{2 \pi} d \theta \psi^*(N) \left ( -\frac{1}{2}\frac{\partial^2}{\partial \theta^2} + \frac{1}{2}g|\psi(N)|^2 \right ) \psi(N),
\label{eq:33}
\end{equation}

and compare it with computer simulations that use fast Fourier transform to
shift the state wave function from position to momentum representation and
vice versa. The graphs in figure \ref{caribe} show in detail the energy 
oscillations for different values of $k$ and $g$ both from numerical simulations 
and from wave function (\ref{eq:24}). Outstandingly, both approaches agree
quite well over a wide range of parameters.

\section{Stability Analysis}

So far we have considered the condensate as an isolated system. However, 
in a more realistic model particles in the ground state interact with 
particles in levels of higher energy and this causes bosons to escape 
from the condensate. Both the repulsion as well as the constant kicking 
determine the system stability. The second quantized analogous 
of the GP equation is given by \cite{CasDum},
 
\begin{equation}
\hat{H}=\int_0^{2\pi} d\theta \left( \hat{\psi^\dagger} \hat{h} \hat{\psi} + \frac{g}{2} \hat{\psi^\dagger}\hat{\psi^\dagger}\hat{\psi}\hat{\psi}   \right),
\label{Seq:1}
\end{equation}

where, 

\begin{eqnarray}
&& \hat{h} = -\frac{1}{2}\frac{\partial^2}{\partial \theta^2} + W(\theta,t), \\ 
&& W(\theta,t) = \left ( - i \frac{\epsilon k}{2}  \cos \theta + \frac{\epsilon k^2}{2} \sin^2 \theta  \right )   \left ( \sum_{n=0}^\infty {\delta(t-n(T))} \right ).
\nonumber
\label{Seq:2}
\end{eqnarray}

In a perturbative approach the field operator can be written as,

\begin{equation}
\hat{\psi}(\theta,t) = \psi(\theta,t) \hat a_0 + \delta \hat{\psi}(\theta,t),
\label{Seq:3}
\end{equation}

where $\psi(\theta,t)$ is the ground state wave function of the condensate, 
given by equation (\ref{eq:24}). $\hat{a}_0$ and $\delta \hat{\psi}(\theta,t)$ 
are the bosonic operators for particles in and out the condensate respectively. 
Hamiltonian (\ref{Seq:1}) must be inserted in the Heisenberg equations of 
motion and then these equations must be analytically integrated to get the
Heisenberg dynamical operators. Ideally, the number of non-condensed 
particles (NNP) can be obtained by adding up the number of particles in
excited modes. However, this can be rarely done directly due to the 
non-linearity of the original Hamiltonian. For this reason we use the
method of reference \cite{CasDum}, where practical equations in the linear 
response regime are proposed for the non-condensate cloud. From this work we 
now borrow the following identity,

\begin{equation}
\left(
\begin{array}{c}
\hat \Lambda(\theta,t)  \\
\hat \Lambda^\dagger(\theta,t)
\end{array}
\right)
=
\sum_{j} 
\left \{
\hat b_j \left(
\begin{array}{c}
U_j (\theta,t)  \\
V_j (\theta,t)
\end{array}
\right)
+
\hat b^\dagger_j \left(
\begin{array}{c}
V^*_j (\theta,t)  \\
U^*_j (\theta,t)
\end{array}
\right)
\right \},
\label{Seq:4}
\end{equation}
 
where,

\begin{equation}
\hat \Lambda = \frac{1}{\sqrt{\hat N}} \hat a^\dagger_0 \delta \hat{\psi}.
\end{equation}

Although $\hat \Lambda$ is not the non-condensate field operator itself, it is 
relevant to the number of particles off the condensate since, 

\begin{equation}
\delta \hat{\psi}^\dagger \delta \hat{\psi} \approx \hat \Lambda^\dagger \hat \Lambda,  
\end{equation}

in the low temperature limit \cite{CasDum}. The number of excitations for each 
mode in equation (\ref{Seq:4}) is completely determined by the temperature 
through the well known Bose-Einstein distribution, 

\begin{equation}
\langle \hat b_j^\dagger \hat b_j \rangle = \frac{1}{e^{\beta E_j}-1}, 
\end{equation}

where $\beta = \frac{1}{k_B T}$. At zero temperature 
$\langle \hat b_j^\dagger \hat b_j \rangle = 0 $ and NNP 
is dictated only by the components of  $U_j(\theta,t)$ and $V_j(\theta,t)$ in 
the space orthogonal to $\psi(\theta,t)$. The evolution of $U_j(\theta,t)$ 
and $V_j(\theta,t)$ is dictated by, 

\begin{eqnarray}
& &
i\frac{\partial}{\partial t}
\left(
\begin{array}{c}
U_j(\theta,t) \\
V_j(\theta,t)   
\end{array}
\right)
= \label{Seq:5} \\
& &
\left(
\left(
\begin{array}{cc} 
-\frac{1}{2} \frac{\partial^2}{\partial \theta^2}& 0 \\
0 & \frac{1}{2} \frac{\partial^2}{\partial \theta^2}
\end{array}
\right)
+
\left(
\begin{array}{cc} 
W & 0 \\
0 & -W
\end{array}
\right)
+
\left(
\begin{array}{cc} 
2 g |\psi|^2 & g\psi^2 \\
-g{\psi^*}^2 & -2 g |\psi|^2
\end{array}
\right)
\right)
\left(
\begin{array}{c}
U_j(\theta,t) \\
V_j(\theta,t)   
\end{array}
\right), \nonumber
\end{eqnarray}

with $W=W(\theta,t)$ and $\psi = \psi(\theta,t)$. Additionally, NNP can be 
written as,

\begin{equation}
NNP = \sum_j \langle V_j |\hat Q \hat Q | V_j \rangle,
\label{Seq:6}
\end{equation}

with,

\begin{equation}
\hat Q = \hat 1 - | \psi \rangle \langle \psi |,
\end{equation}

in the limit of very low temperature. In order to carry out our numerical 
analysis, we first integrate the GP equation (\ref{eZq:1}) using a fast
Fourier transform algorithm and 
get the wave function $\psi(\theta,t)$. Then we insert this $\psi(\theta,t)$ 
into equation (\ref{Seq:5}) and integrate for the initial conditions 
(\ref{Seq:26}). Finally, we calculate NNP using equation (\ref{Seq:6}). 

\section{Evolution of the number of non-condensed particles}

Equation (\ref{Seq:6}) is the same as

\begin{equation}
NNP = \sum_j \left \{ \langle V_j | V_j  \rangle   - | \langle \psi | V_j   \rangle |^2 \right \}.
\label{Seq:7}
\end{equation}

Moreover, as $\psi$ is normalized to 1, NNP is always positive and dependent on 
$\langle V_j | V_j  \rangle$. Correspondingly, we focus on finding the {\it norm} 
of $V_j(\theta,t)$ since it provides us with quality information regarding the NNP. For $g=0$, 
$U_j(\theta,t)$ and $V_j(\theta,t)$ become decoupled in equation (\ref{Seq:5}) 
and both functions undergo unitary evolution that leaves the norm unchanged. 
This implies that norm diffusion is exclusively due to the non-linear term of 
equation (\ref{Seq:5}). In order to explore how the norm of $V_j(\theta,t)$ 
behaves in time, we highlight how the non-linear term acts on the state
vector, namely,

\begin{equation}
\left(
\begin{array}{c}
U_j(\theta,t+\Delta) \\
V_j(\theta,t+\Delta)   
\end{array}
\right)
= 
e^{i \Delta \hat G(\theta,t)  }
\left(
\begin{array}{c}
U_j(\theta,t) \\
V_j(\theta,t)   
\end{array}
\right),
\label{Seq:8}
\end{equation}

where,

\begin{equation}
\hat G (\theta,t) = 
\left(
\begin{array}{cc} 
2 g |\psi (\theta,t)|^2 & g\psi(\theta,t)^2 \\
-g{\psi(\theta,t)^*}^2 & -2 g |\psi(\theta,t)|^2
\end{array}
\right).
\label{Seq:9}
\end{equation}

As can be seen, operator $\hat G$ is not hermitian, which is by no means 
inconsistent, since further exploration shows that the eigenvalues of $\hat G$ 
are themselves real. This is an important issue to consider when exploring
the development of NNP. In fact, according to \cite{Castin}, the norm of the 
$U_j$-$V_j$ vector (global norm) is given by,

\begin{equation}
\left( \langle U_j |,- \langle V_j | \right)
\left(
\begin{array}{c}
| U_j \rangle \\
| V_j \rangle   
\end{array}
\right)
= 
\int_0^{2 \pi} d\theta |U_j(\theta,t)|^2 - \int_0^{2 \pi} d\theta |V_j(\theta,t)|^2 = Constant,
\label{Seq:10}
\end{equation}

with $Constant = 1$ for all $j$. As can be seen, the norm of $V_j(\theta,t)$ 
can grow indefinitely as long as the difference of the two norms remains 
constant. Notice that norm growing depends heavily on the definition of 
inner product 
in the Hilbert space. If we had $+$ instead of $-$ above the norm 
of $V_j(\theta,t)$ and $U_j(\theta,t)$ would be bounded by $Constant$. 
Making use of norm conservation properties we now write the state vector
as,

\begin{equation}
\left(
\begin{array}{c}
U_j(\theta,t) \\
V_j(\theta,t)   
\end{array}
\right)
= 
\left(
\begin{array}{c}
\sinh \left ( \alpha_j(t) \right ) u_j(\theta,t) \\
\cosh \left ( \alpha_j(t) \right ) v_j(\theta,t)   
\end{array}
\right),
\label{Seq:11}
\end{equation}

$u_j(\theta,t)$ and $v_j(\theta,t)$ are 
complex functions normalized to unity while $\alpha_j(t)$ is a real function 
that characterizes the norm of $V_j(\theta,t)$ and $U_j(\theta,t)$. In order 
to gain insight into the evolution of the norm of $U_j(\theta,t)$ and 
$V_j(\theta,t)$, or equivalently, the evolution of $\alpha(t)$, we 
explicitly perform the operation indicated in equation (\ref{Seq:8}) 
following a similar procedure as in reference \cite{Zhang}. In doing so
we find that matrix $\hat G$ can be written as (let us omit the dependence 
with $\theta$ and $t$ for simplicity),

\begin{equation}
\hat G = 
\left(
\begin{array}{cc} 
2 g |\psi|^2 & g |\psi|^2 e^{i \phi} \\
-g|\psi|^2 e^{-i \phi}  & -2 g |\psi|^2
\end{array}
\right),
\label{Seq:12}
\end{equation}

with the following eigenvectors (not normalized),

\begin{eqnarray}
\langle \theta | E_+ ) = 
\left(
\begin{array}{c}
e^{2 i \phi(\theta,t)} \\
\sqrt{3}-2 
\end{array}
\right), \hspace{0.5cm}
\lambda_+ = \sqrt{3} g |\psi(\theta,t)|^2,
\label{Seq:13} \\
\langle \theta | E_- ) = 
\left(
\begin{array}{c}
e^{2 i \phi(\theta,t)} \\
-(\sqrt{3}+2 )
\end{array}
\right), \hspace{0.5cm}
\lambda_- = -\sqrt{3} g |\psi(\theta,t)|^2.
\label{Seq:14}
\end{eqnarray}

From equation (\ref{Seq:8}) we can infer,

\begin{equation}
\left(
\begin{array}{c}
| U_j \rangle\\
| V_j \rangle 
\end{array}
\right)_{t+\Delta}
= \left ( \mu_j e^{i \lambda_+ \Delta} | E_+ ) + \nu_j e^{i \lambda_- \Delta} | E_- )\right)_t.
\label{Seq:15}
\end{equation}

Scalars $\mu_j(\theta,t)$ and $\nu_j(\theta,t)$ can be written in terms 
of $U_j(\theta,t)$ and $V_j(\theta,t)$ making use of this equation with 
$\Delta = 0$. In doing so we find,

\begin{eqnarray}
\mu_j(\theta,t) = \frac{(\sqrt{3}-2)V_j(\theta,t) - e^{-2 i \phi(\theta,t)}U_j(\theta,t)}{2(3-2\sqrt{3})}, \label{Seq:16} \\
\nu_j(\theta,t) = -\frac{(\sqrt{3}+2)V_j(\theta,t) + e^{-2 i \phi(\theta,t)}U_j(\theta,t)}{2(3+2\sqrt{3})}. \label{Seq:17}
\end{eqnarray}

At this point it is worth noticing that global-norm conservation is a 
direct consequence of,

\begin{equation}
(E_-|E_+)=0.
\label{Seq:18}
\end{equation}

Now, let us introduce the symbol $[...|...]$ to describe an inner product as in 
equation (\ref{Seq:10}) but with $+$ instead of $-$, so that we can address 
the following ``global-norm'',

\begin{eqnarray}
&&\left( \langle U_j | U_j \rangle + \langle V_j | V_j \rangle \right )_{t+\Delta} = \label{Seq:19} \\
&& \left ( [ E_+ |  |\mu_j|^2 | E_+] + [ E_- ||\nu_j|^2| E_-] + [E_+ |\mu_j^* \nu_j e^{-2i\lambda_+ \Delta }| E_-] +  [E_- |\nu_j^* \mu_j e^{2i\lambda_+ \Delta }| E_+] \right )_t, \nonumber
\end{eqnarray}

where use has been made of equation (\ref{Seq:15}). If we expand exponentials to 
first order in $\Delta$ we obtain,

\begin{eqnarray}
\left( \langle U_j | U_j \rangle + \langle V_j | V_j \rangle \right )_{t+\Delta} = \left( \langle U_j | U_j \rangle + \langle V_j | V_j \rangle \right )_t + && \label{Seq:20} \\
2 i \Delta  \left ( [E_- |\lambda_+ \nu_j^* \mu_j| E_+] -  [E_+ |\lambda_+ \mu_j^* \nu_j| E_-] \right )_t,  && \nonumber
\end{eqnarray}

hence it follows,

\begin{eqnarray}
\frac{\left( \langle U_j | U_j \rangle + \langle V_j | V_j \rangle \right )_{t+\Delta} - \left( \langle U_j | U_j \rangle + \langle V_j | V_j \rangle \right )_t}{\Delta} = && \label{Seq:21}  \\
2 i  \left ( [E_- |\lambda_+ \nu_j^* \mu_j| E_+] -  [E_+ |\lambda_+ \mu_j^* \nu_j| E_-] \right )_t, && \nonumber
\end{eqnarray}

which amounts to,

\begin{equation}
-8 \sqrt{3} g \int_0^{2\pi} d\theta |\psi|^2 Im( \nu_j^* \mu_j).
\label{Seq:22}
\end{equation}

So, using (\ref{Seq:16}) and (\ref{Seq:17}) to write the integral above in terms of $U_j$ and $V_j$ we get,

\begin{equation}
\frac{d}{dt} \left( \langle U_j | U_j \rangle + \langle V_j | V_j \rangle \right ) = -4 g \int_0^{2\pi} d\theta |\psi|^2 Im(e^{2 i \phi} V_j U_j^* ).
\label{Seq:23}
\end{equation}

This expression can be further simplified by writing it in terms of 
$\alpha_j(t)$ as defined in equation (\ref{Seq:11}),

\begin{equation}
\frac{d}{dt} \left( \sinh^2 \alpha_j + \cosh^2 \alpha_j \right) = -4 g  \sinh \alpha_j \cosh \alpha_j \int_0^{2\pi} d\theta Im( \psi^2 u_j^* v_j ).
\label{Seq:24}
\end{equation}

Finally, using well known identities we arrive to 

\begin{equation}
\frac{d\alpha_j }{dt} =  -g \int_0^{2\pi} d\theta Im( \psi^2 u_j^* v_j ).
\label{Seq:25}
\end{equation}

This equation determines the evolution of the state norm in our problem. 
Let us point out that this equation is valid for arbitrary $t$ since both 
the dynamical term as well as the static potential in equation (\ref{Seq:5}) 
do not affect the state norm. Now equation (\ref{Seq:25}) can be used to 
understand some aspects of NNP evolution. Indeed, for $t=0$ all the variables
in the integral are known, namely,

\begin{equation}
\psi = \frac{1}{\sqrt{2 \pi}}, \hspace{0.5 cm} u_j = \frac{e^{ij\theta}}{\sqrt{2 \pi}}, \hspace{0.5 cm} v_j = \frac{e^{ij\theta}}{\sqrt{2 \pi}},
\label{Seq:26}
\end{equation}

from which it follows,

\begin{equation}
\left. \frac{d\alpha_j }{dt} \right |_{t=0} =  0.
\label{Seq:27}
\end{equation}

\begin{figure}
\begin{center}
\includegraphics[width=0.3\textwidth,angle=-90]{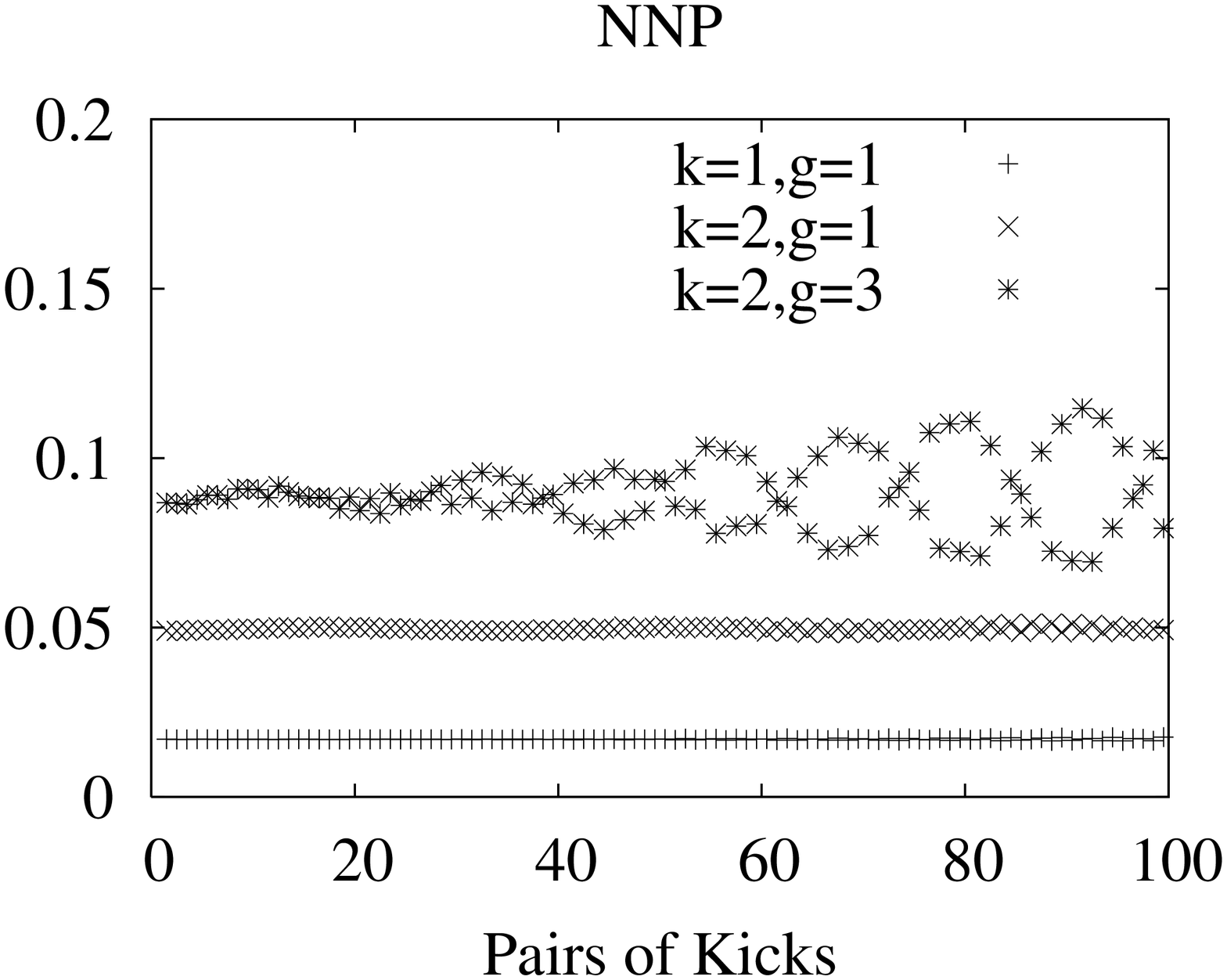}\includegraphics[width=0.3\textwidth,angle=-90]{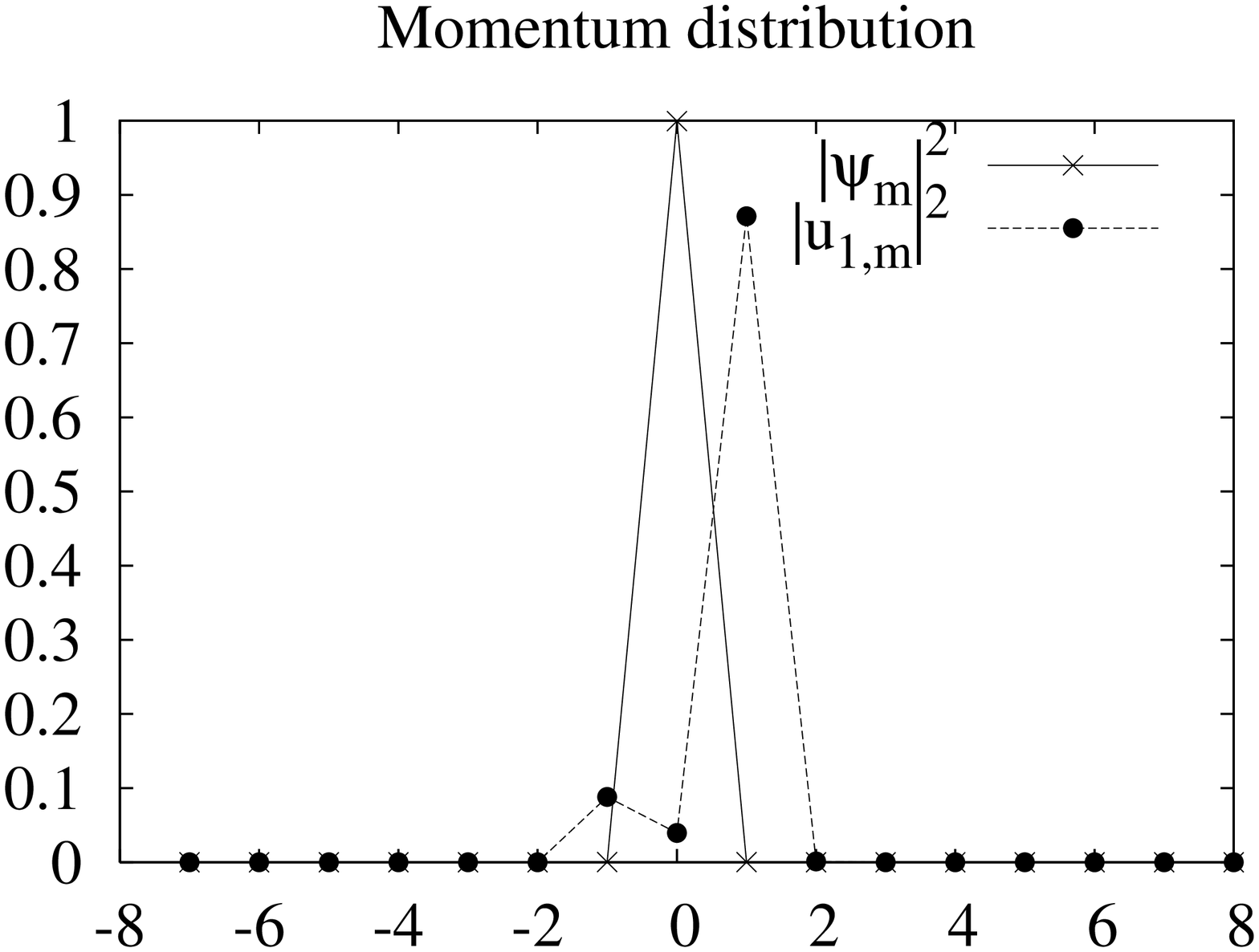}
\caption{Left. $\beta=1.96$, $\epsilon=0.04$. It is seen how NNP is hardly 
altered for weak kicking.  Right. $k=1$, $g=1$. Momentum distribution for 
$\psi$ and $u_1$ after 160 kicks (80 pairs of kicks). After several kicks 
the momentum distributions still look very similar to the initial ones. 
The condensate and the modes develop opposite phases and the integral 
in equation (\ref{Seq:25}) vanishes.} 
\label{fig:4}
\end{center}
\end{figure}

Consequently, as long as the kicking is not strong enough to drive the 
condensate and the modes out their initial states, the NNP will remain 
constant (not necessarily zero). This behaviour is positively corroborated 
by the graphs that we show in figure \ref{fig:4}-left, where the NNP is 
plotted against the number of kicks. As can be seen from this graph, 
dissipation is quite moderate even after several tens of kicks.

\begin{figure}
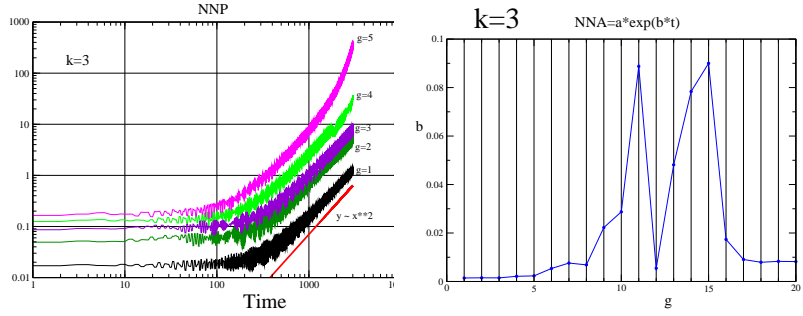

\begin{center}
\includegraphics[width=0.4\textwidth,angle=-0]{ho.eps}\includegraphics[width=0.38\textwidth,angle=-0]{stability2.eps}
\caption{Left. $\beta=1.96$, $\epsilon=0.04.$ In the strong repulsion regime 
NNP grows exponentially and the condensate becomes unstable.
Time is measured in numbers of pairs of kicks}. Right. 
Slope $b$ characterizes the intensity of particle leakage in the condensate.
As can be seen, diffusion peaks around some values of $g$, evidencing the
effect of resonances.
\label{ho}
\end{center}
\end{figure}

Figure \ref{ho} shows the behaviour of NNP for different values of the 
parameters $g$ and $k$. Initially, NNP does not respond to the kicking,
which corresponds to the time that the system takes to evolve away from
the initial conditions, in which case particle diffusion is given by 
equation (\ref{Seq:27}). Once this initial unresponsiveness is over, 
NNP starts to develop in a characteristic manner. In the small 
interaction regime, NNP-growth becomes linear with a slope equal to 2. 
Empirically, this dissipation is tolerable and the condensate can be
considered as stable \cite{Raizen1,Rancon}. In contrast, for $g=5$  
NNP-growth becomes 
exponential, which unequivocally characterizes the unstable phase. 
Figure \ref{ho}-right shows the slopes of several
curves of NNP as they are all fitted to exponential behaviour. Although
exponential behaviour may not be the most suitable fitting for every
curve, it allows us to put our results in a comparative perspective.
Figure \ref{ho}-right indicates that the system response is notably
intense around some sections of the horizontal axis. This strong response arises 
when the kick frequency coincides or is close to one of the natural 
frequencies of the condensate, which are given by $\omega_1$ and 
$\omega_2$ in equation (\ref{eq:24}) \cite{Dana}. This effect is analogous to
the resonances of mechanical systems, which occur when the driving
frequency coincides with one of the system's natural frequencies.
In this way, the strong resonance driving triggers particle
dissipation and hence destroys the condensate. We point out 
that this is different from instability due to linear resonances,
which take place when the contributions from different kicks 
interact coherently. In fact, these latter resonances are known to lead 
to chaos in the classical realm \cite{Fishman,Grempel,Stocklin,Creffield2}.

Here we have studied the wave function of a double-kicked BEC as well
as the stability of the system. We have derived analytical expressions
for the condensate wave function in the linear response regime. From 
the solution of the Gross-Pitaevskii equation, we identified resonance
conditions for which particle dissipation is maximum. Our findings were
verified by numerical simulations using a fast Fourier transform
algorithm. We also investigated the second quantized Hamiltonian 
and established useful formulae regarding the dynamics of
the NNP. Our study is relevant to experiments of cold atoms in 
optical lattices and our results are interesting from a
purely theoretical viewpoint.

\section{Advancements in the field}

Since the presentation of our results in 2007, there has been further
progress on the subject of condensate stability. Here we want to 
summarize the main results of these interesting works in order to 
provide a complete view of the actual state of the topic.

A complete characterization of the condensate stability has been
reported in \cite{Rancon}, including a map of the stability
parameters showing the position of the resonances for weak and
strong interactions. In such reference the authors study how the 
non-linear response due to the interaction among atoms affects
the well understood linear response, for which linear resonances
have been studied and characterized. In this latter case, it is
known that resonances are located at integer multiples of the
so called Talbot time, $T=4\pi$. It was found that the
non-linear dynamics not only displaces the positions of the
linear resonances but also affects the nature of the 
response. Specifically, it is shown that in the range 
$1 \le g \le 20$ the resonance response is characterized by a sharp
cut-off of the NNP and therefore the condensate revives after
it has lost stability. The non-linear resonances manifest 
individually for small values of $g$ and $k$, but they 
progressively proliferate and overlap as the parameters 
are slowly tuned up. Likewise, the results of simulations
using the GP equation are compared against the 
results of a second quantized model constructed from the
insight provided by previous studies of the kicked-rotor dynamics
\cite{Creffield3}. 
It is shown that by including Beliaev and Landau terms coming
from phonon-phonon interactions it is possible to reproduce 
the sharp fall in NNP close to a resonance as well as the shift 
in the position of the resonance in regimes where depletion of 
the ground state is $\le 10\%$. Conversely, the same approach
fails to reproduce the condensate dynamics when such terms
are not taken into consideration. The
second quantized model clarifies many issues regarding
the depletion process and helps understand how resonances
overlap and also how the cut-off process takes place.
In the second quantized approach the condensate revival
appears as a combination of Beliaev and Landau processes,
in a way that is very similar to non-linear self-trapping of
a BEC in a double-well potential. Certainly, the cut-off
phenomenon can be understood by the following first principle
analysis. When the condensate is being kicked slightly 
off-resonance, the interaction drives particles from the
condensate towards the first excited mode. This triggers 
particle migration to the second mode, which at the same
time induces a phase shift that brings the first mode closer 
into resonance. This is known as {\it non-linear feedback} and
it is specially responsible for the strong condensate 
depletion seen before the cut-off. However, if the 
kicking is not sufficiently close to resonance, the 
synchronization process among excited modes does not occur
and depletion is suppressed, which results in the abrupt
cut-off reported in reference \cite{Rancon}. Further inspection
suggests that the intensity of the cut-off is much sharper
than in double-well condensates. Moreover, the resonances
are classified in two basic categories. First, 
linear resonances, which derive from the Talbot resonances
and are seen to be parametrically displaced by the non-linearity. 
These lead to strong, high amplitude response. Second, non-linear 
resonances, which disappear when $g=0$. Such resonances 
can arise individually, as Bogoliubov modes, or in combination
of two or more modes. Non-linear response is weaker than
linear response, but surprisingly, a Lyapunov analysis 
reveals that non-linear resonances are unstable, contrary
to linear resonances. A possible explanation of this
is that non-linear resonances are linked to exponential
oscillations and therefore the Lyapunov stability criterion
becomes questionable. Finally, the stability profile could be 
used in applications where very sharp excitation thresholds are 
needed to carry out high accuracy measurements, as for example
measuring gravity or detecting small changes of frequency in
rotating BEC's.

\chapter*{Conclusions}

\addcontentsline{toc}{chapter}{Conclusions}

In this work we have adjusted recently developed numerical methods 
to probe the ground state as well as the dynamics of bosonic systems. 
In addition, we have introduced a numerical method that displays some 
advantages over standard approaches, and we applied it to solve
challenging problems.

In the first part of the document, we presented the concepts and
ideas that we subsequently used along most of the work. As numerics
is an important part of this research, we extensively discussed how
our programs were constructed and how they work. We then discussed 
alternative uses of MPS in cases where analytical solutions for the 
Heisenberg operators are available. The efficiency and reliability
of our method was tested, first in simple scenarios, and then in more
complex situations. The most important application of the method
in this work was to find the state of a system of bosons initially 
prepared in a separable state after a many particle collision has 
taken place in the centre of the chain. 
This calculation would have been much more difficult to 
do using TEBD or tDMRG as both of these methods are suitable only
for short-time intervals. Likewise, we also derived analytically the 
conditions for an efficient collection of bosons on the chain terminals
after the particle waves collide. This is an important point as
in the absence of controlling mechanisms boson waves flatten and
entanglement diffuses. Conversely, in our approach particles
are collected so that the entanglement contained in the system
as a result of the collision can be used. Our graphs show that the 
entanglement generated in this way is quite substantial and 
it is shown that such entanglement can be made to grow by just
adding more particles to the process.

We next focus on applications of TEBD to the BH model, aiming at enhancing 
the amount of entanglement between the ends of the chain, either in 
the ground state or as a result of dynamics. We show that EEE in
the ground state scales logarithmically with a positive coefficient
when a hopping profile with perfect transmission properties is
incorporated in the Hamiltonian. We conclude that such effect is
due to a marked change in the tunnelling profile, which undergoes 
a transition from local to global scope.
We use the real-time version of TEBD to simulate the dynamics
of a chain with high repulsion in intermediate sites. The amount
of entanglement generated out of dynamics alone is smaller than
what can be obtained from the ground state. We then use a
scheme that combines both ground state and dynamics to generate
entanglement. Our results indicate that it is in fact possible to
use dynamics to increase the amount of entanglement contained
in the ground state by taking advantage a perturbative scheme
that we have introduced.

In the last part we concentrate on kicked bosonic systems. We
utilized a perturbative approach to obtain an analytical expression
for the bosonic cloud in the linear response regime. Additionally, 
we derived useful formulae regarding the number of non-condensed
particles. We find that the condensate is highly depleted by a
kicking with a driving frequency that matches any of the natural 
frequencies of the system. Similarly, we have reviewed the latest 
advances in the field.

Among the several potential research extensions of the present work 
we would like to mention the possibility of studying chains with
periodic boundary conditions, where vortices are expected to appear.
Also, the numerical method presented here is liable of improvement.
Indeed, it seems plausible that the method can be formulated 
without the need for an explicit solution of the Heisenberg 
equations of motion for the operators. The procedure can be applied
to a wide variety of problems, not only in linear arrangements,
but also in lattices, since in our method the geometry of the system 
does not interfere with the reduction process. Potential extensions
of the method to spin and fermionic systems are quite direct. 
Additionally, it would be very interesting to implement the
configuration proposed in chapter \ref{cart} in an actual experiment,
since such configuration can be realized in optical
lattices using state-of-the-art technology. One of the main
conclusions of this work is the enhancement of entanglement in
chains with PTH hopping. Similarly, as repulsion constants are
turned up in the middle, we can think of intermediate places as
forming a quantum wire. Therefore, it should be interesting to study
a system made up of two particle-reservoirs connected by
a quantum wire. As the infinity-repulsion BH model is known to
derive in the $XX$ model, it is likely that a problem like this
can be solved analytically. If this is the case, it should be
possible to establish the actual relation between the hopping
profile and entanglement.

The present has been an interdisciplinary investigation that 
covered areas such as numerical methods, many-body systems, 
quantum information and so on. Our findings help understand how 
entanglement relates to some physical processes that occur in 
bosonic systems. Such understanding can be used in different ways,
as for example, to propitiate the circumstances that are convenient 
to produce entanglement in an experiment, or just as a way of appreciating 
the phenomenology of bosonic systems from a different perspective.
Altogether, we hope our contribution to be sufficiently relevant
so that it can inspire further investigation in any of the very
exciting scientific fields with which this research overlaps.

\chapter*{Appendix A: Publications}
\addcontentsline{toc}{chapter}{Appendix A: Publications}

\begin{itemize}

\item Jose Reslen and Sougato Bose, {\bf Long Range Free Bosonic Models in Block Decimation Notation: Applications and Entanglement}, arXiv:0907.4315. {\it Submitted}.

\item Jose Reslen and Sougato Bose, {\bf End-to-end entanglement in Bose-Hubbard chains}. {\it Physical Review A}, {\bf 80}: 012330, (2009).

\item J. Reslen, C.E. Creffield and T.S. Monteiro, {\bf Dynamical instability in kicked Bose-Einstein condensates: Bogoliubov resonances}. {\it Physical Review A}, {\bf 77}: 043621, (2008).

\end{itemize}

\bibliography{thesis}

\nocite{*}

\end{document}